%% file: main.tex
\documentclass[aip, pop, amsmath, amssymb, reprint,]{revtex4-1}

\usepackage{graphicx}
\usepackage{dcolumn}
\usepackage{bm}
\usepackage[utf8]{inputenc}
\usepackage[T1]{fontenc}
\usepackage{mathptmx}

\usepackage[dvipsnames]{xcolor}

\newcommand{\eg}{e.g., }
\newcommand{\ie}{i.e., }

\begin{document}

\title{RF plugging of multi-mirror machines}

\author{Tal Miller}
\affiliation{Racah Institute of Physics, The Hebrew University of Jerusalem, Jerusalem, 91904 Israel}
\affiliation{Rafael Plasma Laboratory, Rafael Advanced Defense Systems, POB 2250, Haifa, 3102102 Israel}

\author{Ilan Be'ery}
\affiliation{Rafael Plasma Laboratory, Rafael Advanced Defense Systems, POB 2250, Haifa, 3102102 Israel}

\author{Eli Gudinetsky}
\affiliation{Racah Institute of Physics, The Hebrew University of Jerusalem, Jerusalem, 91904 Israel}
\affiliation{Physics Department, Nuclear Research Center, Negev, PO Box 9001, Beer Sheva, Israel}

\author{Ido Barth}
\email{ido.barth@mail.huji.ac.il}
\affiliation{Racah Institute of Physics, The Hebrew University of Jerusalem, Jerusalem, 91904 Israel}

\date{\today}

\input{abstract}

\maketitle

\input{intro}

\input{body}
\input{conclusions}

\bibliographystyle{aipnum4-1}
\bibliography{references}

\end{document}

%% file: abstract.tex
\begin{abstract}

One of the main challenges of fusion reactors based on magnetic mirrors is the axial particle loss through the loss cones.
In multi-mirror (MM) systems, the particle loss is addressed by adding mirror cells on each end of the central fusion cell. 
Coulomb collisions in the MM sections serve as the retrapping mechanism for the escaping particles.
Unfortunately, the confinement time in this system only scales linearly with the number of cells in the MM sections and requires an unreasonably large number of cells to satisfy the Lawson criterion.
Here, it is suggested to reduce the outflow by applying a traveling RF electric field that mainly targets the particles in the outgoing loss cone.
The Doppler shift compensates for the detuning of the RF frequency from the ion cyclotron resonance mainly for the escaping particles resulting in a selectivity effect.
The transition rates between the different phase space populations are quantified via single-particle calculations and then incorporated into a semi-kinetic rate equations model for the MM system, including the RF effect.
It is found that for optimized parameters, the confinement time can scale exponentially with the number of MM cells, orders of magnitude better than a similar MM system of the same length but without the RF plugging, and can satisfy the Lawson criterion for a reasonable system size.

\end{abstract}

%% file: intro.tex
\section{Introduction}

Axial confinement is one of the main challenges for sustainable fusion in linear magnetically confined systems based on the mirroring effect.
These open trap systems are advantageous for their simplicity, high-$\beta$ (the ratio of plasma to magnetic pressures), and continuous mode of operations but suffer from radial instabilities and axial losses. 
Radial instabilities can be mitigated via passive, \cite{tajima1991instabilities, beklemishev2010vortex, ryutov2011magneto} radio frequency (RF),\cite{hershkowitz1982dynamic, ferron1983rf, ryutov2011magneto, seemann2018stabilization} or active \cite{zhil1975plasma, be2015feedback, ryutov2011magneto} methods, while solutions for the outgoing flux through the loss-cone commonly involve geometrical modifications of the mirroring magnetic field.
The variety of design concepts to reduce the loss-cone flux include tandem plugs with thermal barriers,\cite{inutake1985thermal, grubb1984thermal, katanuma1986thermal, pratt2006global, tamano1995tandem, ivanov2013gas, ivanov2017gas, anderson2020introducing, forest2022physics, egedal2022fusion}, diamagnetic confinement,\cite{beklemishev2016diamagnetic, kotelnikov2020structure} multi-mirrors (MM) systems,\cite{post1967confinement, logan1972multiple, logan1972experimental, mirnov1972gas, makhijani1974plasma, tuszewski1977transient, burdakov2016multiple, budker1971influence, mirnov1996multiple, kotelnikov2007new, mirnov1972gas} moving multiple mirrors,\cite{tuck1968reduction, budker1982gas}, helical mirror with rotating plasma, \cite{beklemishev2013helicoidal, postupaev2016helical, sudnikov2019first, ivanov2021long, sudnikov2022plasma}, ponderomotive RF plugging,\cite{motz1967radio, watari1974theory, hatori1975critical, watari1978radio, uehara1978radio, hiroe1978experiment, fader1981rf, fisch2003current, dodin2004ponderomotive, dodin2005ponderomotive} and plugging using field reversed configuration (FRC) at the mirror throats.\cite{shi2019magnetic}

MM systems are characterized by the two sections of multiple magnetic mirrors attached to each side of the central mirror cell, where fusion occurs.
As a particle escapes the central cell into the right or left MM sections, it can be collisionally scattered out of the loss cone of one of the MM cells and later be scattered again back into the central cell.
This process can be modeled as a one-dimensional diffusion dynamics, where we expect the axial flux to scale inversely with the number of cells in the MM section.
The collisionality, which depends on the temperature and density profiles along the system, determines the number of cells needed for reducing the loss-cone flux to a desired level.
Moreover, the collisionality in this scheme is required to be high enough \ie Coulomb mean free path of the order of the mirror cell length, resulting in a requirement for lower temperatures and higher densities, which are sub-optimal for fusion.\cite{post1967confinement, kotelnikov2007new, miller2021rate}
 
MM machines are commonly considered isothermal systems due to fast electron thermalization across the system,\cite{mirnov1972gas} resulting in a linear scaling of the confinement time with the system length.
In a recent study, we developed a semi-kinetic rate equations model for MM systems that divides the ions in each mirror cell into three populations (trapped, escaping, and returning) and includes the processes of Coulomb scattering within each cell and the transmission between neighboring cells.
The steady-state solution of the rate equations yields the density profile within the MM section and the outgoing axial flux.
It was found that the scaling of the outgoing flux with the system's size depends on the thermodynamic scenario.
The best confinement was obtained for isentropic systems, where the plasma adiabatically cools down with the density decreases within the MM section, and the mean free path drops along the MM section.
However, even in the most optimistic thermodynamic scenario, the scaling of the confinement time with the system's length requires an impractically large number of cells to satisfy the Lawson criterion.\cite{lawson1957some, miller2021rate, wurzel2022progress}
Therefore, new confinement enhancement methods should be developed in the quest for sustainable fusion in MM machines.

RF fields are widely used in magnetically confined plasma systems, including axial plugging of mirror systems by the aforementioned ponderomotive effect \cite{motz1967radio, watari1974theory, hatori1975critical, watari1978radio, uehara1978radio, hiroe1978experiment, fader1981rf, fisch2003current, dodin2004ponderomotive, dodin2005ponderomotive}, 
radial stabilization of magnetic mirror systems\cite{seemann2018stabilization, guo2005stabilization, jhang2005influence, kwon2005progress, zhu2019new, akimune1981dynamic, majeski1987effect, yasaka1984rf, yasaka1985rf}, stabilization of toroidal systems\cite{dippolito1987rf, knowlton1988stabilization, jin2020pulsed}, plasma heating,\cite{england1989power, diaz2001overview, li2011recent} and current drive\cite{hooke1984review, fisch1987theory}.

In this work, we suggest a new RF plugging method for MM systems. 
The idea is to apply an external RF electric field resonantly coupled with the ion cyclotron frequency in the moving frame of the outgoing particles.
To this end, we employ a radially rotating electric field with a frequency slightly detuned from the exact ions' Larmor frequency and with a non-zero axial wave vector such that the Doppler shift compensates for the resonance mismatch only for outgoing particles.
As a result, the RF field resonantly increases the perpendicular energy of the escaping particles only so they can be recaptured in the magnetic mirror cell. 
In contrast, it mildly affects the returning particles with the opposite axial velocities.
This selection effect yields a significant confinement enhancement effect, which we study via single particle simulations and the semi-kinetic rate equations model.

The structure of the paper is as follows.
Sec.~\ref{Fields configurations} introduces the considered static magnetic mirror field and the external RF field.
Sec.~\ref{single particle simulations} studies the effect of the RF fields on the time evolution of particles in phase space via single particle simulations.
Sec.~\ref{RF trapping} integrates the simulation results over the distribution of the fuel particles and evaluates the transition rates between different populations in phase space.
Sec.~\ref{Rate Equations Model and RF plugging} incorporates the effect of the RF fields into an extended rate equations model for MM systems and calculates the confinement enhancement resulting from the RF fields.
Sec.~\ref{conclusions} summarizes the findings.

%% file: body.tex
\section{Fields configurations}
\label{Fields configurations}

\begin{figure}[b]
    \includegraphics[clip, trim=0.0cm 0.0cm 1.cm 0.0cm,
    width=0.95\linewidth]{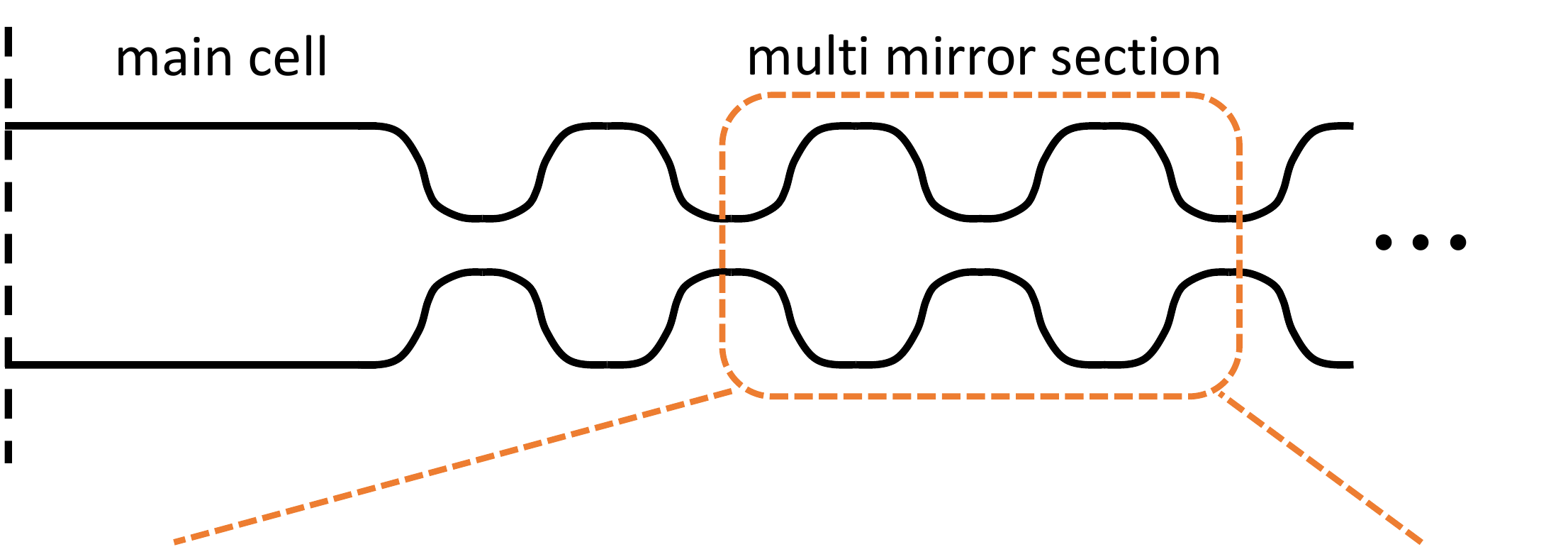}
    \includegraphics[clip, trim=.5cm 0.4cm .0cm 0.4cm,
    width=0.95\linewidth]{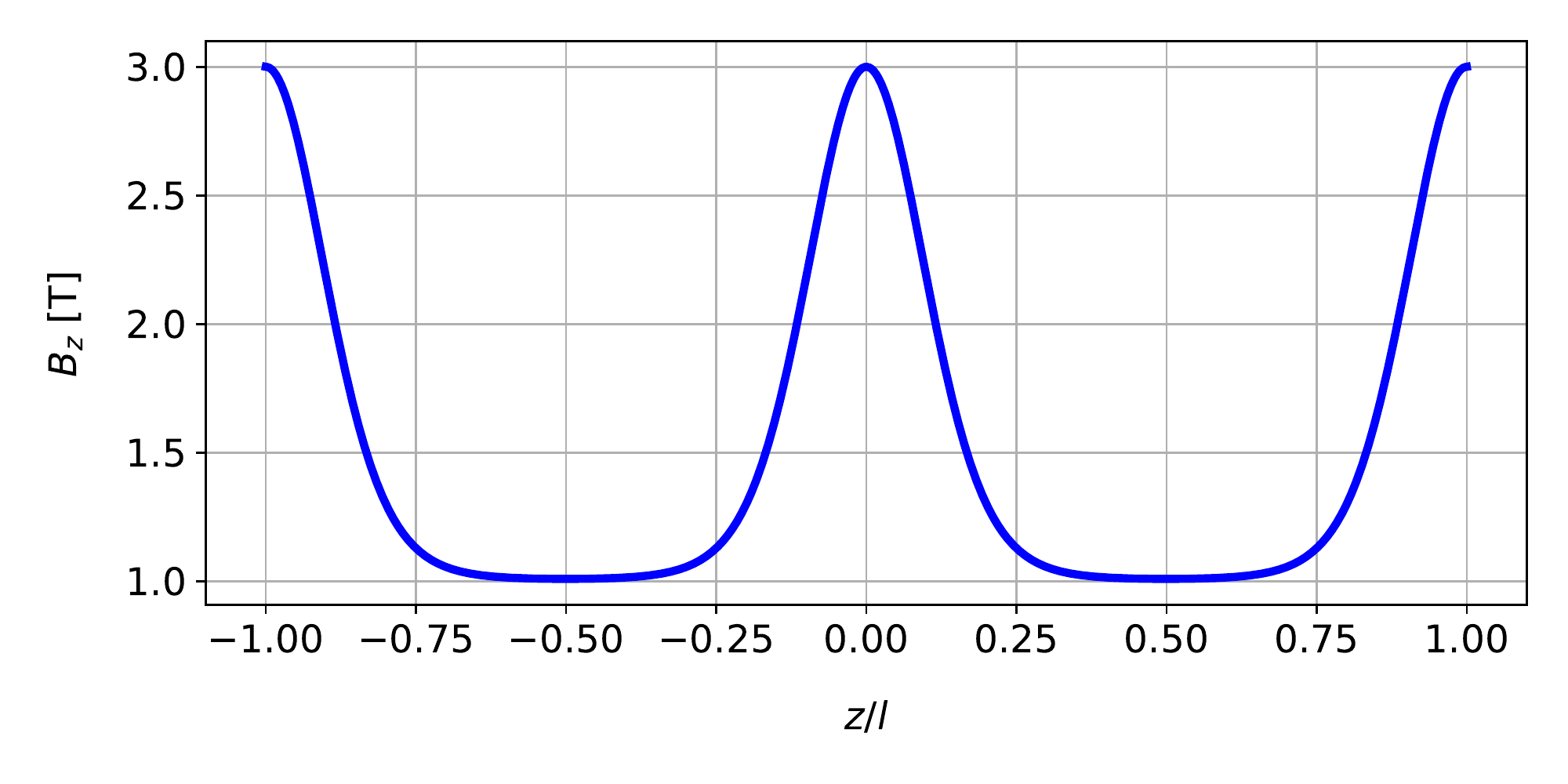}
    \caption{An illustration of one side of a MM system (top) and the axial magnetic field of two MM cells used in this study (bottom).}
    \label{fig: axial magnetic field prfile}
\end{figure}

The full MM system composes one central fusion cell and two MM sections.\cite{post1967confinement, logan1972multiple, logan1972experimental, mirnov1972gas, makhijani1974plasma, tuszewski1977transient, burdakov2016multiple, budker1971influence, mirnov1996multiple, kotelnikov2007new, mirnov1972gas}
To study the confinement improvement of adding the MM sections, we consider only one side of the MM sections, say the right one, and model it by a periodic magnetic field with an axial component of the form\cite{post1967confinement}
\begin{eqnarray}
    \label{eq: B_z}
    B_{z}=B_{0}\left[1+\left(R_{m}-1\right)\exp\left(-5.5\sin^{2}\frac{\pi z}{l} \right)\right]
\end{eqnarray}
\noindent
where $B_0$ is the minimal magnetic field,
$R_m = B_{max} / B_0$ is the mirror ratio, and $l$ is the length of each MM cell.
The other magnetic field components in cylindrical coordinates, ($r,\theta,z$), are $B_{\theta}=0$ and $B_{r}=-r\frac{\partial}{\partial z}B_{z}$, to satisfy $\nabla \cdot \mathbf{B} = 0$.
Fig.~\ref{fig: axial magnetic field prfile} illustrates the MM system and the axial component of the magnetic field, $B_z$.
The magnetic mirror trapping criterion is $\left( v_{\perp}/v \right)^2 > R_m^{-1}$, where
$v=\sqrt{v_z^2+v_\perp^2}$ is the total velocity with axial ($v_z$) and transverse ($v_{\perp}$) components that are measured in the minimum of $B_z$, \ie at the center of the magnetic mirror cell $(z=l/2)$. 
Therefore, manipulating the velocities ratio, $v_{\perp}/v$, for example, by RF fields, is an appealing path to enhance trapping, especially in MM systems where the effect on the pitch angle can be accumulated along the MM section.

The simplest approach is to apply an RF field near the ion cyclotron resonance on each MM.
This approach acts symmetrically on both right- and left-going particles.
However, in MM machines, an asymmetric drive would be preferable since one direction is inward (toward the fusion cell) while the other is outward (outside the system). 
Moreover, unlike in the central (fusion) cell, plasma heating in the MM sections does not contribute to the fusion rate.
It may be even deleterious for confinement as it reduces the collision rate, which is the salient re-trapping mechanism in standard MM systems.\cite{logan1972multiple, makhijani1974plasma, miller2021rate}
Therefore, it is favorable to find ways to deliver transverse energy to outgoing particles while minimizing plasma heating.

An avenue that, as far as we know, has not yet been explored in MM systems is to apply traveling RF fields that are resonant only with the outgoing particles due to the combination of frequency detuning and Doppler shift.
For each MM cell, we consider an electric field induced by external electrodes of the form
\begin{equation}
    \label{eq: E_RF}
    \mathbf{E} = E_{RF}\left[\cos\left(k_{RF} z-\omega_{RF} t \right)\hat{x}+\sin\left(k_{RF} z-\omega_{RF} t\right)\hat{y}\right]
\end{equation}
where, $E_{RF}$ is the electric field amplitude, $\omega_{RF}$ is the field's frequency, and $k_{RF}$ is the axial wave vector.
Since the relevant used parameters obey $\left(\frac{\omega_{RF} r}{c}\right)^{2}\ll1$ (see Sec. \ref{single particle simulations}),
we employ the electrostatic approximation and neglect all higher-order corrections to the electric and magnetic fields.
Nonetheless, we have verified this assumption numerically by including the first-order correction to the time-dependent magnetic field for some of the calculated trajectories, which resulted in negligible differences.

The rotating electric field travels along the MM cell in velocity  
$v_{RF}=\omega_{RF}/k_{RF}$.
Therefore, particles with axial velocity $v_z$ experience a Doppler-shifted RF frequency in their rest frame
\begin{equation}
    \label{eq: Doppler shift condition}
    \omega_{rest} = \omega_{RF} - k_{RF}v_z.
\end{equation}
For $k_{RF}=0$ (spatially uniform RF field), particles of all velocities experience the same driving frequency. 
However, applying external fields with $k_{RF}\ne0$ allows differentiating between out-going and in-going particles with opposite $v_z$ since the resonance condition, $\omega_{rest}=\omega_{cyc}$, where $\omega_{cyc}=qB/m$ is the ion cyclotron frequency, depends on the particle velocity, $v_z$.
For example, applying driving field with $k_{RF}<0$ and $\omega_{RF}<\omega_{cyc}$ results in Doppler compensation for particles with positive axial velocity, $v_z>0$, such that they can meet the resonance condition, $\omega_{rest}=\omega_{cyc}$, for suitable values of $v_z$.
In contrast, the driving frequency in the rest frame of particles with $v_z<0$ is down-shifted and therefore gets farther away from resonance for the same driving parameters.
Similarly, if we pick $k_{RF}>0$ and $\omega_{RF}>\omega_{cyc}$, particles with $v_z>0$ approach the resonance, while those with $v_z<0$ get farther away. 

Naively thinking, both scenarios would exhibit a similar selectivity effect between ingoing and outgoing particles.
Nonetheless, in the discussion above, we considered particles in a fixed location, \eg in the mirror midplane where the magnetic field is minimal. 
But as a particle moves away from the center, $\omega_{cyc}$ increases while $|v_z|$ decreases with the amplitude of the axial magnetic field that increases when the particle approaches the mirror throat.
Namely, For $k_{RF}>0$ and $\omega_{RF}>\omega_{cyc}$, particles with $v_z<0$ also approach resonance, resulting in a reduction in the desired selection effect.
On the other hand, when $k_{RF}<0$ and $\omega_{RF}<\omega_{cyc}$ only particles with $v_z>0$ can experience Doppler compensation.
Therefore, the latter scenario is more favorable for achieving selectivity. 
In Sec.~\ref{single particle simulations}, we explore this selectivity effect analytically and numerically.
Interestingly, similar compensation between density gradient and frequency detuning for maintaining a selective and directional resonance condition while suppressing unwanted noise amplification was employed in Raman amplifiers.\cite{malkin2000detuned}

Finally, we note that the form of the rotating RF field defined in Eq.~(\ref{eq: E_RF}) is a simplified model for a real RF field configuration inside the plasma. 
In future work, we may study more complex yet more realistic electric and magnetic fields, for example, fields generated by the different types of Nagoya coils\cite{watari1978radio} or helicon antennas\cite{light1995helicon, miljak1998helicon}, including the effects of plasma screening and field penetration.

\section{single particle simulations}
\label{single particle simulations}

To quantify the RF effect on particles in the mirror system, we employ the single-particle approximation and perform a Monte-Carlo analysis, where the initial velocities were sampled from a thermal distribution with an ion temperature of $k_{B} T_i=10 \mathrm{keV}$ and a random direction.
We consider one MM cell with axial magnetic field given in Eq.~(\ref{eq: B_z}), where, $l=1\mathrm{m}$, $B_0=1\mathrm{T}$, and $R_m=3$.
Neglecting collisions and collective effects, we calculate the trajectories of deuterium (D) and tritium (T) ions under the influence of the Lorentz force $\mathbf{F}=q\left(\mathbf{E}+\mathbf{v}\times\mathbf{B}\right)$.
The static magnetic field and the time-dependent RF electric field are given in Eqs.~(\ref{eq: B_z}) and~(\ref{eq: E_RF}), respectively, while we used a symplectic scheme\cite{he2015volume}, which preserves phase space volume.
We calculate for both fusion species because a specific RF field might affect the two hydrogen isotopes differently due to the mass difference, yet both would exist in a D-T fusion reactor.

First, let us demonstrate the dynamics of a charged ion in velocity space under the influence of a traveling rotating electric field and the resulting selectivity effect for four sets of parameters, detailed in Table \ref{table: RF parameters}. 
For all the parameter sets, we picked the RF field amplitude $E_{RF}=50\mathrm{kV/m}$, which is large but realistic and results in an appreciable effect, as will be shown below. 
The values of $\omega_{RF}$ and $k_{RF}$ in the different sets were chosen to represent typical scenarios of the diverse dynamical behavior of the system. 
In Table \ref{table: RF parameters}, the frequencies, $\omega_{RF}$, are normalized by the ion cyclotron frequencies of deuterium and tritium at the mirror midplane, $\omega_{0,D} = eB_0/m_D = 4.75 \cdot 10^7 \mathrm{s}^{-1}$ and $\omega_{0,T} = eB_0/m_T = 3.17 \cdot 10^7 \mathrm{s}^{-1}$, respectively.
The wave numbers, $k_{RF}$, are given in the units of $2\pi m^{-1}$.
\begin{table}[t]
    \medskip{}
    \centering{}
    \setlength{\tabcolsep}{4pt}
    \def\arraystretch{1.4}
    \begin{tabular}{||c || ccc || cccc||c||}
        \hline
        \textbf{set} & $\frac{\omega_{RF}}{\omega_{0,D}}$ & $\frac{\omega_{RF}}{\omega_{0,T}}$ & $\frac{k_{RF}}{2\pi\mathrm{m}^{-1}}$ & $\bar{N}_{rc}$ & $\bar{N}_{lc}$ & $\bar{N}_{cr}$ & $\bar{N}_{cl}$ & $s=\frac{\bar{N}_{rc}}{\bar{N}_{lc}}$ \\
        \hline
        \hline
        1 (D) &  1.12  &  1.68  &  3.0  &  0.31  &  0.39  &  0.02  &  0.02  &  0.8 \\
        1 (T) & & & & 0.63  &  0.16  &  0.02  &  0.03  &  3.8 \\
        \hline
        2 (D) &  0.80  &  1.20  &  -3.0  &  0.64  &  0.12  &  0.01  &  0.02  &  5.3 \\
        2 (T) & & & & 0.38  &  0.40  &  0.03  &  0.02  &  0.9 \\
        \hline
        3 (D) &  0.56  &  0.84  &  -3.0  &  0.68  &  0.08  &  0.02  &  0.02  &  9.0 \\
        3 (T) & & & & 0.59  &  0.14  &  0.02  &  0.03  &  4.2 \\
        \hline
        4 (D) &  0.44  &  0.66  &  -7.0  &  0.50  &  0.05  &  0.02  &  0.02  &  10.7 \\
        4 (T) & & & & 0.36  &  0.06  &  0.01  &  0.02  &  6.2 \\
        \hline
        \hline
        \end{tabular}
    \caption{The four sets of RF parameters (frequency and k-vector) in this work. 
    The rows denoted by D and T refer to the isotope-dependent saturation values of the transition probabilities, $\bar{N}_{rc,lc,cr,cl}$, and to the selectivity parameter, $s$, calculated in Sec.~\ref{RF trapping}.}
    \label{table: RF parameters}
\end{table}
In the simulations, all particles were initialized at the center of a magnetic mirror cell (minimum axial magnetic field).
The simulation time was twice the characteristic passage time of the slower isotope, tritium, across one mirror cell, $2 l/v_{th,T}$, where $v_{th,T}=\sqrt{2 k_B T / m_T}=7.97 \cdot 10^5 \mathrm{m/s}$ is the thermal velocity of tritium. 

\begin{figure}[]

    \quad
    \includegraphics[clip, trim=.2cm .2cm .2cm 0.2cm, 
    width=0.98\linewidth]{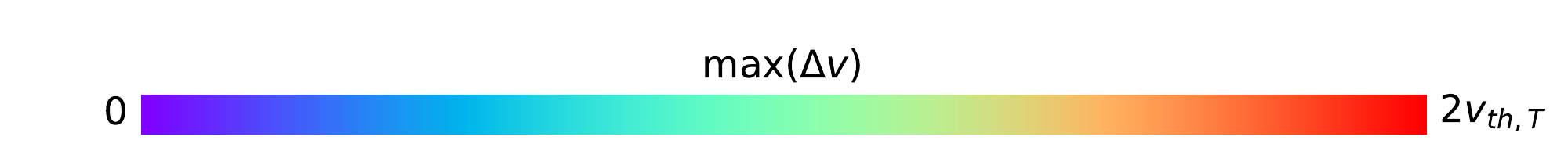}
    
    \includegraphics[clip, trim=.2cm .2cm .2cm 0.2cm, 
    width=0.49\linewidth]{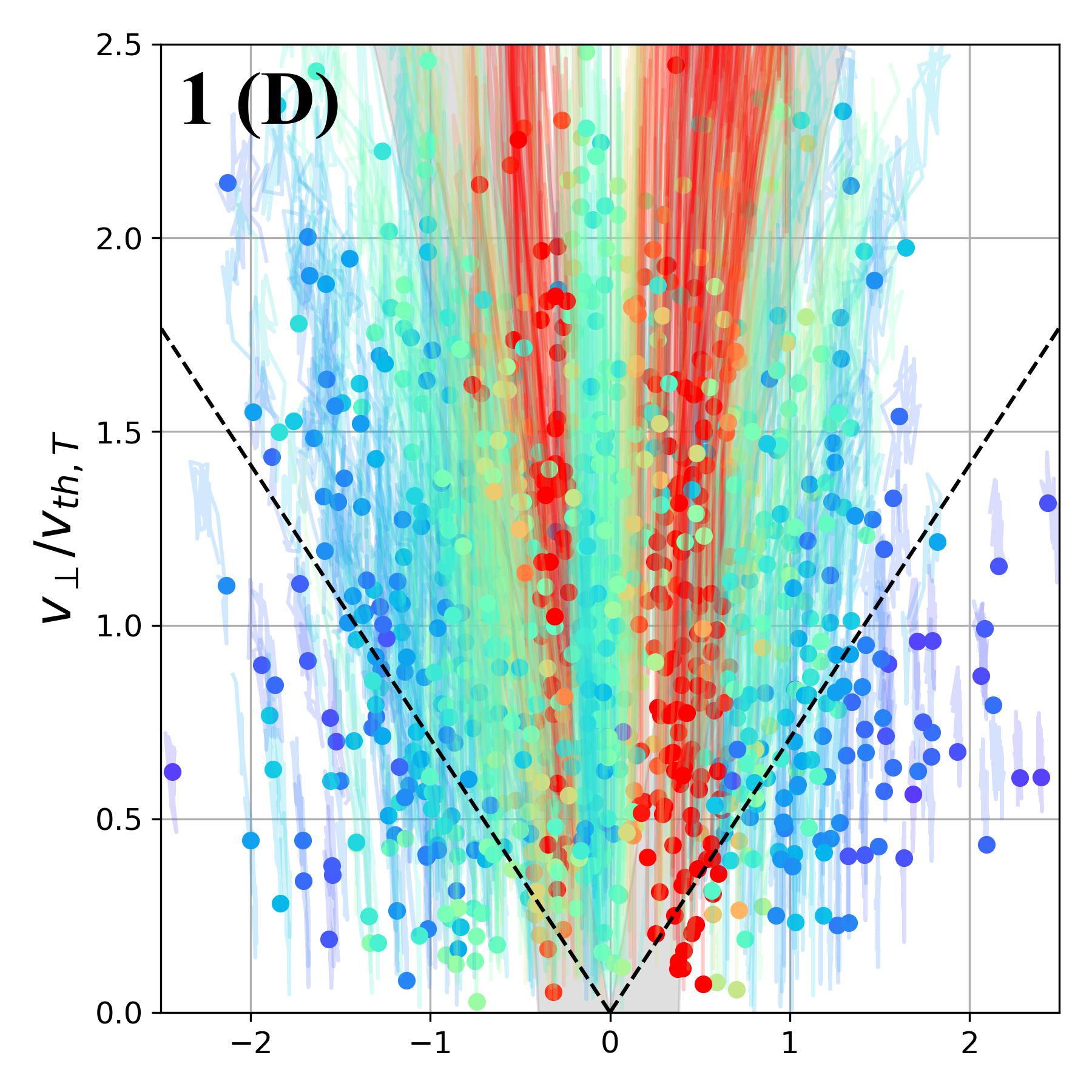}
    \includegraphics[clip, trim=.2cm .2cm .2cm 0.2cm, 
    width=0.49\linewidth]{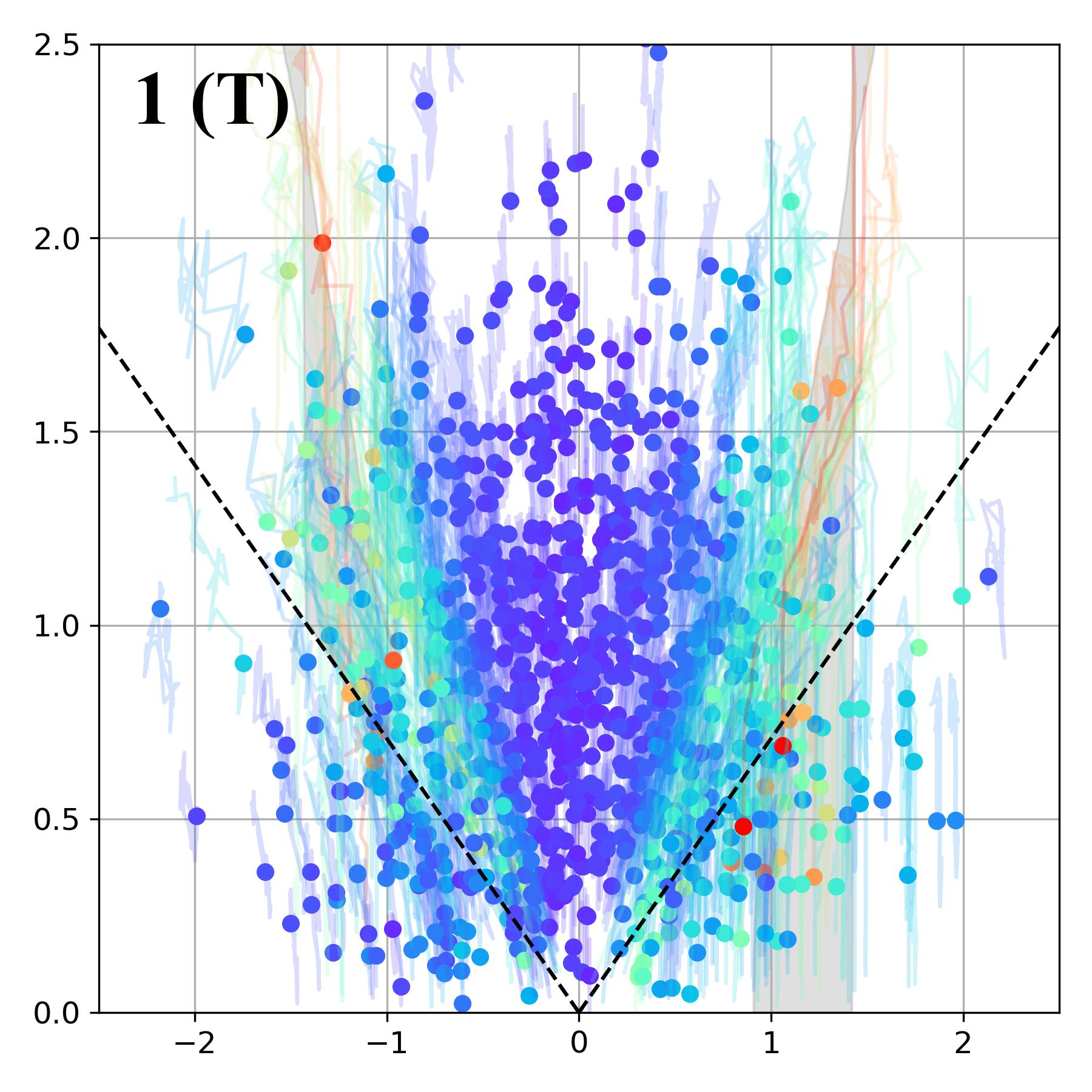}

    \includegraphics[clip, trim=.2cm .2cm .2cm 0.2cm, 
    width=0.49\linewidth]{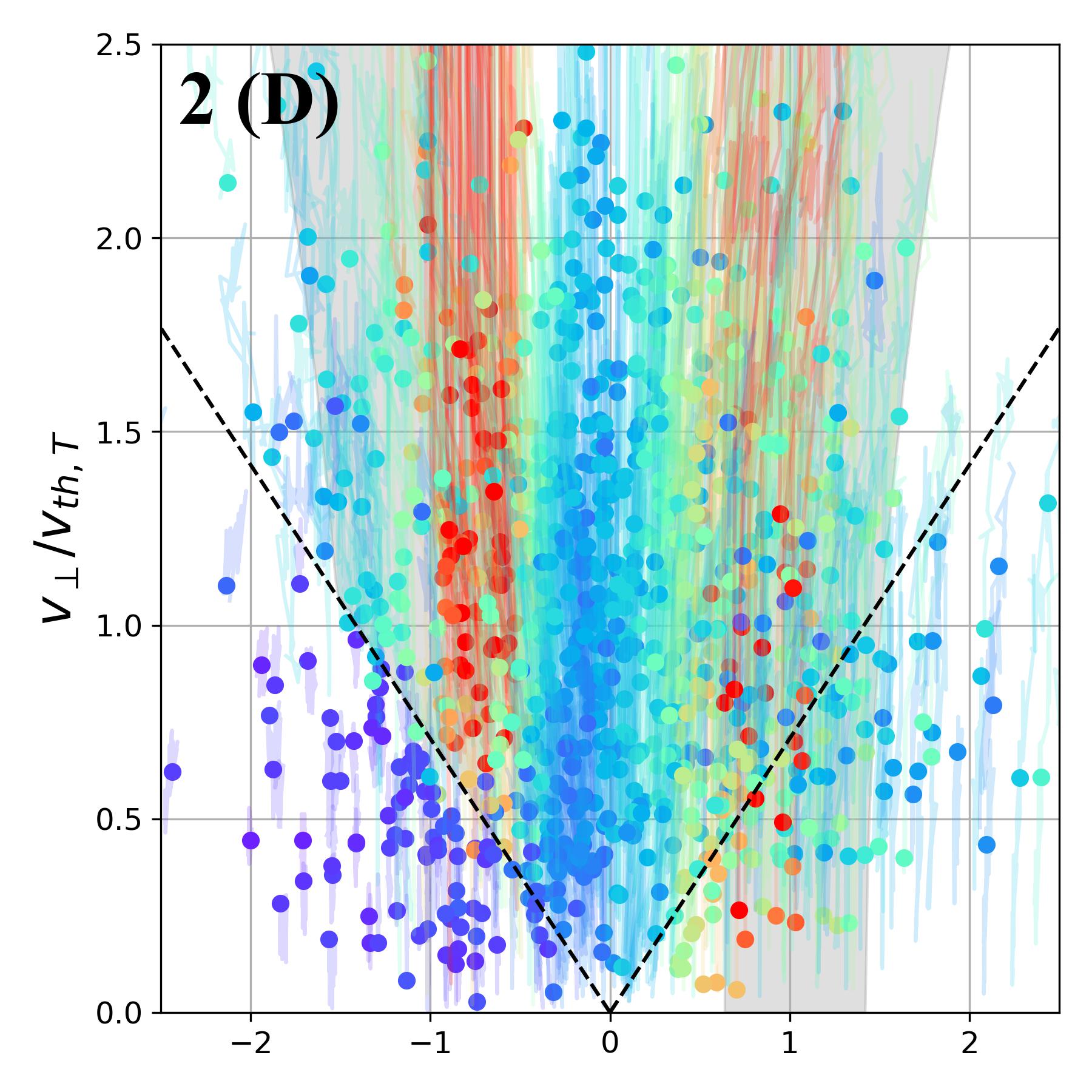}
    \includegraphics[clip, trim=.2cm .2cm .2cm 0.2cm, 
    width=0.49\linewidth]{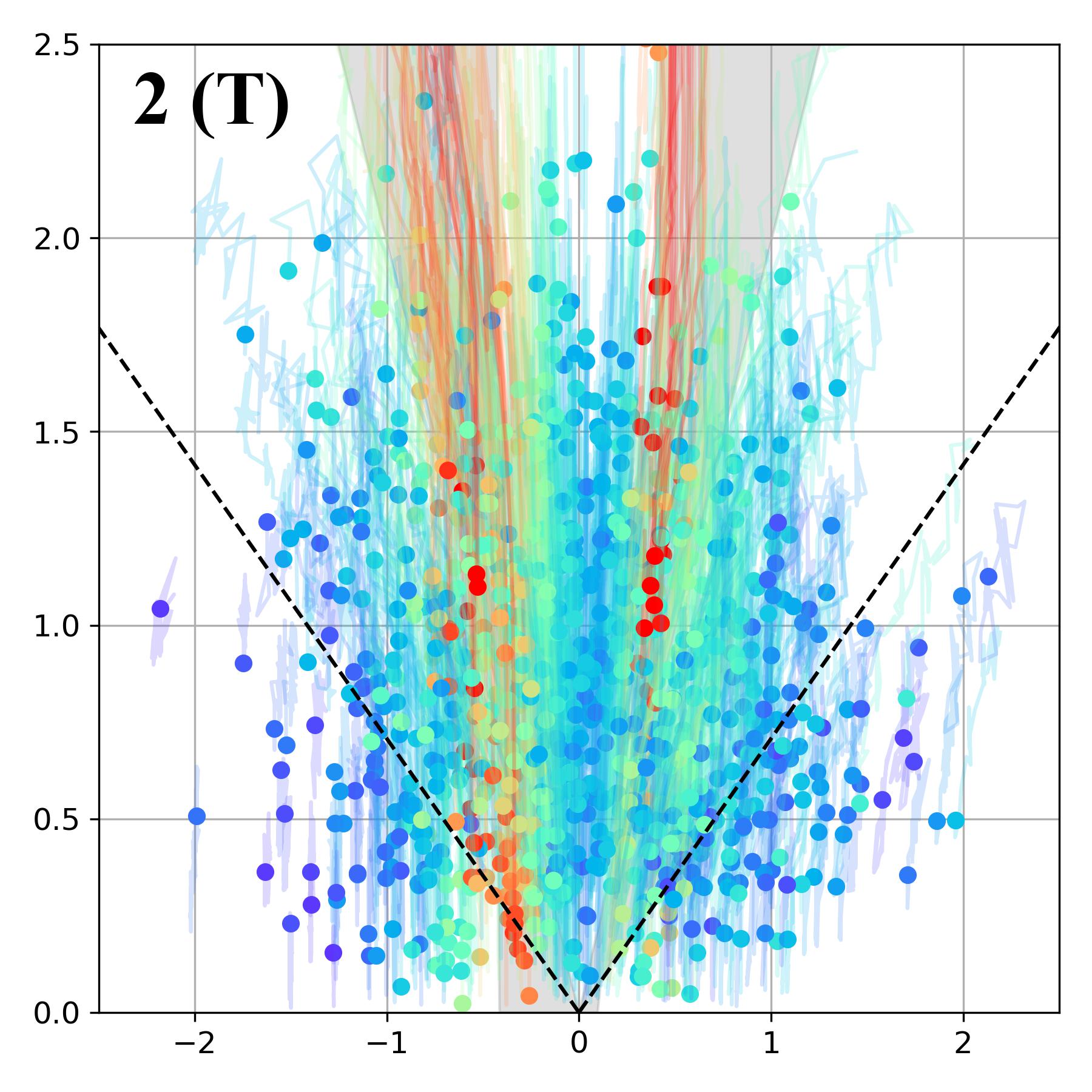}

    \includegraphics[clip, trim=.2cm .2cm .2cm 0.2cm, 
    width=0.49\linewidth]{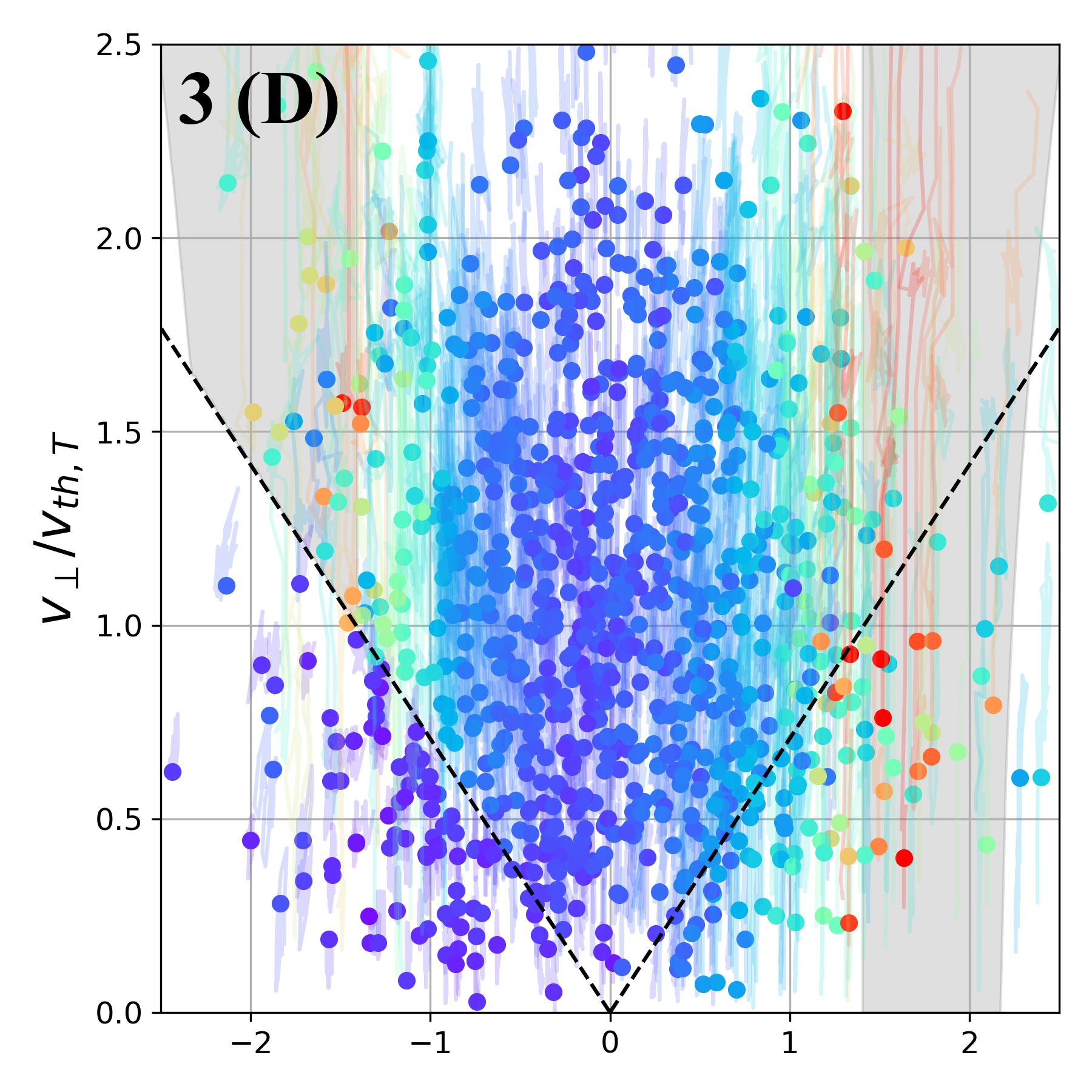}
    \includegraphics[clip, trim=.2cm .2cm .2cm 0.2cm, 
    width=0.49\linewidth]{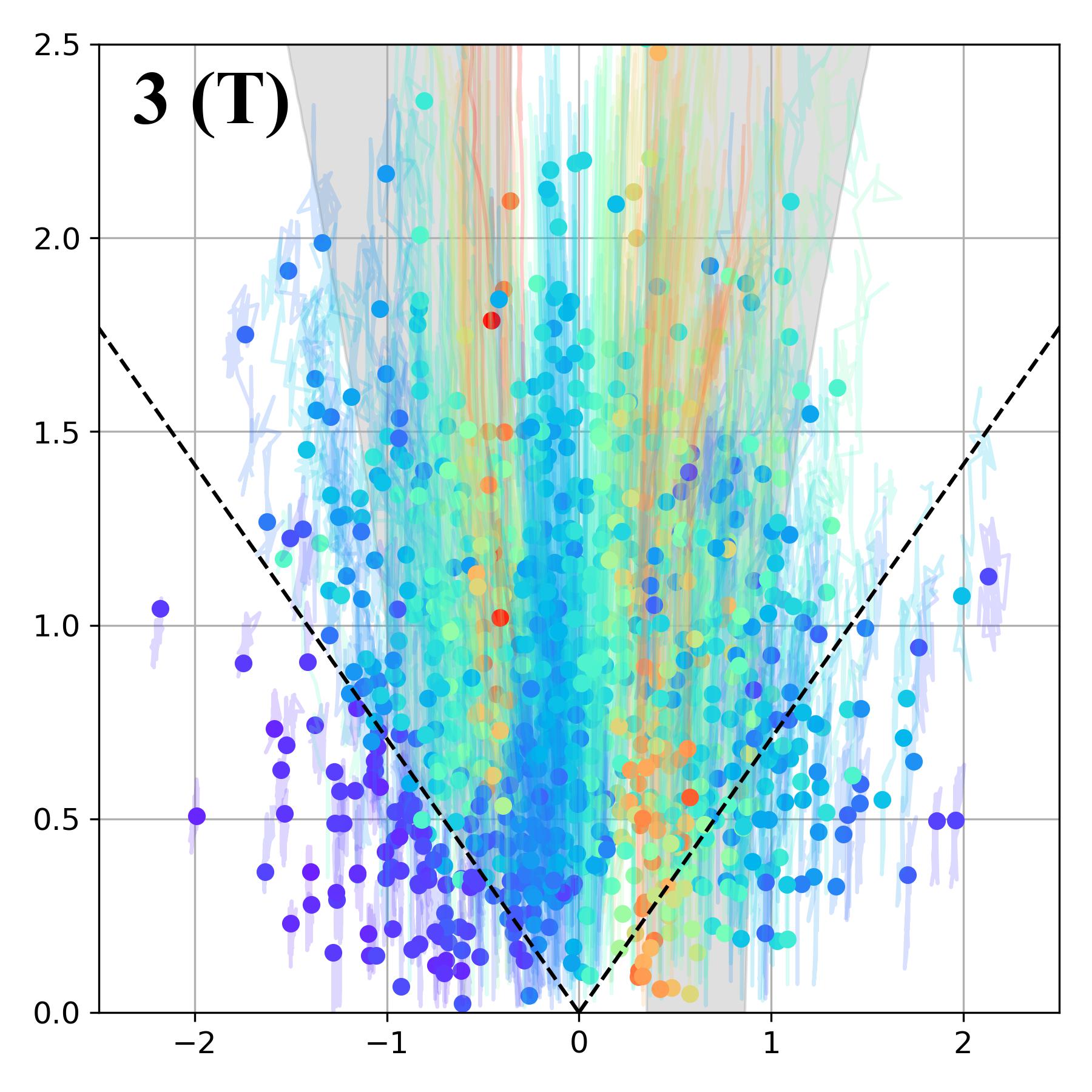}

    \includegraphics[clip, trim=.2cm .2cm .2cm 0.2cm, 
    width=0.49\linewidth]{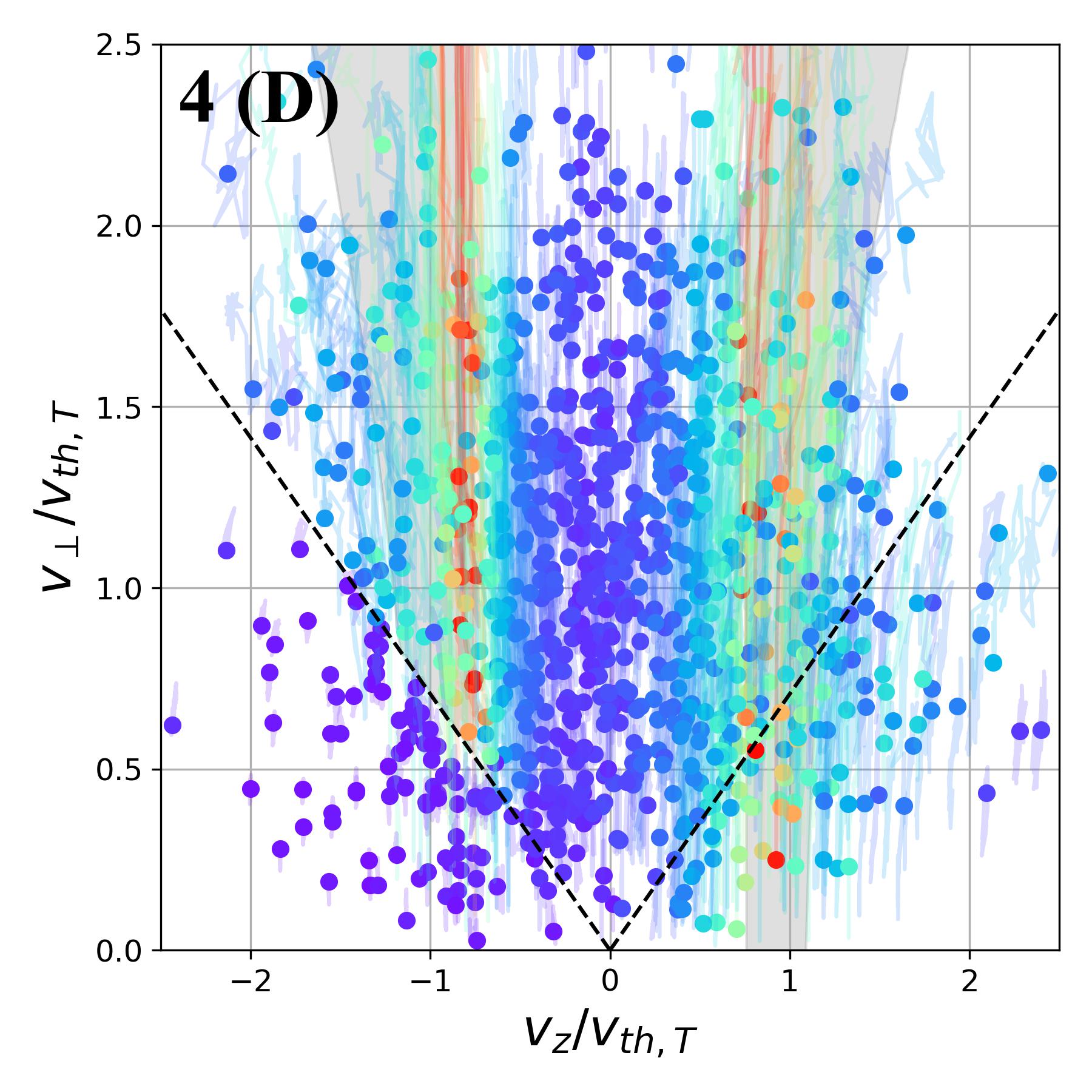}
    \includegraphics[clip, trim=.2cm .2cm .2cm 0.2cm, 
    width=0.49\linewidth]{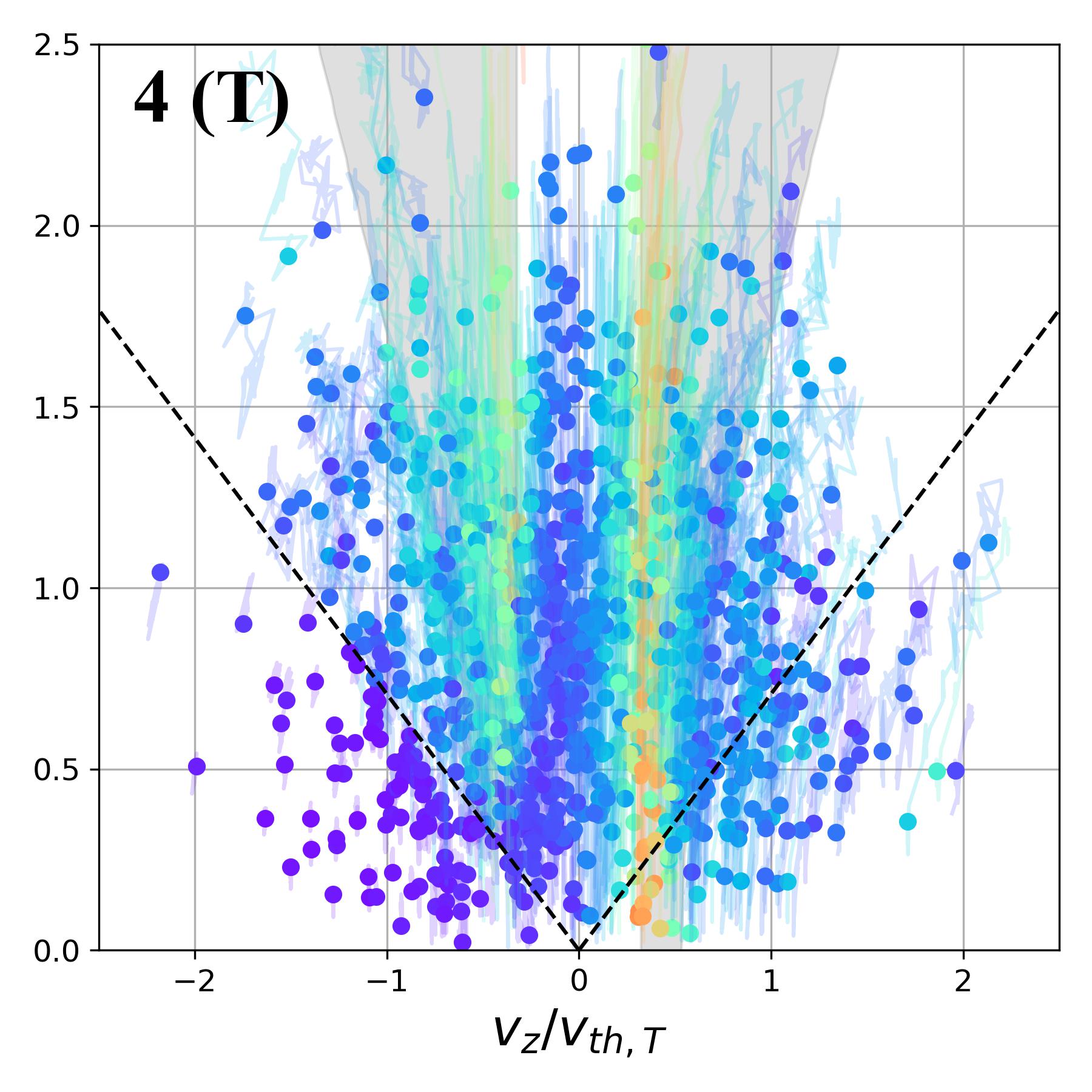}

    \caption{single particle simulations for 1000 different initial conditions sampled from the Maxwell-Boltzmann distribution, according to Eq.~(\ref{eq: velocity transformation}). 
    The upper-left text in each panel indicates the set of RF parameters used (see Table \ref{table: RF parameters}) and particle type used, deuterium (D) or tritium (T). 
    The color (rainbow colormap going from blue to green to red) of each initial condition point represents the magnitude of the RF effect as defined in Eq.~(\ref{Eq: Delta v max}).
    The grey areas are the resonant regions found in Sec.~\ref{single particle simulations}.
    }
    \label{fig: velocity space evolution}
\end{figure}

Since in the absence of the external RF field, the magnetic moment of both bouncing and passing particles is adiabatically conserved, the transverse and axial degrees of freedom exchange energy such that the total energy is conserved.
Therefore when the particle experiences a field $B$ at a position $z$, its perpendicular and axial velocity components, $v_{\perp}$ and $v_{z}$, respectively, can be mapped into those at the center of the mirror, $\tilde{v}_{\perp}$ and $\tilde{v}_{z}$, where the magnetic field is $B_0$ via the transformation,
\begin{eqnarray}
    \tilde{v}_{\perp} &=& v_{\perp}\sqrt{B_{0}/B} \notag \\
    \tilde{v}_{z} &=& s \sqrt{v_{z}^{2}+v_{\perp}^{2}\left(1-B_{0}/B\right)}.
    \label{eq: velocity transformation}
\end{eqnarray}
\noindent Here, we defined $s=\mathrm{sign}\left(v_{z}\right)$ to indicate the initial axial direction.
Although in the presence of an external RF field, this transformation is not precise, we use this approximated transformation in Fig.~\ref{fig: velocity space evolution} to visualize the particle dynamics in velocity phase space under the influence of the RF.
In each panel of the figure, we plot the dynamics in the projected mid-plane, $(\tilde{v}_{z},\tilde{v}_{\perp})$, for many different initial velocities.
The different panels are associated with the different sets of RF parameters of Table \ref{table: RF parameters} for both deuterium (D) and tritium (T).
The dashed black lines indicate the approximated loss cones boundaries, $\tilde{v}_{\perp}/\tilde{v}_{z}=\left(R_{m}-1\right)^{-1/2}$.
The large dots represent the initial velocities $v_{\perp,0}, v_{z,0}$ and the faint lines connected to them depict the time-evolution in the projected mid-plane according to the transformation in Eq.~(\ref{eq: velocity transformation}). 
The color indicates how much the RF fields affect each particle during the simulation time, according to the metric defined as the maximal displacement in the projected velocity plane ($\tilde{v}_{z},\tilde{v}_{\perp}$)
\begin{eqnarray} \label{Eq: Delta v max}
	\Delta v_{max}=\max \left(\sqrt{\left(\tilde{v}_{z}(t) - v_{z,0}\right)^{2}+\left(\tilde{v}_{\perp}(t)-v_{\perp,0}\right)^{2}}\right)
\end{eqnarray}
where the maximum is taken over the simulation time.
In the figure, one can see that the maximal effect (red dots) in all figures is focused mainly around vertical lines around a specific axial velocity $v_z$, associated with the Doppler shifted ion-cyclotron resonance.

Next, we analytically analyze the Doppler compensation mechanism and develop a simple model for the resonant region of initial conditions in the $(v_{z,0},v_{\perp,0})$ plane.
The idea is that particles will be most affected by the RF if they happen to be at a location $z$ with an axial velocity, $v_z$ such that in their rest frame, the Doppler shifted RF field, $\omega_{rest}$, which is defined in Eq.~(\ref{eq: Doppler shift condition}), meets the resonance condition,
\begin{eqnarray}
    \omega_{rest}=\omega_{c}\left(B\right)
\end{eqnarray}
where, $\omega_{c}\left(B\right)=\omega_{c,0}B/B_{0}$ depends on $z$ through Eq.~(\ref{eq: B_z}). 
Here, we neglect the radial components of the magnetic fields that are small for particles near the mirror's axis.

As a particle travels down the mirror, its Larmor frequency increases with the axial mirror field within the range $1\le B/B_0 \le R_m$, but the Doppler shifted RF frequency also changes since it depends on the particle velocity.
Similarly to the transformation (\ref{eq: velocity transformation}), from the conservation of energy and magnetic moment, one finds,
\begin{eqnarray}
   v_{\perp}\left(B\right) &=& v_{\perp,0}\sqrt{B/B_{0}} \\
    v_{z}\left(B\right) &=& s_0\sqrt{v_{z,0}^{2}+v_{\perp,0}^{2}\left(1-B/B_{0}\right)} 
    \label{Eq: v_z}
\end{eqnarray}
where, $s_0=\text{sign}\left(v_{z,0}\right)$.
Therefore, the approximated condition for resonance to take place at some field $B$ reads
\begin{eqnarray}
    \omega_{RF}-k_{RF}v_{z}\left(B\right)=\omega_{c}\left(B\right)=\omega_{c,0}\frac{B}{B_{0}}.
\end{eqnarray}%
Substituting $v_z(B)$ from Eq.~(\ref{Eq: v_z}) and squaring the resonance condition yields a quadratic equation for $B$.
An acceptable solution exists if the determinant is non-negative and at least one of the roots falls in the range $1\leq B/B_{0}\leq R_{m}$. 
It is noted that the interaction time with the RF is larger near the mirror center, where the magnetic field gradient is smaller. 
Therefore, we further constrain the effective resonant region to the range $1\leq B/B_{0}\leq 1.25$ even though in our simulations $R_m=3$.

Note that for a particle with initial velocities in the loss cone, we also require that the resonant axial velocity is in the same direction as the initial axial velocity because such an escaping particle does not have the opportunity to reverse its direction.
In other words, the frequency compensation depends on the direction of the escaping particles (inward or outward in the MM system), giving rise to selectivity.

In Fig.~\ref{fig: velocity space evolution}, we colored (in grey) the acceptable resonant regions in mirror mid-plane velocity space.
Notably, the theoretical resonant regions fit well with the single-particle simulation results, meaning all the most affected particles (colored red) are inside the theoretical (gray) regions. 
However, the magnitude of the RF effect in velocity space, and even more importantly, the RF trapping probability, are not predictable by this simple model and thus will be addressed numerically in the next section.

\section{RF trapping probability}
\label{RF trapping}

At any given time, each particle can be characterized as being inside or outside  the loss cones by checking the local loss cone condition $\left( v_{\perp}/v \right)^2 < B_{max} / B_z$, where $B_z$ is the local axial mirror magnetic field at the particle's location. 
Here, we assume that the particles are near the vicinity of the mirror axis, so the radial field components are small.
Of course, this condition is valid only when there is no external RF, so energy and magnetic moment are conserved.
In other words, it correctly determines the particle's final state if the RF were to be instantly turned off.
We divide the particles into three populations by their initial conditions being either trapped particles or, right or left, non-trapped particles,  where right means escaping outside the MM system, and left means the opposite, \ie towards the central fusion cell (see Fig.~\ref{fig: axial magnetic field prfile}).
We define the number of particles in each population as $N_{c}$ for captured, $N_{r}$ for right-going, and $N_{l}$ for left-going.
Then we track each particle's identity as a function of time \ie whether or not it crosses one of the loss cone lines, as depicted in Fig.~\ref{fig: velocity space evolution} (dashed lines).

The critical parameter required to estimate the overall efficiency of RF plugging in MM machines is the number of "converted" particles, say, that originated at the right loss cone but ended up trapped. 
We define four transitions: $N_{r \rightarrow c}$, $N_{l \rightarrow c}$, $N_{c \rightarrow r}$, and $N_{c \rightarrow l}$.
The other two possible transitions, left to right and right to left, were negligible under the influence of the external RF field, and therefore are omitted from the generalized rate equation model (see Sec.~\ref{Rate Equations Model and RF plugging}).

\begin{figure}[t]
    \includegraphics[clip, 
    width=0.49\linewidth]{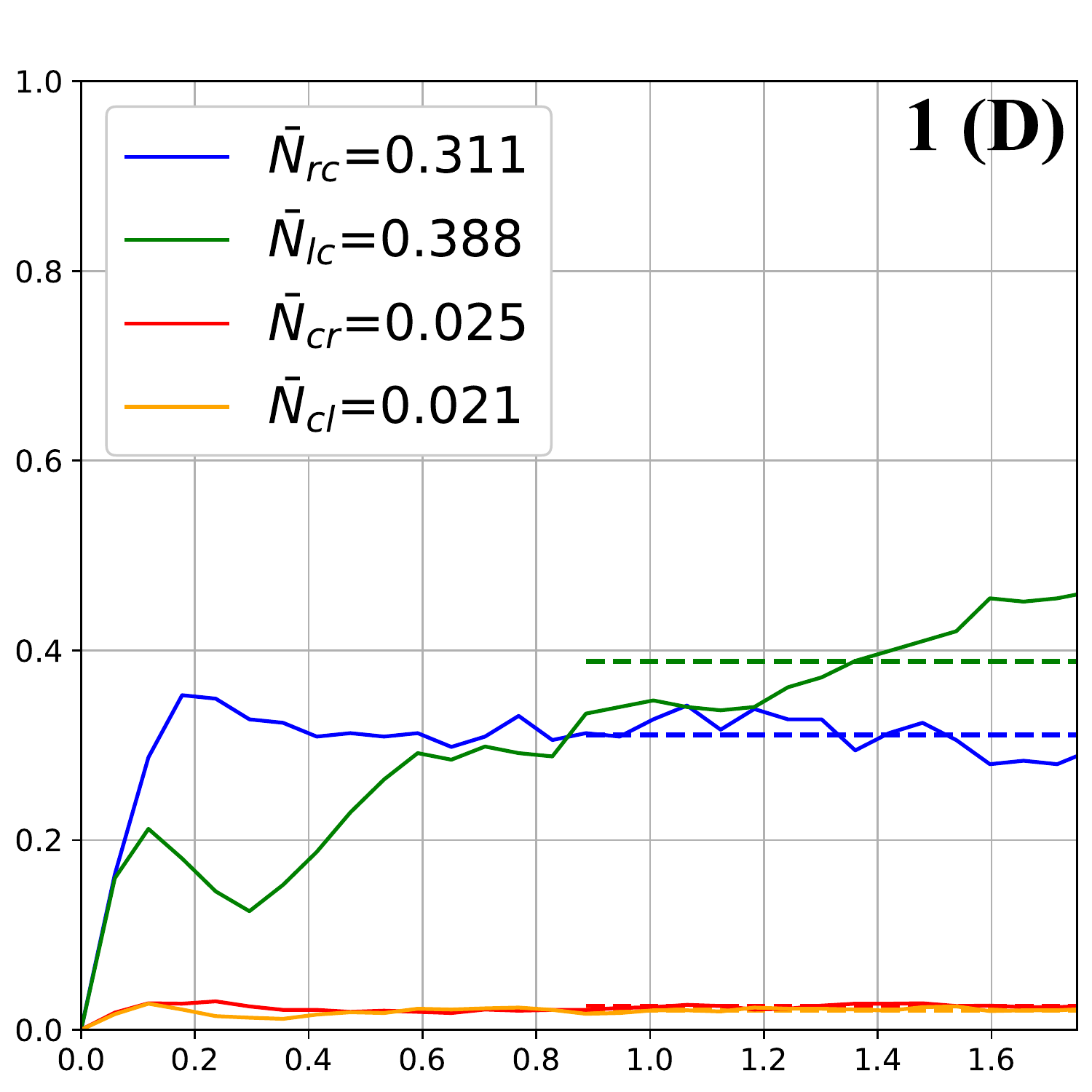}
    \includegraphics[clip, 
    width=0.49\linewidth]{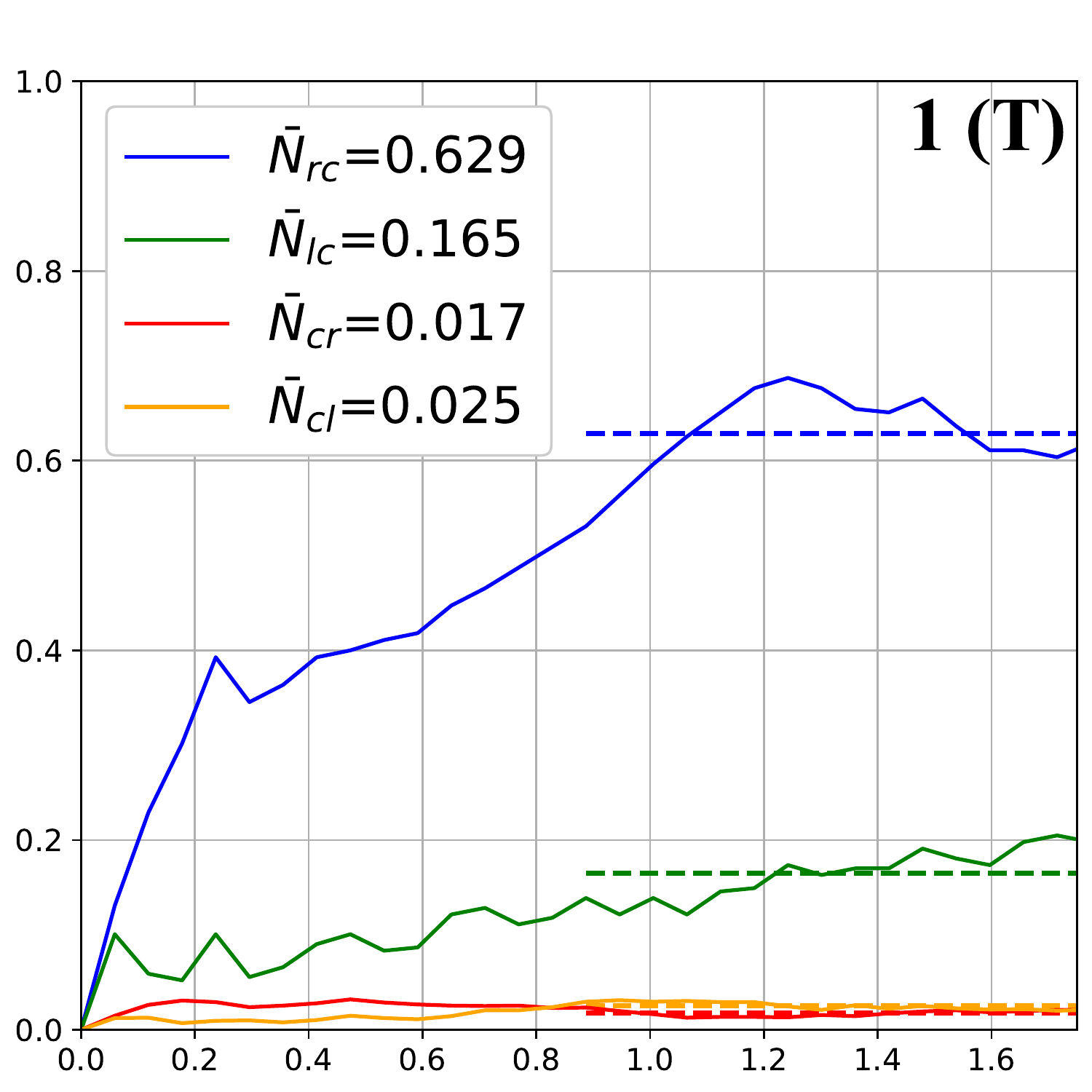}

    \includegraphics[clip, 
    width=0.49\linewidth]{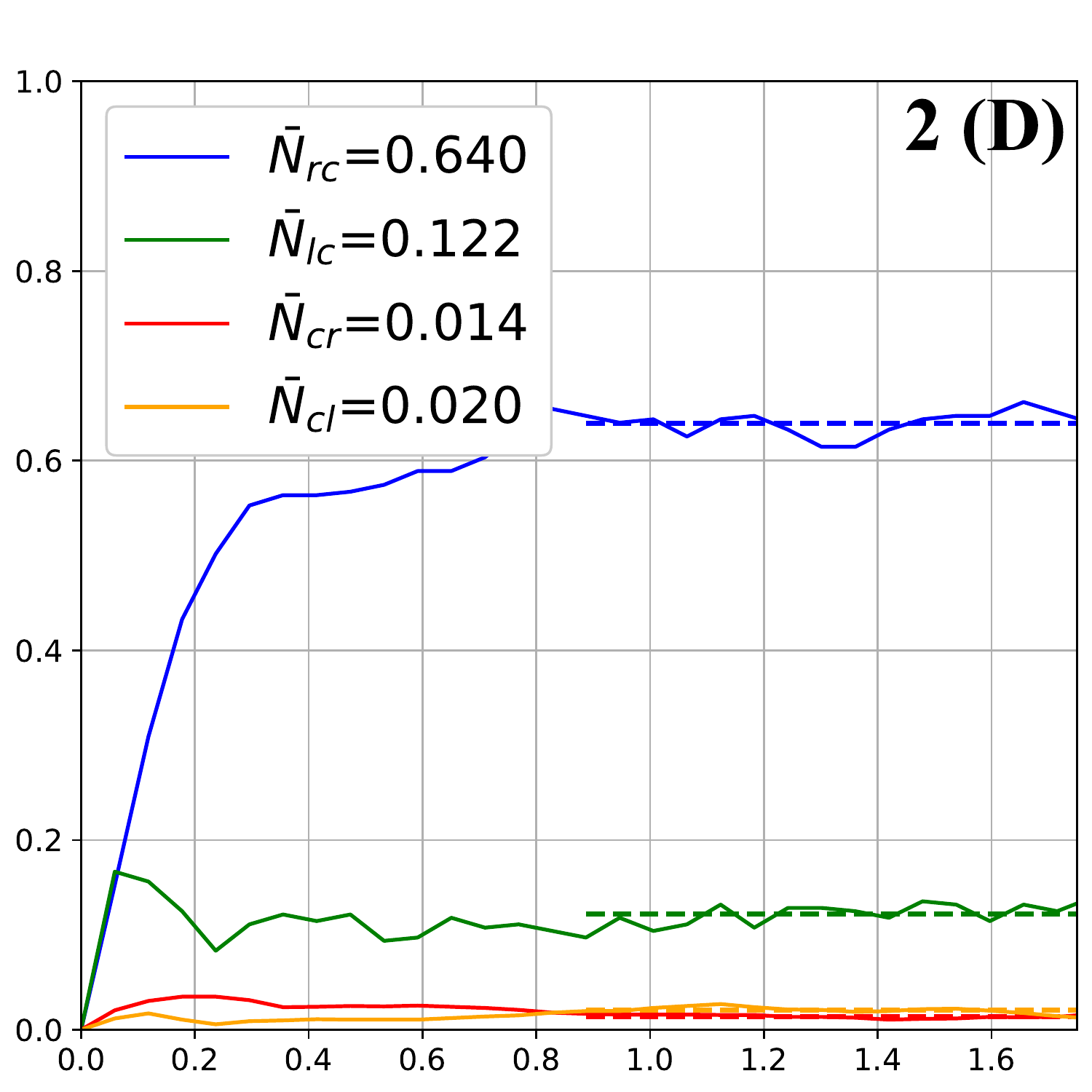}
    \includegraphics[clip,
    width=0.49\linewidth]{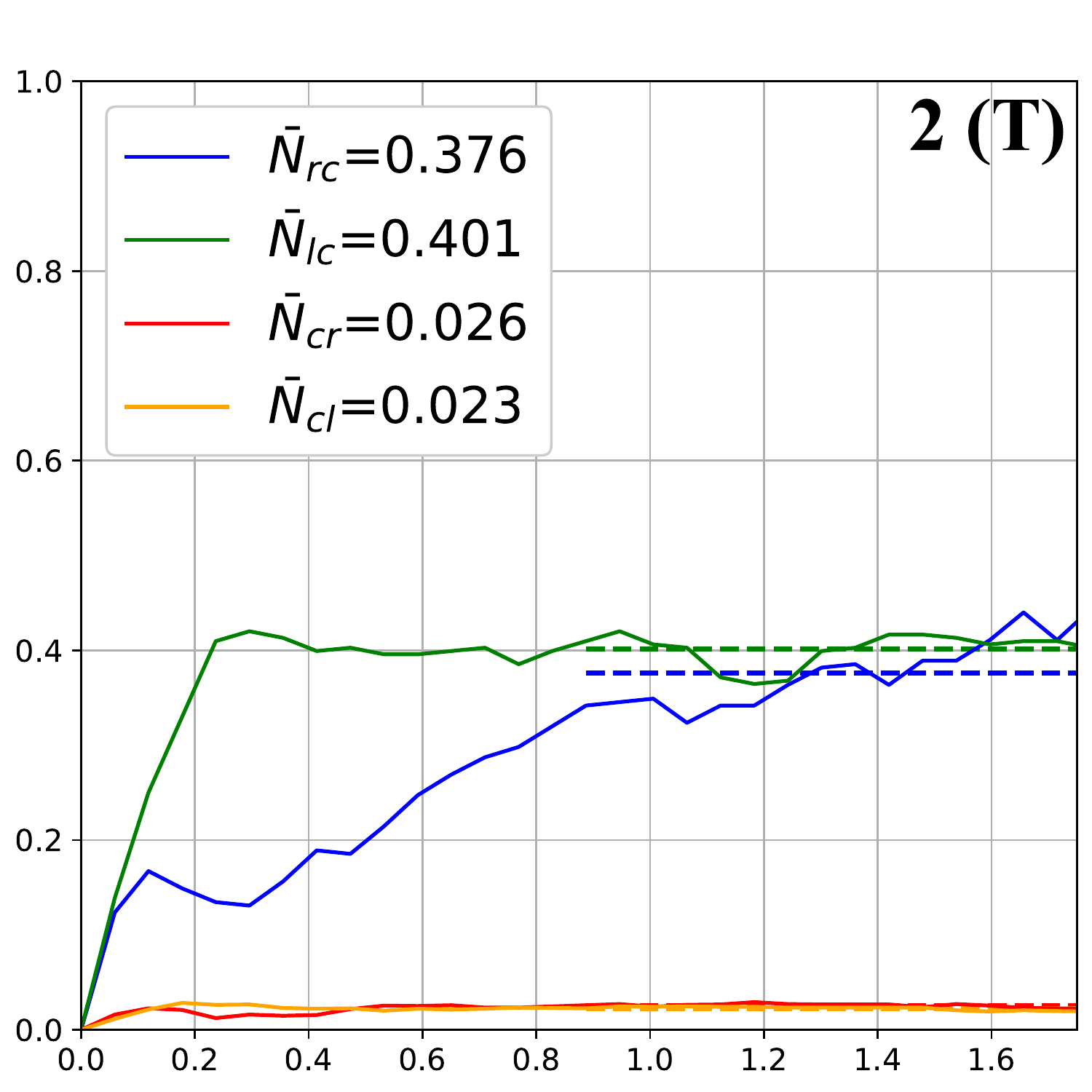}

    \includegraphics[clip,  
    width=0.49\linewidth]{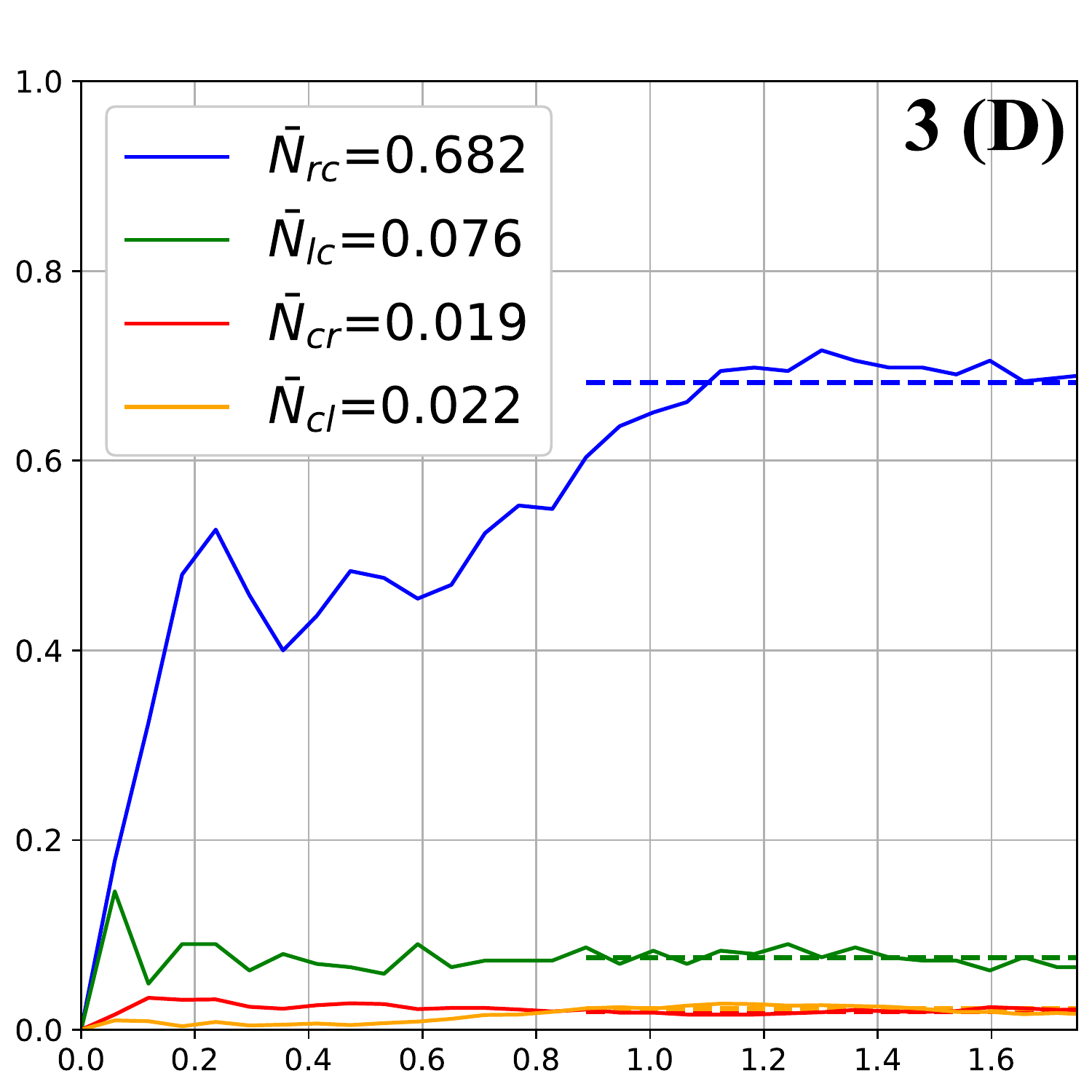}
    \includegraphics[clip, 
    width=0.49\linewidth]{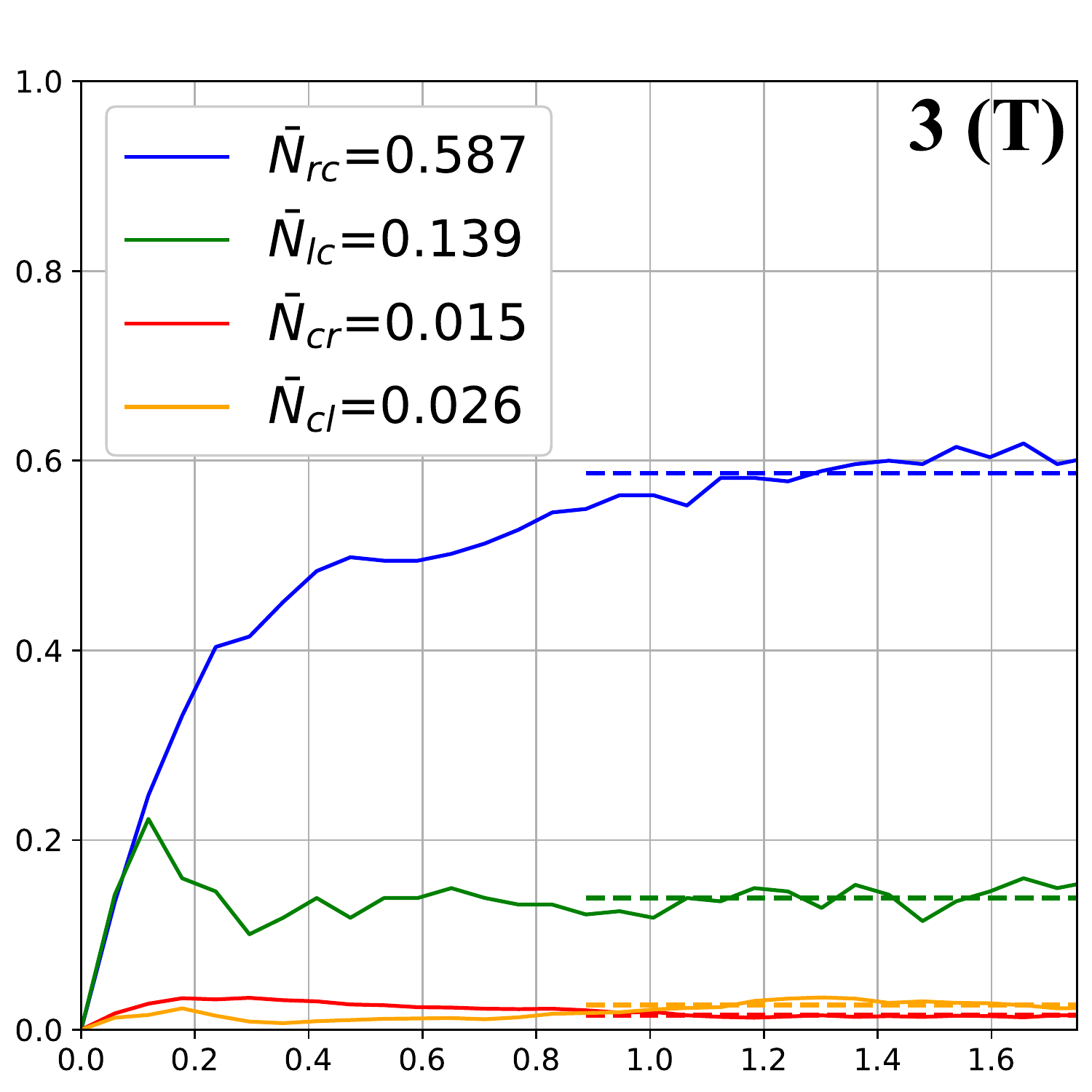}

    \includegraphics[clip,
    width=0.49\linewidth]{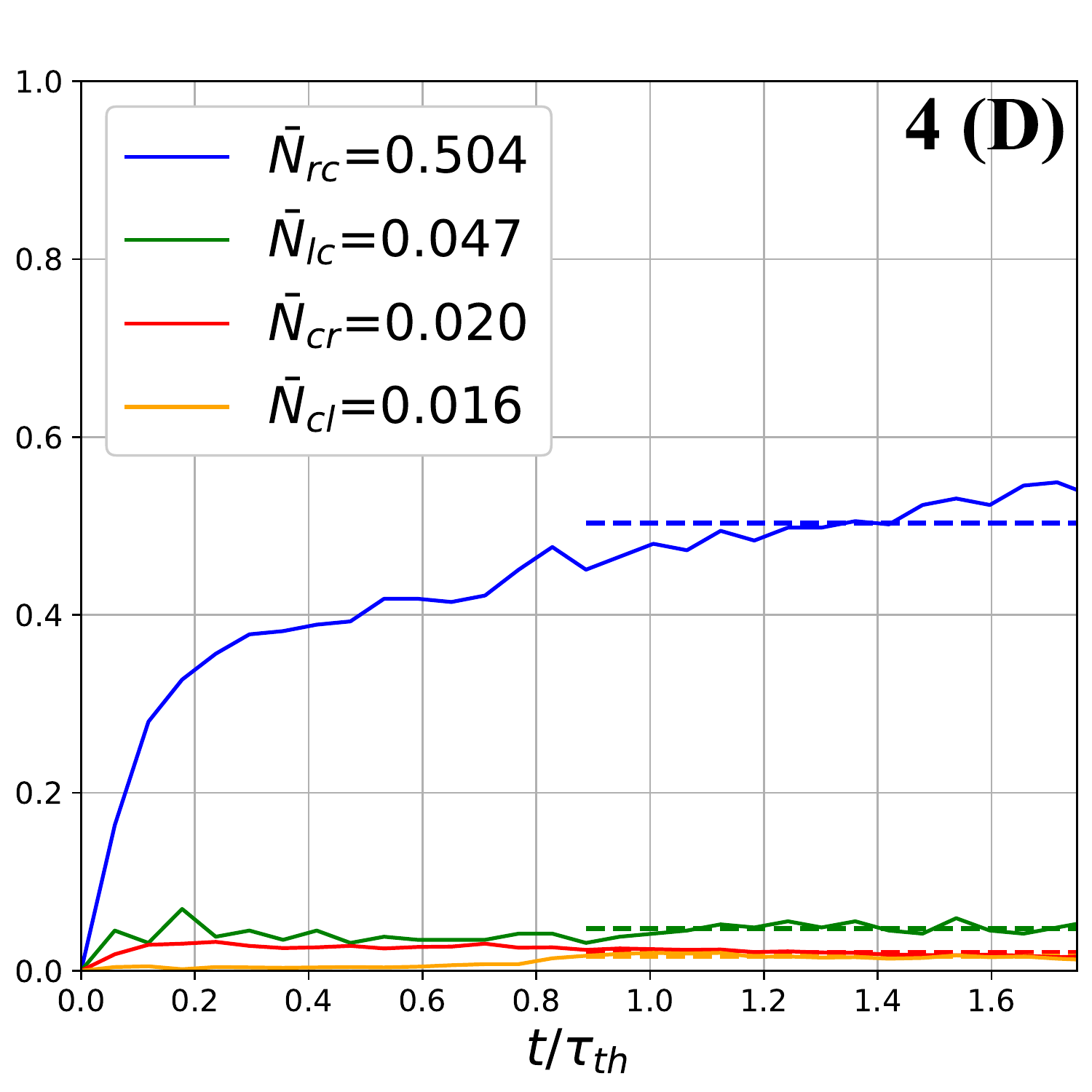}
    \includegraphics[clip,
    width=0.49\linewidth]{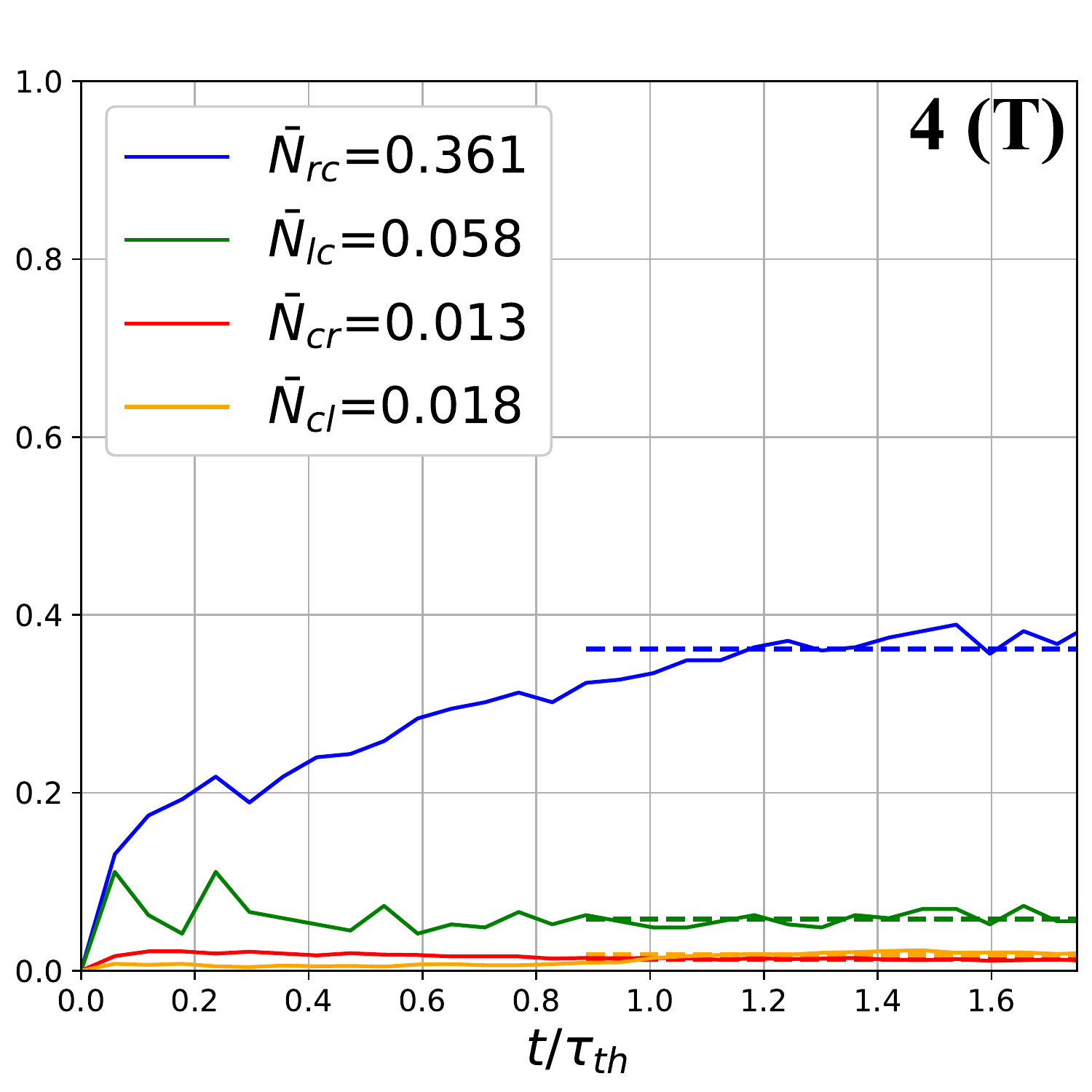}

    \caption{The relative number of converted particles as a function of time (solid lines): $N_{r \rightarrow c}/N_{r,0}$ (blue),  $N_{l \rightarrow c}/N_{l,0}$ (green), $N_{c \rightarrow r}/N_{c,0}$ (red) and $N_{c \rightarrow l}/N_{c,0}$ (orange).
    The dashed lines indicate the mean of each curve in the second half of the simulation time. 
    These saturation values, $\bar{N}_{rc}$, $\bar{N}_{lc}$, $\bar{N}_{cr}$, and $\bar{N}_{cl}$ are recorded in the legend and summarized in Table \ref{table: RF parameters}.
    The upper-right text in each panel indicates the RF parameters set number and the particle type (see Table \ref{table: RF parameters}).}
    
    \label{fig: saturation}
\end{figure}

To get relative transition quantities, we normalize each converted particle quantity by the initial number of particles in that population. 
These normalized transition quantities are plotted in Fig.~\ref{fig: saturation} for the different RF parameter sets of Table \ref{table: RF parameters}.
For example, the blue line represents the number of particles that originated in the right lose cone that at time $t$ were in the capture region in velocity space, \ie $N_{r \rightarrow c}/N_{r,0}$, where $N_{r,0}$ is the initial number of particles in the right lose cone in the simulation.
It can be seen from the figure that, typically, the number of converted particles grows with time at early times but may fluctuate as particles can change their identity multiple times. 
In most cases, the number of converted particles saturates at some value after the transient increase.
Therefore, for each case, we define and calculate the saturated value by averaging over the second half of the simulated time. 
It is noted that in some of the cases, the saturation has not been reached yet, so the approximation is not as good as in other cases.
We will address this caveat in the next section when studying the overall RF plugging effect.
We denote the saturation values by $\bar{N}_{rc}, \bar{N}_{lc}, \bar{N}_{cr}, \bar{N}_{cl}$ and plot them by dashed lines in Fig.~\ref{fig: saturation}, where the values are recorded in the legend of each panel.

\begin{figure}[t]
    \includegraphics[clip, trim=.4cm .5cm .5cm 0.5cm, 
    width=0.49\linewidth] {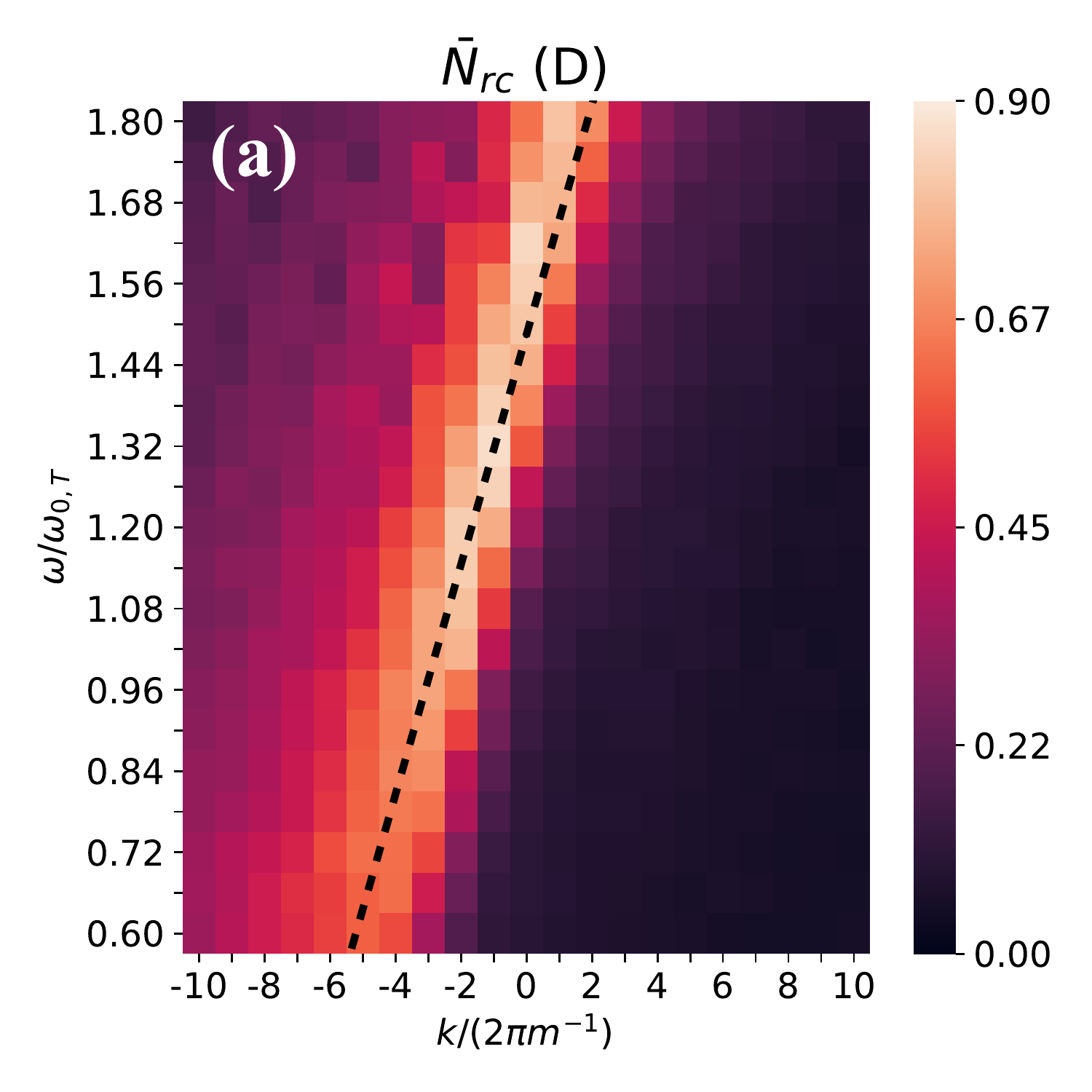}
    \includegraphics[clip, trim=.4cm .5cm .5cm 0.5cm, 
    width=0.49\linewidth] {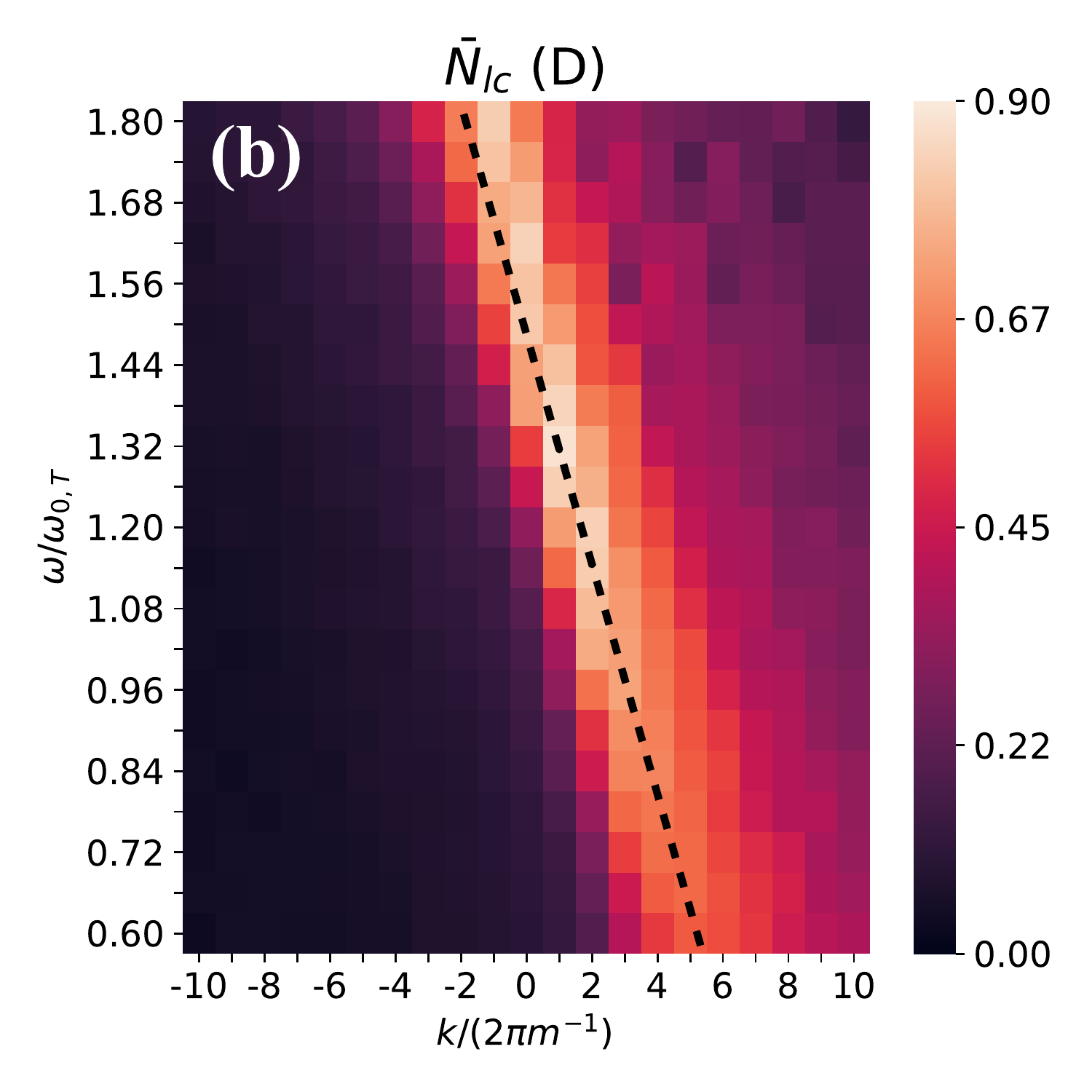}
    \includegraphics[clip, trim=.4cm .5cm .5cm 0.5cm, 
    width=0.49\linewidth] {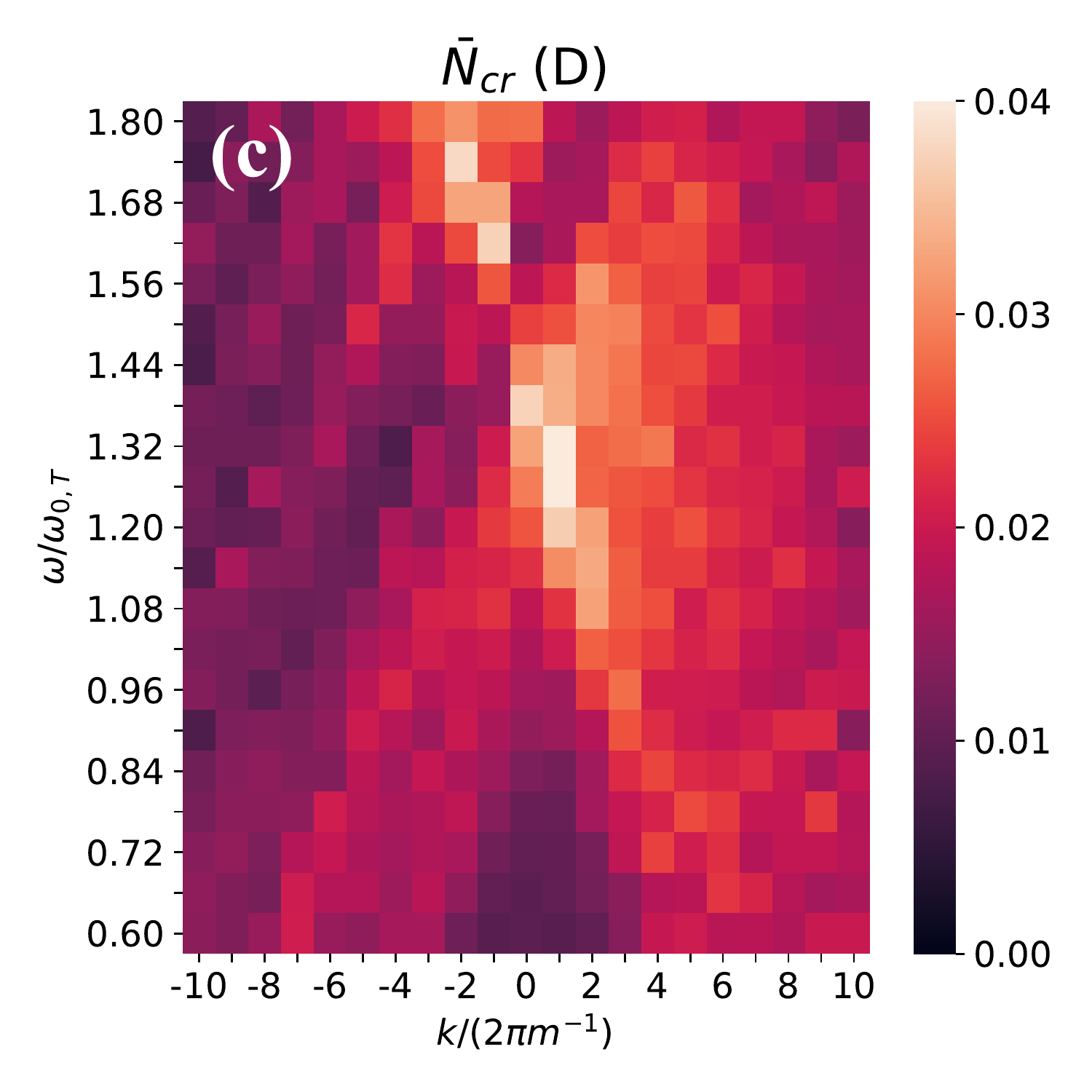}
    \includegraphics[clip, trim=.4cm .5cm .5cm 0.5cm, 
    width=0.49\linewidth] {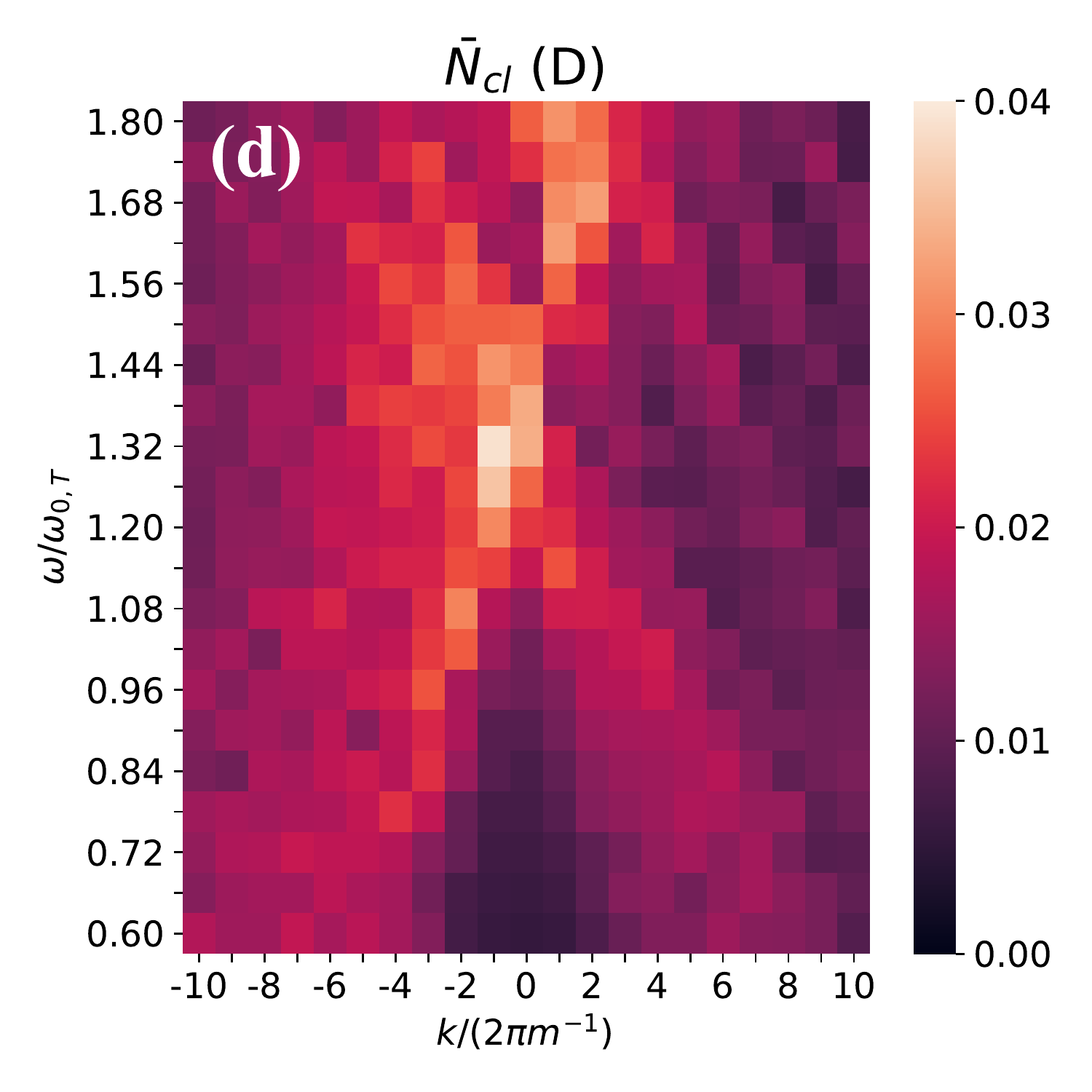}
    \includegraphics[clip, trim=.4cm .4cm .4cm 0.4cm, 
    width=0.49\linewidth] {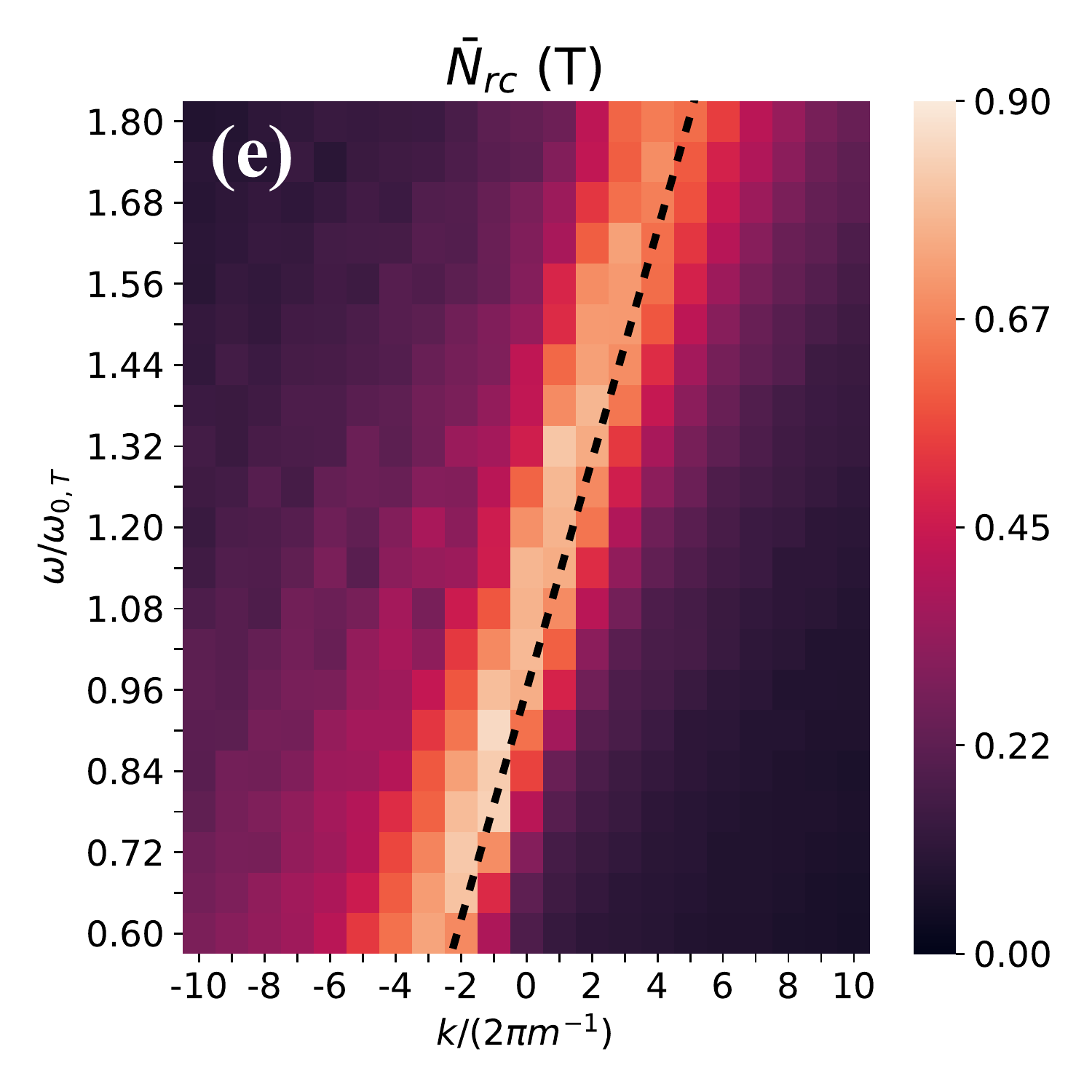}
    \includegraphics[clip, trim=.4cm .4cm .4cm 0.4cm, 
    width=0.49\linewidth] {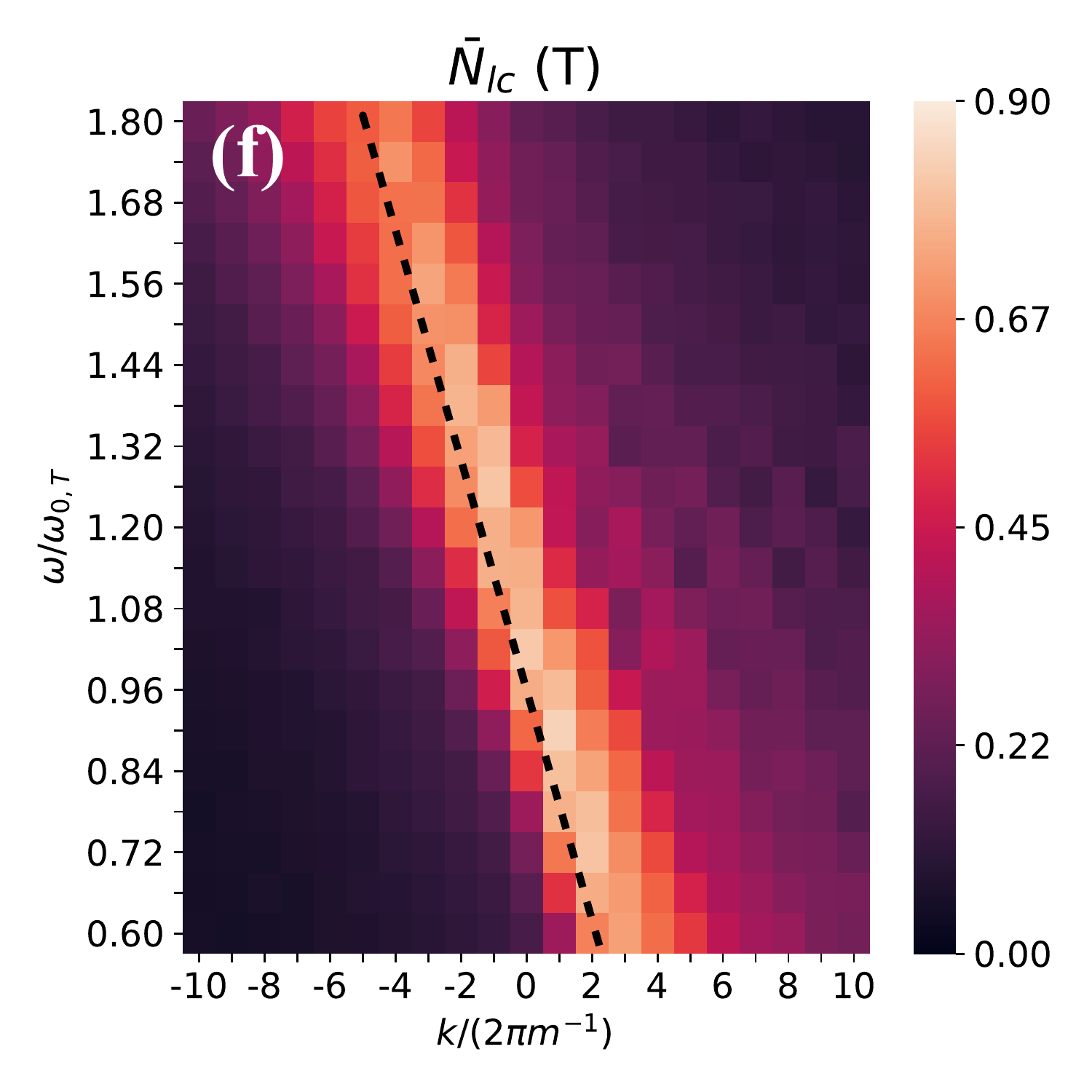}
    \includegraphics[clip, trim=.4cm .4cm .4cm 0.4cm, 
    width=0.49\linewidth] {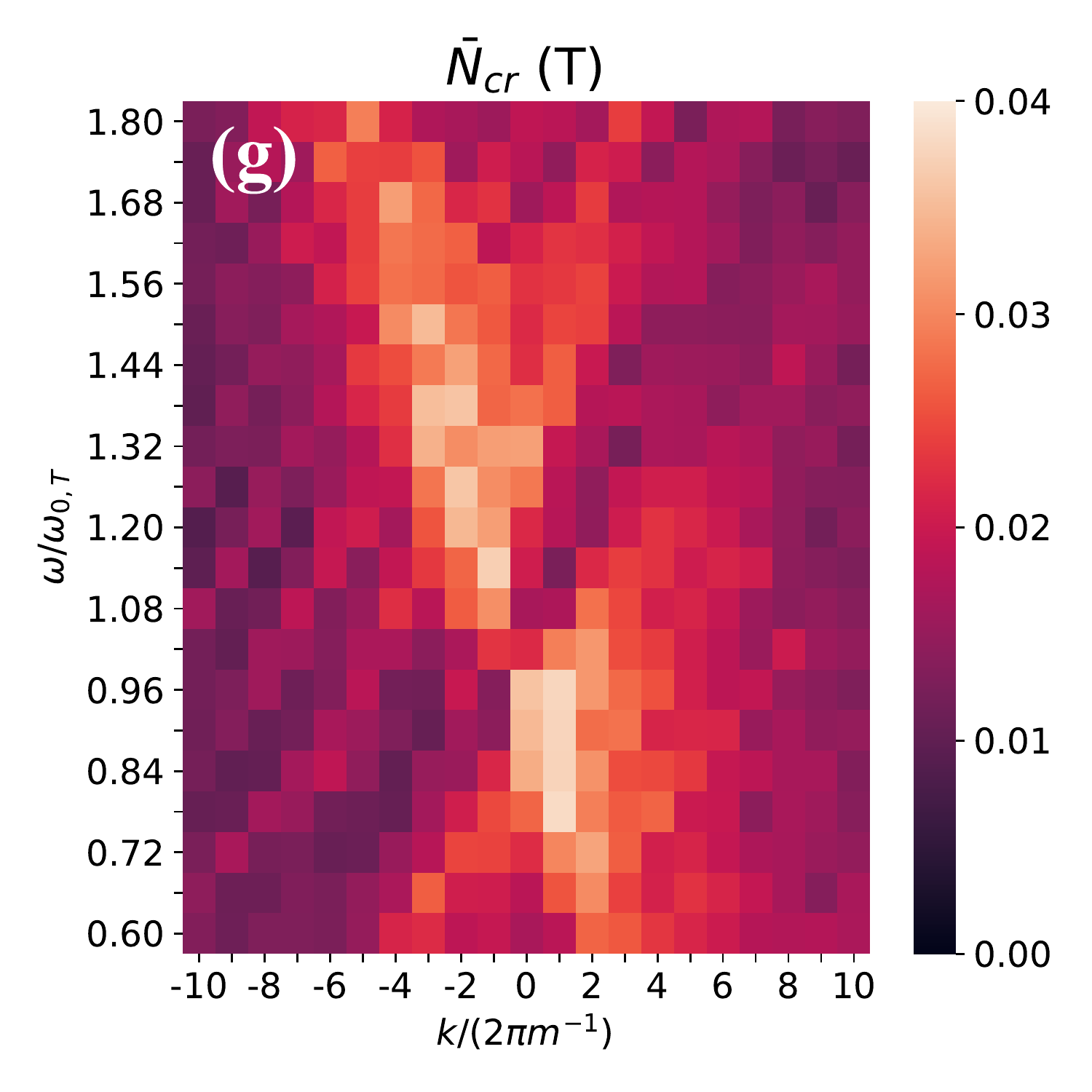}
    \includegraphics[clip, trim=.4cm .4cm .4cm 0.4cm, 
    width=0.49\linewidth] {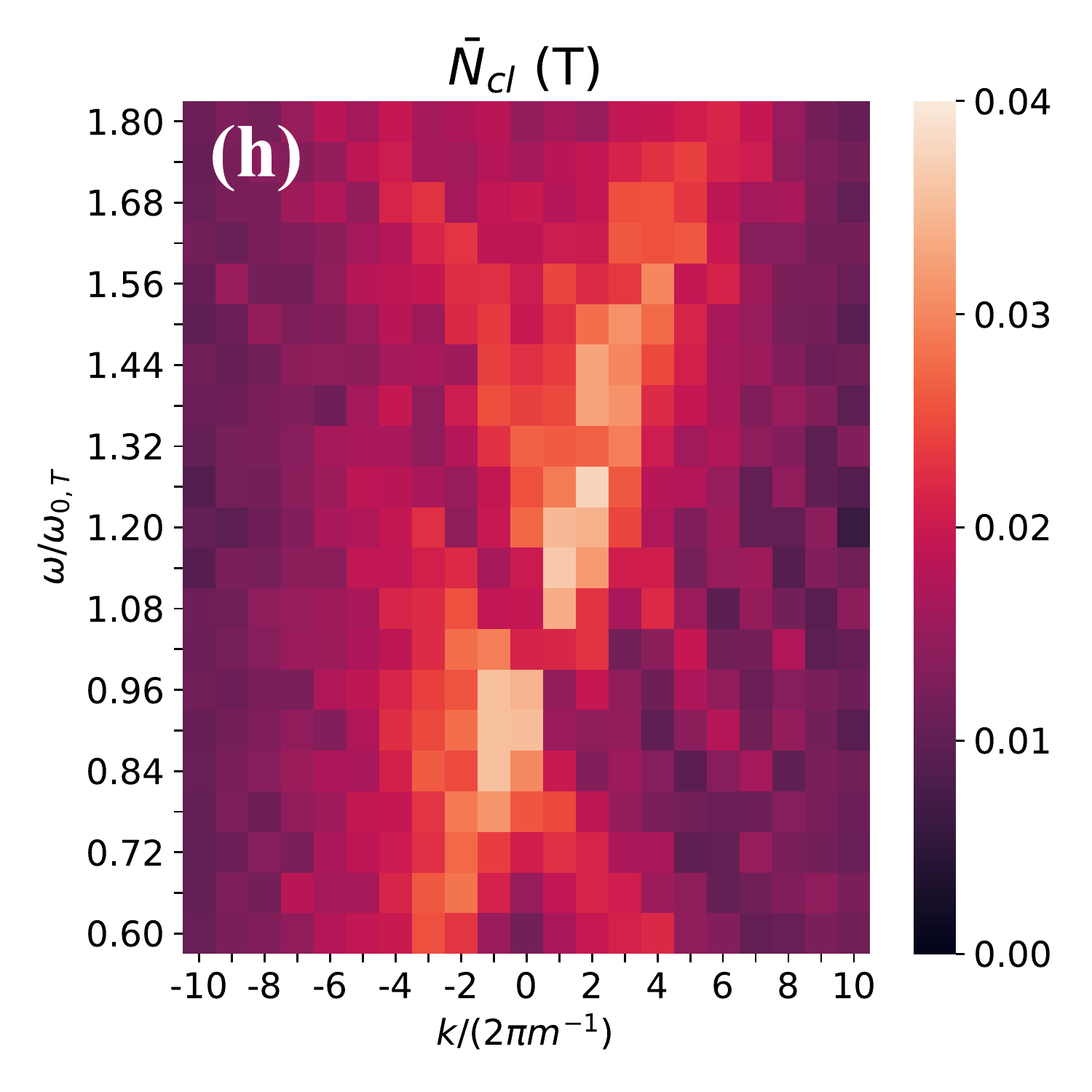}
    \caption{The saturation values (as in Fig.~\ref{fig: saturation}) for a wide range of $k_{RF},\omega_{RF}$ values for both deuterium and tritium. $\bar{N}_{rc}$ in panels (a, e), $\bar{N}_{lc}$  in panels (b, f), $\bar{N}_{cr}$  in panels (c, g), and $\bar{N}_{cl}$ in panels (d, h). Note the color scale is different for the saturation metrics pairs $\bar{N}_{rc}, \bar{N}_{lc}$ and $\bar{N}_{cr}, \bar{N}_{cl}$.  The black dashed lines in panels (a,b,e,f) indicate the theoretical resonant lines described in Sec. \ref{RF trapping}.}
     \label{fig: saturation heatmaps}
\end{figure}

The next step is to study the RF effect for a wide $k_{RF},\omega_{RF}$ parameter space.
We calculate the saturation values of converted particles for the four relevant transitions and plot the results in Fig.~\ref{fig: saturation heatmaps} for both deuterium and tritium.
It can be seen that the maximal transitions from the two loss-cones populations to the captured population, \ie $\bar{N}_{rc}$ and $\bar{N}_{lc}$ (panels (a), (b), (e), and (f)) are concentrated along a straight line in $k_{RF},\omega_{RF}$ space. 
These lines correspond to the cyclotron resonance condition that depends on the Doppler compensation between the frequency detuning and the RF wave velocity as described in Sec.~\ref{Fields configurations}.
In dashed black lines we overlaid the resonance condition of Eq.(\ref{eq: Doppler shift condition}), where $v_z$ is replaced by the mean axial velocity in the loss cones 
\begin{eqnarray}
    \bar{v}_{z,LC}=\frac{\intop_{LC}\,f_{MB}\left(\mathbf{v}\right)\,v_{z} \,d^{3}v}{\intop_{LC}f_{MB}\,\left(\mathbf{v}\right)\,d^{3}v}. 
\end{eqnarray}%
Here, $f_{MB}\left(\mathbf{v}\right)=\pi^{-3/2}\, v_{th}^{-3} \exp(-\mathbf{v}^2/v_{th}^2)$ is the Maxwell-Boltzmann distribution function and the integral is taken only over the loss cone section of the velocity space.
One finds 
\begin{eqnarray}
    \bar{v}_{z,LC}=\frac{1+\sqrt{1-\frac{1}{R_m}}}{\sqrt{\pi}}\, v_{th}, 
\end{eqnarray}%
where for the considered mirror ratio, $R_m=3$, the result is $\bar{v}_{z,LC}=1.025 \,v_{th}$. 
In panels (a,b) of Fig.~\ref{fig: saturation heatmaps}, the thermal velocity, $v_{th}=\sqrt{2k_B T/m}$, is calculated with the deuterium mass, and in panels (e,f) with the tritium mass.
Notably, the right and left loss cones have the opposite mean axial velocity directions and therefore the opposite slope of the resonance line in the figure.
We can see that the theoretical lines match quite well with the peak transition rates found in simulations.
Also, one can see that the results for deuterium (a-d) and tritium (e-h) look very similar, except the latter are downshifted toward lower values of $\omega_{RF}$. 
This is because the resonance condition is associated with the cyclotron frequency that is inversely related to the ion's mass, and tritium is heavier than deuterium.

\begin{figure}[tb]
    \includegraphics[clip, trim=.4cm .4cm .4cm 0.4cm, 
    width=0.49\linewidth] {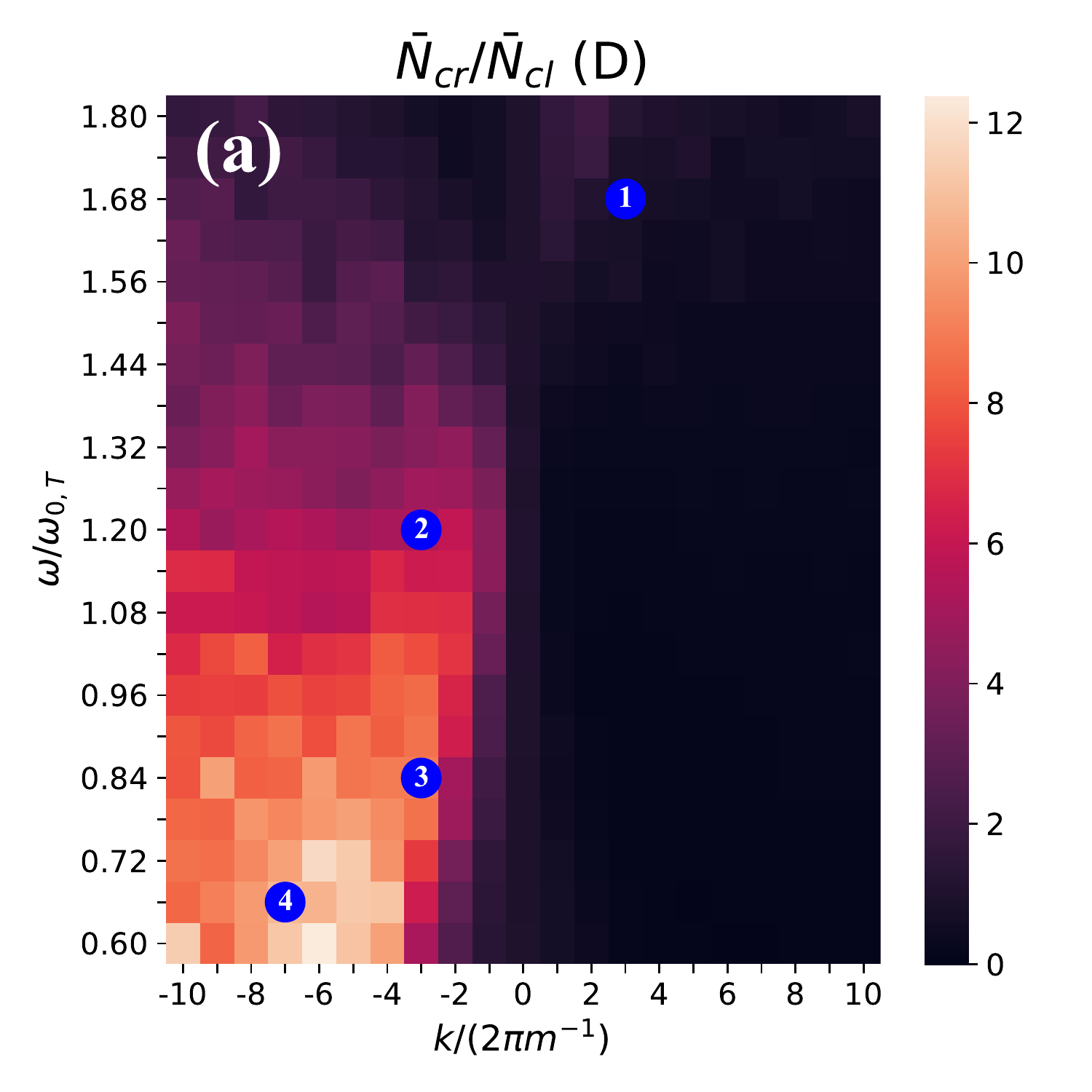}
    \includegraphics[clip, trim=.4cm .4cm .4cm 0.4cm, 
    width=0.49\linewidth] {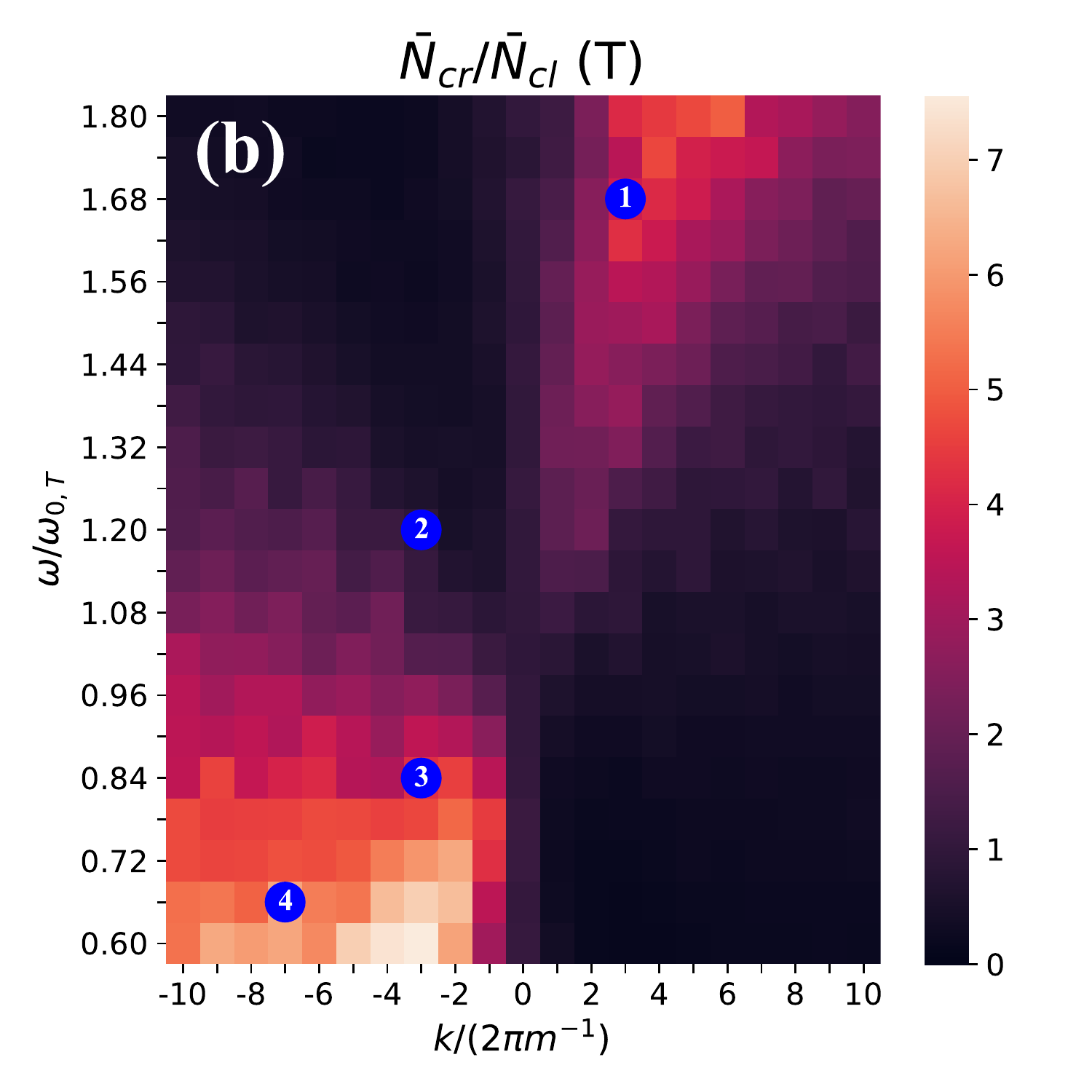}
    \caption{Right-left selectivity, $\bar{N}_{rc}/\bar{N}_{lc}$, as a function of the RF parameters $k_{RF}$ and $\omega_{RF}$ for deuterium (a) and tritium (b). 
    The locations of the parameter sets of table \ref{table: RF parameters} are indicated by numbered (from one to four) blue dots.}
    \label{fig: selectivity}
\end{figure}

Since RF plugging is based on the asymmetric effect of left- and right-going particles, it is elucidating to focus on the ratio, $\bar{N}_{rc}/\bar{N}_{lc}$.
We call this ratio the right-left selectivity, and we draw it for both deuterium and tritium in Fig.~\ref{fig: selectivity} for the same parameter space as Fig.~\ref{fig: saturation heatmaps}. 
One can again see that, due to the mass difference, the results for tritium (Fig.~\ref{fig: selectivity}b) look similar but downshifted in frequency compared to those of deuterium (Fig.~\ref{fig: selectivity}a).
In the figure, we also indicated by numbered blue circles the locations (in $k_{RF},\omega_{RF}$ plane) of the parameter sets used in the examples of Table \ref{table: RF parameters} and Figs.~\ref{fig: velocity space evolution} and~\ref{fig: saturation}.

Finally, we repeated the single particle simulations for a half-amplitude RF electric field, \ie $25\mathrm{kV/m}$.
We found that although the transition rates in the optimal parameters regions decrease by a factor of two, the selectivity, $\bar{N}_{rc}/\bar{N}_{lc}$, changed by a few percent only.
We conclude that the selectivity effect is not sensitive to the RF amplitude because of its  resonant nature. 
However, the overall plugging efficiency might be more sensitive and should be studied and optimized for a specific system in future research.

\section{A Generalized Rate Equations Model}
\label{Rate Equations Model and RF plugging}

In previous work, we developed a semi-kinetic rate equations model for the MM system, which includes Coulomb scattering within each cell and transmissions between neighboring cells through the loss cones.\cite{miller2021rate} 
Next, we review the rate equation model's definitions and assumptions and introduce new terms for a driven transport induced by the external RF fields.
The ion-ion Coulomb scattering rate is denoted by $\nu_s$ and roughly scales with density and temperature as $\propto n/T^{3/2}$.
The inter-cell transmission rate is approximated by $\nu_t = v_{th}/l$, where $v_{th}$ is the ion thermal velocity and $l$ is the mirror cell length.
The particles are divided into three populations per cell, captured particles due to the mirroring effect, and right- and left-going particles through the right and left loss cones, respectively. 
The densities of the three populations in the $i$'th cell are denoted by $n_c^i$, $n_r^i$, and $n_l^i$.
The steady-state solution is defined by $\dot{n}=0$ for all cell populations. 
The outgoing flux between two neighboring cells (\eg $i\rightarrow i+1$) is proportional to $\phi_{i,i+1} \propto v_{th}^i n_{r}^i -v_{th}^{i+1} n_{l}^{i+1}$. 
In a steady state, by definition, all inter-cell fluxes are the same and denoted by $\phi_{ss}$. 
The overall system confinement time is inversely related to the steady-state flux, namely, $\tau \propto 1/\phi_{ss}$, which we aim to maximize for satisfying the Lawson criterion in a fusion system.\cite{lawson1957some, wurzel2022progress}
In the absence of RF fields, the confinement time 
is expected to scale linearly with the number of MM cells $N$, in agreement with simple one-dimensional diffusion models.\cite{logan1972multiple, makhijani1974plasma, miller2021rate}
However, even for optimized system parameters and optimistic thermodynamical regimes, fusion time scales require an impractical number of MM cells.\cite{miller2021rate}
Therefore, as discussed above, we propose to apply an RF field that asymmetrically affects the particle transport in MM systems, \ie recaptures more escaping (right-going) than returning (left-going) particles.

Here, we study the accumulative plugging effect by including the RF-induced transition terms in an extended rate equation model.
In Sec.~\ref{RF trapping}, we quantified the amounts of particles that transport between the different populations of one MM cell as $\bar{N}_{rc}, \bar{N}_{lc}, \bar{N}_{cr}, \bar{N}_{cl}$ under the influence of traveling rotating electric field for an extensive parameter range (see Fig.~\ref{fig: saturation heatmaps}). 
The characteristic time scale for a non-trapped particle to pass across one mirror cell depends on its thermal velocity via $\tau_{th} = l/v_{th}$. 
Therefore, we estimate the RF conversion rates by the ratio between the relative conversion amounts and the characteristic time scale, \ie $\nu_{RF,rc}=\bar{N}_{rc}/\tau_{th}$,  $\nu_{RF,lc}=\bar{N}_{lc}/\tau_{th}$, $\nu_{RF,cr}=\bar{N}_{cr}/\tau_{th}$, and $\nu_{RF,cl}=\bar{N}_{cl}/\tau_{th}$.
We have generalized the rate equations model,\cite{miller2021rate} to include these four induced transition rates such that the generalized model reads
\begin{eqnarray}
\label{Eq: dn_c_dt} \dot n_{c}^i &=&  \nu_{s}^i \left[(1-2 \alpha) (n_{l}^i + n_{r}^i) - 2 \alpha n_{c}^i\right]  \\
                        &&  - (\nu_{RF,cl} + \nu_{RF,cr}) n_{c}^i + \nu_{RF,lc} n_{l}^i + \nu_{RF,rc} n_{r}^i \nonumber \\ 
\label{Eq: dn_tL_dt} \dot n_{l}^i &=&  \nu_{s}^i\left[\alpha (n_{r}^i+n_{c}^i) - (1-\alpha) n_{l}^i\right] - \nu_{t}^i n_{l}^i + \nu_{t}^{i+1} n_{l}^{i+1} \\ 
                        &&  - \nu_{RF,lc} n_{l}^i + \nu_{RF,cl} n_{c}^i \nonumber \\ 
\label{Eq: dn_tR_dt} \dot n_{r}^i &=& \nu_{s}^i \left[\alpha (n_{l}^i +n_{c}^i) - (1-\alpha) n_{r}^i\right]- \nu_{t}^i n_{r}^i +  \nu_{t}^{i-1} n_{r}^{i-1} \\
                        &&  - \nu_{RF,rc} n_{r}^i + \nu_{RF,cr} n_{c}^i  \nonumber \; 
\end{eqnarray}
\noindent The normalized loss-cone solid angle is $\alpha = \sin^{2} \left(\theta_{LC}/2\right)$, where the loss-cone angle satisfies $\sin\theta_{LC}=v_{\perp}/v=R_{m}^{-1/2}$. 
Our convention is that the left side of the MM section is connected to the main cell, and the right side is the exit.
The boundary conditions that close the rate equations are a constant density in the left cell ($n_c^1+n_{l}^1+n_{r}^1=n_0=$const., which practically reads $n_{r}^1=n_0-n_c^1-n_l^1$ since $n_l^1$ and $n_c^1$ are determined by the rate equations themselves), where  $n_0$ is the central fusion cell density, and a free flow boundary condition on the right cell ($\nu_t^{N+1}n_l^{N+1}=0$), \ie no left going flux from outside the system.
The Coulomb scattering rate, $\nu_s$, depends on the thermodynamic scenarios along the MM section.\cite{miller2021rate}
Here, for simplicity, we focus on the common isothermal scenario, which can be justified by electron thermalization since electrons are much faster than ions.

The classic MM operates best when the Coulomb scattering mean free path (MFP), $\lambda=v_{th} / \nu_s$, is of the order of the mirror cell length, $l$, where the assumption of one-dimensional diffusion between cells is valid. 
However, for plasma parameters commonly considered for D-T fusion machines, $n=10^{21}\,\mathrm{m^{-3}}$ and $k_{B} T=10 \mathrm{keV}$, the MFP is of the order of kilometers, making MM systems impractical. 
It is possible to tweak the plasma parameters (increasing the density and reducing the temperature) to get a lower MFP for the same magnetic pressure.
However, the price is a higher required confinement time to meet the Lawson criterion.\cite{lawson1957some, wurzel2022progress, miller2021rate}
It is notable that for symmetric RF transition rates, \ie when $\nu_{RF,rc}=\nu_{RF,lc}$, the role of the RF terms in Eqs.~(\ref{Eq: dn_c_dt})$-$(\ref{Eq: dn_tR_dt}) is equivalent to that of the Coulomb scattering.
Therefore, by using RF, one can circumvent the difficulty of choosing suitable plasma parameters (\ie temperature and density), as the RF transition rates depend on the externally applied field rather than on the plasma parameters.
Furthermore, when the RF trapping rates are not symmetrical, we can get a right-left selectivity such that the trapping becomes more favorable for escaping particles than for returning particles.
Exploiting this effect, the plugging of the MM section can be based on the RF effect rather than on collisions.
As a result, one can choose plasma parameters that are preferable for fusion in the central cell without worrying about the collision rates in the MM sections.

\begin{figure}[tb]
    \includegraphics[clip, trim=.4cm 0.4cm .4cm 0.4cm,
    width=0.8\linewidth]{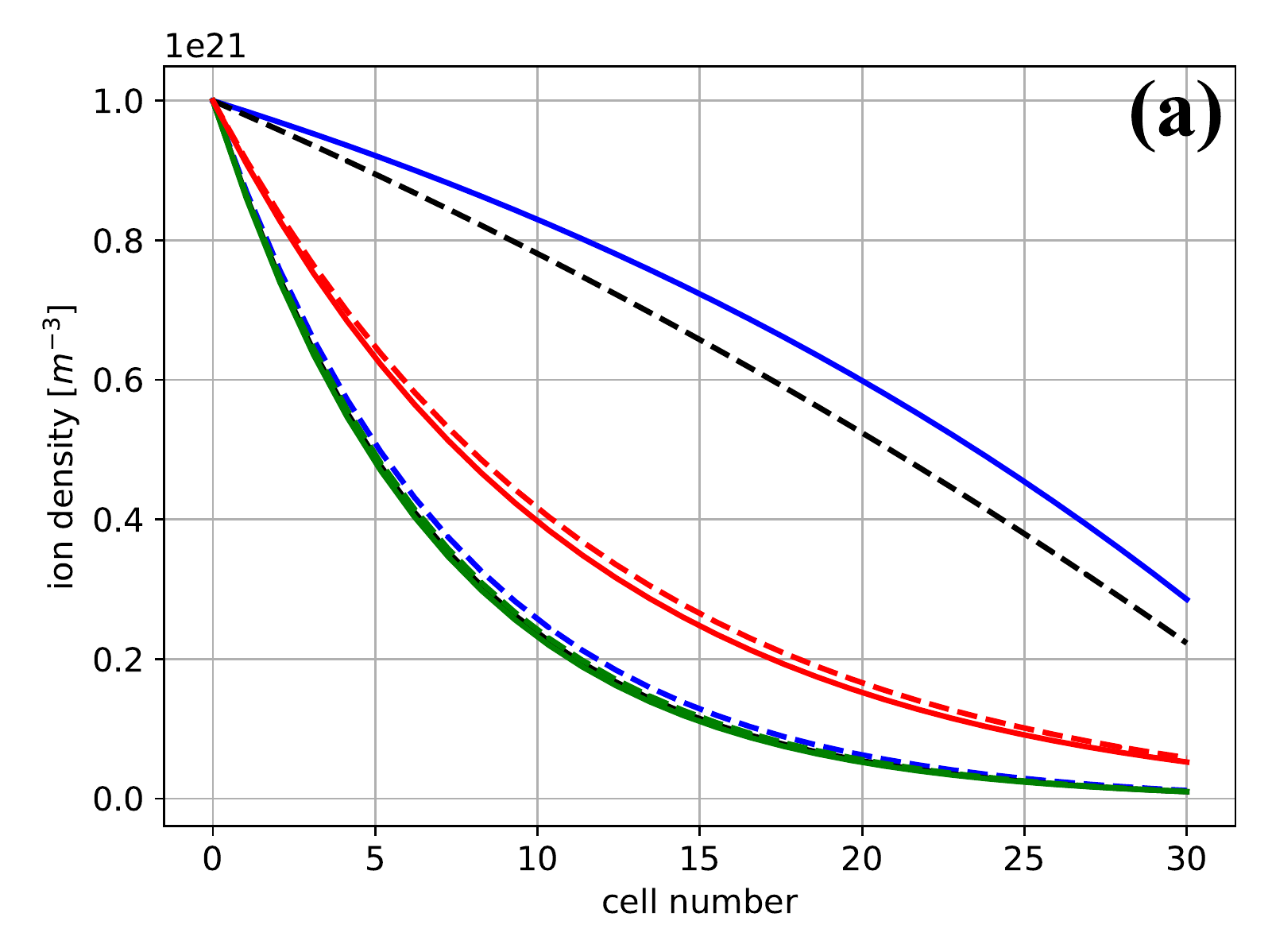}
    \includegraphics[clip, trim=.4cm 0.4cm .4cm 0.4cm,
    width=0.8\linewidth]{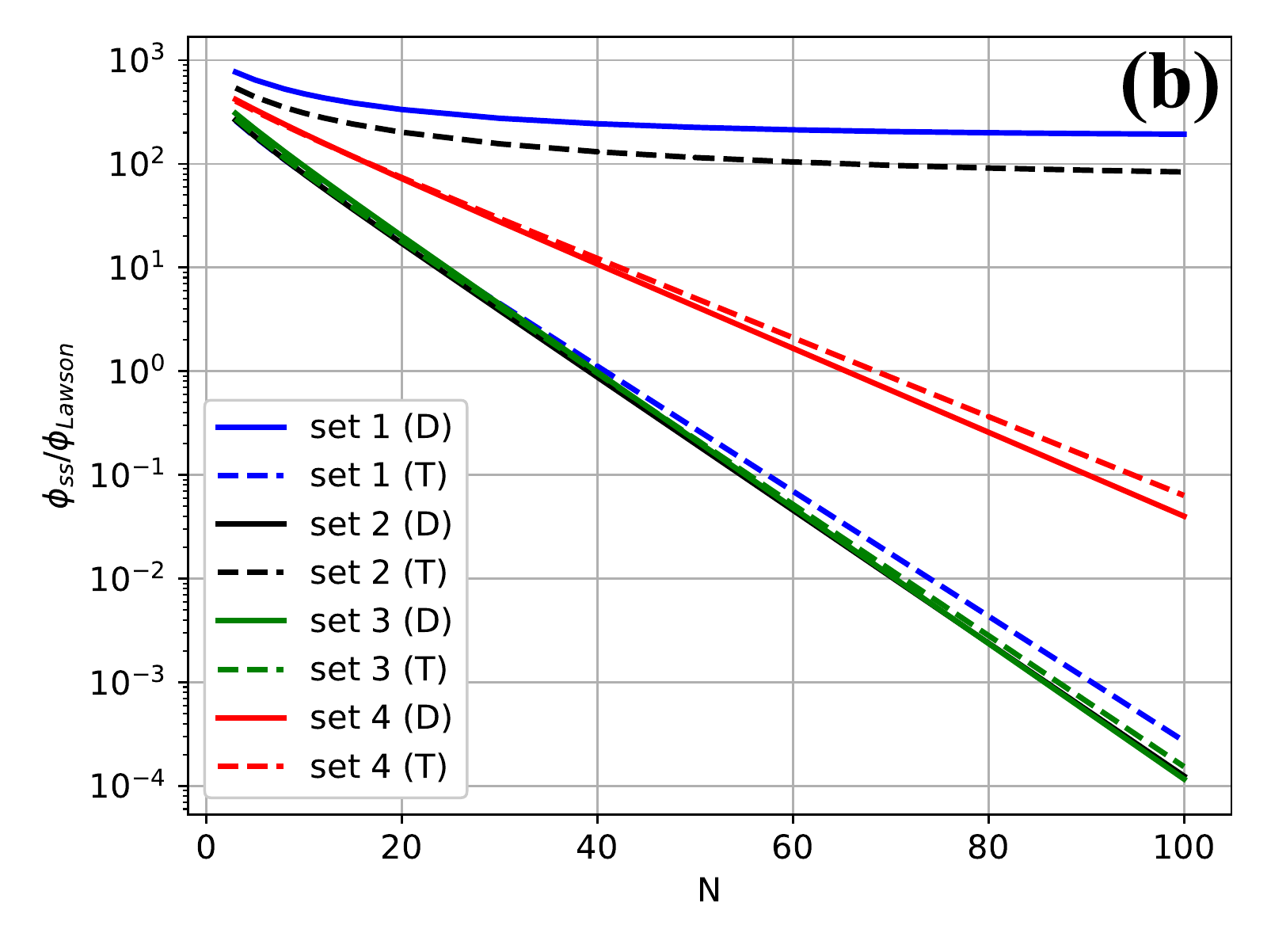}
    \caption{Steady-state simulation results of the MM rate equations models, including (a) the ion density profiles as a function of cell number for an MM section with $N=30$ cells, and (b) flux normalized by the maximal Lawson flux $\phi_{ss}/\phi_{\rm Lawson}$ as a function of system size.
    Presented are different sets of RF parameters and different isotopes (D and T), as described in the legend.}
    \label{fig: mm rate eqs solutions}
\end{figure}

We demonstrate the RF plugging effect by solving Eqs.~(\ref{Eq: dn_c_dt}-\ref{Eq: dn_tR_dt}) for a steady-state flux.
In Fig.~\ref{fig: mm rate eqs solutions}, we present the density profiles and the steady state flux obtained in simulations for different numbers of cells $N$ and different sets of RF parameters.
Each set of RF parameters corresponds to four different values of the RF rates that also vary for deuterium and tritium (see Table \ref{table: RF parameters}).
We note that in the rate equations model, we can solve only for a single particle type, which does not describe well a D-T system that composes both deuterium and tritium. 
However, in the cases where the Coulomb scattering rates are negligible compared to the RF, we can assume both species do not interact.
We, therefore, solve the rate equations for deuterium and tritium separately.
Fig.~\ref{fig: mm rate eqs solutions}(a) shows the steady state density profiles for MM systems with $N=30$ cells.
Fig.~\ref{fig: mm rate eqs solutions}(b) shows the steady state flux as a function of the total number of cells, $N$. 
The flux in the figure is normalized by the Lawson flux, 
\begin{eqnarray} 
    \label{eq: Lawson flux}
    \phi_{Lawson}=\frac{nV}{2\tau_{\rm Lawson}},
\end{eqnarray}
where $V$ is the volume of the central fusion cell and 
\begin{eqnarray}
    \tau_{\rm Lawson}=\frac{3k_{B}T / n }{\left\langle \sigma v\right\rangle E_{ch}/4 - C_B T^{1/2} }
\end{eqnarray}
is the minimal required confinement time due to Lawson's criterion for ignition.
$E_{ch}$ is the charged fusion products energy (\eg\,$3.5$  MeV for D-T reactions), $\left\langle \sigma v\right\rangle$ is the Maxwell-Boltzmann averaged fusion reactivity, and $C_B=5.34\cdot 10^{-37} \mathrm{W} \mathrm{m}^3 \mathrm{keV}^{-1/2}$ is the bremsstrahlung radiation power coefficient assuming fully ionized plasma and equal ion and electron temperatures.\cite{lawson1957some, miller2021rate, wurzel2022progress}
The factor of $1/2$ in Eq.~(\ref{eq: Lawson flux}) stands for the two sides of the systems.
Further higher-order adjustments to the Lawson criterion for RF-plugged MM systems, such as the RF energy deposited in the escaping particles, are left to a future study.

For a practical example, if the main cell length is $100m$ and its diameter is $D=0.5m$ (plasma volume in the central cell of $V\approx 20 m^3$), then $\phi_{Lawson}\approx 3\cdot10^{22}\text{s}^{-1}$.
One can see that the flux drops exponentially with $N$ in the cases with high right-left selectivity.
This exponential drop allows the system to reach the fusion relevant regime $\phi_{ss} \ll \phi_{\rm Lawson}$ with a reasonable MM section size, \ie tens of cells and less than $100m$ total system length.
This should be contrasted with Fig.~5 in Ref.~\onlinecite{miller2021rate}, which shows that without the RF terms, the steady-state flux drops inversely linear with $N$, leading to an impractical number of cells for reducing the flux to a required value.

Finally, the mass difference between deuterium and tritium results in a frequency shift in the particles' response to the RF fields, as shown in Figs.~\ref{fig: saturation heatmaps}, \ref{fig: selectivity}.
Consequently, the optimal parameters for RF plugging are also different, as illustrated in the examples of Table \ref{table: RF parameters} and in the rate equations results presented in Fig.~\ref{fig: mm rate eqs solutions}.
Importantly, there is an overlap region in parameter space, so a single RF field can plug both isotopes.
For example, in sets 3 and 4, the selectivities for D and T are of the same order (less than a factor of two). 
The differences in the flux suppression between the two species (solid and dashed lines for D and T, respectively) are minor in these examples (green and red sets in Fig.~\ref{fig: mm rate eqs solutions}).
On the other hand, for set 1 (blue), the flux suppression effect is more dominant in tritium than in deuterium and vice versa for set 2 (black). 
Interestingly, the scenario of set 1 can be advantageous for fusion reactors fueled and heated by injecting a high-energy neutral deuterium beam.
This is because most of the fusion power comes from the reactions between the high-energy, injected deuterium and the thermal tritium.
Therefore, it might be desirable in this case to remove the slowed-down thermal deuterium particles  while plugging the tritium solely.\cite{ivanov2017gas, fowler2017new, egedal2022fusion}

%% file: conclusions.tex
\section{Conclusions}
\label{conclusions}

Axial confinement in MM fusion machines can be significantly enhanced  by applying external RF fields that are resonant mostly with the outgoing particles.
The considered plugging field is a rotating electric field with a frequency slightly detuned from the ion cyclotron frequency and with a non-zero axial wave vector, such that in the particles' rest frame, the Doppler shift compensates the frequency detuning.
The resonant interaction would then increase the transverse energy of the escaping fuel particles and thus move them outside of the loss cone, \ie recapture them.
On the other hand, the incoming particles move in the opposite direction, from the MM section toward the central fusion cell.
Therefore, they experience an opposite Doppler shift, \ie away from resonance with the external field, resulting in an insignificant change in their transverse energy.

We studied this asymmetric effect by sampling initial conditions from a thermal distribution and employing single particle simulations to find the trajectories under the mutual influence of the mirror magnetic field and the RF electric field.
We developed theoretical criteria for the resonant region in velocity space, which agreed well with the single particle calculations.
From the Monte Carlo simulations results, we extracted the net transport rate of particles between the three populations of the rate equations model, trapped, escaping, and returning particles.
Because the re-trapping effect depends on the particle's mass and electric charge, we calculate the transition rates for each set of RF parameters for both species of D-T plasma.
We incorporated the RF transition rates into an extended semi-kinetic rate equations model for MM and solved it for steady state.

The main result is that adding RF plugging significantly changes the scaling of the axial confinement time with the number of MM cells from linear (without RF) to exponential (with RF).
Remarkably, this flux reduction could help satisfy the Lawson criterion for a reasonable system size.
We also demonstrated that although the plugging effect is different for deuterium and tritium, there are RF parameters for which the results are similar for both species. 
However, it is noted that a confinement enhancement of tritium alone could be advantageous for fusion machines that include high-energy, neutral deuterium beams.

Despite the promising results, a few serious caveats remain for us to address in future studies. 
First, the rotating electric field, which increases the kinetic energy of the particles, also heats the plasma to temperatures that might be deleterious for confinement and stability in some scenarios. 
Second, the measure of the external electric field penetration into the plasma, the excitation of various collective modes and waves, and the issue of wave backscattering from the plasma should be carefully examined theoretically and experimentally.
Third, the problem of radial stability, which is crucial for linear machines, is out of the scope of this paper.
Finally, the amount of plugging that can be induced by RF fields generated by realistic antennas like Nagoya coils\cite{watari1978radio} or helicon antennas\cite{light1995helicon, miljak1998helicon} and the influence of interactions between the particles on the RF plugging are also left for future study along with other types of RF fields, such as a rotating magnetic field.

\section*{Data Availability Statement}
The data that support the findings of this study are available from the corresponding author upon reasonable request.

\section*{Acknowledgments}
This work was supported by the PAZI Foundation, Grant No. 2020-191.


%% file: main.bbl
\begin{thebibliography}{72}%
\makeatletter
\providecommand \@ifxundefined [1]{%
 \@ifx{#1\undefined}
}%
\providecommand \@ifnum [1]{%
 \ifnum #1\expandafter \@firstoftwo
 \else \expandafter \@secondoftwo
 \fi
}%
\providecommand \@ifx [1]{%
 \ifx #1\expandafter \@firstoftwo
 \else \expandafter \@secondoftwo
 \fi
}%
\providecommand \natexlab [1]{#1}%
\providecommand \enquote  [1]{``#1''}%
\providecommand \bibnamefont  [1]{#1}%
\providecommand \bibfnamefont [1]{#1}%
\providecommand \citenamefont [1]{#1}%
\providecommand \href@noop [0]{\@secondoftwo}%
\providecommand \href [0]{\begingroup \@sanitize@url \@href}%
\providecommand \@href[1]{\@@startlink{#1}\@@href}%
\providecommand \@@href[1]{\endgroup#1\@@endlink}%
\providecommand \@sanitize@url [0]{\catcode `\\12\catcode `\$12\catcode
  `\&12\catcode `\#12\catcode `\^12\catcode `\_12\catcode `\%12\relax}%
\providecommand \@@startlink[1]{}%
\providecommand \@@endlink[0]{}%
\providecommand \url  [0]{\begingroup\@sanitize@url \@url }%
\providecommand \@url [1]{\endgroup\@href {#1}{\urlprefix }}%
\providecommand \urlprefix  [0]{URL }%
\providecommand \Eprint [0]{\href }%
\providecommand \doibase [0]{http://dx.doi.org/}%
\providecommand \selectlanguage [0]{\@gobble}%
\providecommand \bibinfo  [0]{\@secondoftwo}%
\providecommand \bibfield  [0]{\@secondoftwo}%
\providecommand \translation [1]{[#1]}%
\providecommand \BibitemOpen [0]{}%
\providecommand \bibitemStop [0]{}%
\providecommand \bibitemNoStop [0]{.\EOS\space}%
\providecommand \EOS [0]{\spacefactor3000\relax}%
\providecommand \BibitemShut  [1]{\csname bibitem#1\endcsname}%
\let\auto@bib@innerbib\@empty
\bibitem [{\citenamefont {Tajima}\ \emph {et~al.}(1991)\citenamefont {Tajima},
  \citenamefont {Horton}, \citenamefont {Morrison}, \citenamefont {Schutkeker},
  \citenamefont {Kamimura}, \citenamefont {Mima},\ and\ \citenamefont
  {Abe}}]{tajima1991instabilities}%
  \BibitemOpen
  \bibfield  {author} {\bibinfo {author} {\bibfnamefont {T.}~\bibnamefont
  {Tajima}}, \bibinfo {author} {\bibfnamefont {W.}~\bibnamefont {Horton}},
  \bibinfo {author} {\bibfnamefont {P.~J.}\ \bibnamefont {Morrison}}, \bibinfo
  {author} {\bibfnamefont {J.}~\bibnamefont {Schutkeker}}, \bibinfo {author}
  {\bibfnamefont {T.}~\bibnamefont {Kamimura}}, \bibinfo {author}
  {\bibfnamefont {K.}~\bibnamefont {Mima}}, \ and\ \bibinfo {author}
  {\bibfnamefont {Y.}~\bibnamefont {Abe}},\ }\href {\doibase
  https://doi.org/10.1063/1.859850} {\bibfield  {journal} {\bibinfo  {journal}
  {Phys. Fluids B}\ }\textbf {\bibinfo {volume} {3}},\ \bibinfo {pages} {938}
  (\bibinfo {year} {1991})}\BibitemShut {NoStop}%
\bibitem [{\citenamefont {Beklemishev}\ \emph {et~al.}(2010)\citenamefont
  {Beklemishev}, \citenamefont {Bagryansky}, \citenamefont {Chaschin},\ and\
  \citenamefont {Soldatkina}}]{beklemishev2010vortex}%
  \BibitemOpen
  \bibfield  {author} {\bibinfo {author} {\bibfnamefont {A.~D.}\ \bibnamefont
  {Beklemishev}}, \bibinfo {author} {\bibfnamefont {P.~A.}\ \bibnamefont
  {Bagryansky}}, \bibinfo {author} {\bibfnamefont {M.~S.}\ \bibnamefont
  {Chaschin}}, \ and\ \bibinfo {author} {\bibfnamefont {E.~I.}\ \bibnamefont
  {Soldatkina}},\ }\href {\doibase https://doi.org/10.13182/FST10-A9497}
  {\bibfield  {journal} {\bibinfo  {journal} {Fusion Sci. Technol.}\ }\textbf
  {\bibinfo {volume} {57}},\ \bibinfo {pages} {351} (\bibinfo {year}
  {2010})}\BibitemShut {NoStop}%
\bibitem [{\citenamefont {Ryutov}\ \emph {et~al.}(2011)\citenamefont {Ryutov},
  \citenamefont {Berk}, \citenamefont {Cohen}, \citenamefont {Molvik},\ and\
  \citenamefont {Simonen}}]{ryutov2011magneto}%
  \BibitemOpen
  \bibfield  {author} {\bibinfo {author} {\bibfnamefont {D.}~\bibnamefont
  {Ryutov}}, \bibinfo {author} {\bibfnamefont {H.}~\bibnamefont {Berk}},
  \bibinfo {author} {\bibfnamefont {B.}~\bibnamefont {Cohen}}, \bibinfo
  {author} {\bibfnamefont {A.}~\bibnamefont {Molvik}}, \ and\ \bibinfo {author}
  {\bibfnamefont {T.}~\bibnamefont {Simonen}},\ }\href {\doibase
  https://doi.org/10.1063/1.3624763} {\bibfield  {journal} {\bibinfo  {journal}
  {Phys. Plasmas}\ }\textbf {\bibinfo {volume} {18}},\ \bibinfo {pages}
  {092301} (\bibinfo {year} {2011})}\BibitemShut {NoStop}%
\bibitem [{\citenamefont {Hershkowitz}\ \emph {et~al.}(1982)\citenamefont
  {Hershkowitz}, \citenamefont {Breun}, \citenamefont {Callen}, \citenamefont
  {Chan}, \citenamefont {Ferron}, \citenamefont {Golovato}, \citenamefont
  {Pew}, \citenamefont {Nelson}, \citenamefont {Sing},\ and\ \citenamefont
  {Yujiri}}]{hershkowitz1982dynamic}%
  \BibitemOpen
  \bibfield  {author} {\bibinfo {author} {\bibfnamefont {N.}~\bibnamefont
  {Hershkowitz}}, \bibinfo {author} {\bibfnamefont {R.}~\bibnamefont {Breun}},
  \bibinfo {author} {\bibfnamefont {J.}~\bibnamefont {Callen}}, \bibinfo
  {author} {\bibfnamefont {C.}~\bibnamefont {Chan}}, \bibinfo {author}
  {\bibfnamefont {J.}~\bibnamefont {Ferron}}, \bibinfo {author} {\bibfnamefont
  {S.}~\bibnamefont {Golovato}}, \bibinfo {author} {\bibfnamefont
  {J.}~\bibnamefont {Pew}}, \bibinfo {author} {\bibfnamefont {B.}~\bibnamefont
  {Nelson}}, \bibinfo {author} {\bibfnamefont {D.}~\bibnamefont {Sing}}, \ and\
  \bibinfo {author} {\bibfnamefont {L.}~\bibnamefont {Yujiri}},\ }\href@noop {}
  {\bibfield  {journal} {\bibinfo  {journal} {Phys. Rev. Lett.}\ }\textbf
  {\bibinfo {volume} {49}},\ \bibinfo {pages} {1489} (\bibinfo {year}
  {1982})}\BibitemShut {NoStop}%
\bibitem [{\citenamefont {Ferron}\ \emph {et~al.}(1983)\citenamefont {Ferron},
  \citenamefont {Hershkowitz}, \citenamefont {Breun}, \citenamefont
  {Golovato},\ and\ \citenamefont {Goulding}}]{ferron1983rf}%
  \BibitemOpen
  \bibfield  {author} {\bibinfo {author} {\bibfnamefont {J.~R.}\ \bibnamefont
  {Ferron}}, \bibinfo {author} {\bibfnamefont {N.}~\bibnamefont {Hershkowitz}},
  \bibinfo {author} {\bibfnamefont {R.~A.}\ \bibnamefont {Breun}}, \bibinfo
  {author} {\bibfnamefont {S.~N.}\ \bibnamefont {Golovato}}, \ and\ \bibinfo
  {author} {\bibfnamefont {R.}~\bibnamefont {Goulding}},\ }\href {\doibase
  https://doi.org/10.1103/PhysRevLett.51.1955} {\bibfield  {journal} {\bibinfo
  {journal} {Phys. Rev. Lett.}\ }\textbf {\bibinfo {volume} {51}},\ \bibinfo
  {pages} {1955} (\bibinfo {year} {1983})}\BibitemShut {NoStop}%
\bibitem [{\citenamefont {Seemann}, \citenamefont {Be’ery},\ and\
  \citenamefont {Fisher}(2018)}]{seemann2018stabilization}%
  \BibitemOpen
  \bibfield  {author} {\bibinfo {author} {\bibfnamefont {O.}~\bibnamefont
  {Seemann}}, \bibinfo {author} {\bibfnamefont {I.}~\bibnamefont {Be’ery}}, \
  and\ \bibinfo {author} {\bibfnamefont {A.}~\bibnamefont {Fisher}},\ }\href
  {\doibase 10.1017/S0022377818000971} {\bibfield  {journal} {\bibinfo
  {journal} {J. of Plasma Phys.}\ }\textbf {\bibinfo {volume} {84}},\ \bibinfo
  {pages} {BO6.011} (\bibinfo {year} {2018})}\BibitemShut {NoStop}%
\bibitem [{\citenamefont {Zhil'tsov}\ \emph {et~al.}(1975)\citenamefont
  {Zhil'tsov}, \citenamefont {Likhtenshtejn}, \citenamefont {Panov},
  \citenamefont {Kosarev}, \citenamefont {Chuyanov},\ and\ \citenamefont
  {Shcherbakov}}]{zhil1975plasma}%
  \BibitemOpen
  \bibfield  {author} {\bibinfo {author} {\bibfnamefont {V.~A.}\ \bibnamefont
  {Zhil'tsov}}, \bibinfo {author} {\bibfnamefont {V.~K.}\ \bibnamefont
  {Likhtenshtejn}}, \bibinfo {author} {\bibfnamefont {D.~A.}\ \bibnamefont
  {Panov}}, \bibinfo {author} {\bibfnamefont {P.~M.}\ \bibnamefont {Kosarev}},
  \bibinfo {author} {\bibfnamefont {V.~A.}\ \bibnamefont {Chuyanov}}, \ and\
  \bibinfo {author} {\bibfnamefont {A.~G.}\ \bibnamefont {Shcherbakov}},\
  }\href@noop {} {\bibfield  {journal} {\bibinfo  {journal} {Nucl. Fusion}\ }
  (\bibinfo {year} {1975})}\BibitemShut {NoStop}%
\bibitem [{\citenamefont {Be’ery}\ and\ \citenamefont
  {Seemann}(2015)}]{be2015feedback}%
  \BibitemOpen
  \bibfield  {author} {\bibinfo {author} {\bibfnamefont {I.}~\bibnamefont
  {Be’ery}}\ and\ \bibinfo {author} {\bibfnamefont {O.}~\bibnamefont
  {Seemann}},\ }\href {\doibase 10.1088/0741-3335/57/8/085005} {\bibfield
  {journal} {\bibinfo  {journal} {Plasma Phys. Controlled Fusion}\ }\textbf
  {\bibinfo {volume} {57}},\ \bibinfo {pages} {085005} (\bibinfo {year}
  {2015})}\BibitemShut {NoStop}%
\bibitem [{\citenamefont {Inutake}\ \emph {et~al.}(1985)\citenamefont
  {Inutake}, \citenamefont {Cho}, \citenamefont {Ichimura}, \citenamefont
  {Ishii}, \citenamefont {Itakura}, \citenamefont {Katanuma}, \citenamefont
  {Kiwamoto}, \citenamefont {Kusama}, \citenamefont {Mase}, \citenamefont
  {Miyoshi}, \citenamefont {Nakashima}, \citenamefont {Saito}, \citenamefont
  {Sakasai}, \citenamefont {Sawada}, \citenamefont {Wakaida}, \citenamefont
  {Yamaguchi},\ and\ \citenamefont {Yatsu}}]{inutake1985thermal}%
  \BibitemOpen
  \bibfield  {author} {\bibinfo {author} {\bibfnamefont {M.}~\bibnamefont
  {Inutake}}, \bibinfo {author} {\bibfnamefont {T.}~\bibnamefont {Cho}},
  \bibinfo {author} {\bibfnamefont {M.}~\bibnamefont {Ichimura}}, \bibinfo
  {author} {\bibfnamefont {K.}~\bibnamefont {Ishii}}, \bibinfo {author}
  {\bibfnamefont {A.}~\bibnamefont {Itakura}}, \bibinfo {author} {\bibfnamefont
  {I.}~\bibnamefont {Katanuma}}, \bibinfo {author} {\bibfnamefont
  {Y.}~\bibnamefont {Kiwamoto}}, \bibinfo {author} {\bibfnamefont
  {Y.}~\bibnamefont {Kusama}}, \bibinfo {author} {\bibfnamefont
  {A.}~\bibnamefont {Mase}}, \bibinfo {author} {\bibfnamefont {S.}~\bibnamefont
  {Miyoshi}}, \bibinfo {author} {\bibfnamefont {Y.}~\bibnamefont {Nakashima}},
  \bibinfo {author} {\bibfnamefont {T.}~\bibnamefont {Saito}}, \bibinfo
  {author} {\bibfnamefont {A.}~\bibnamefont {Sakasai}}, \bibinfo {author}
  {\bibfnamefont {K.}~\bibnamefont {Sawada}}, \bibinfo {author} {\bibfnamefont
  {I.}~\bibnamefont {Wakaida}}, \bibinfo {author} {\bibfnamefont
  {N.}~\bibnamefont {Yamaguchi}}, \ and\ \bibinfo {author} {\bibfnamefont
  {K.}~\bibnamefont {Yatsu}},\ }\href {\doibase
  https://doi.org/10.1103/PhysRevLett.55.939} {\bibfield  {journal} {\bibinfo
  {journal} {Phys. Rev. Lett.}\ }\textbf {\bibinfo {volume} {55}},\ \bibinfo
  {pages} {939} (\bibinfo {year} {1985})}\BibitemShut {NoStop}%
\bibitem [{\citenamefont {Grubb}\ \emph {et~al.}(1984)\citenamefont {Grubb},
  \citenamefont {Allen}, \citenamefont {Casper}, \citenamefont {Clauser},
  \citenamefont {Coensgen}, \citenamefont {Correll}, \citenamefont {Cummins},
  \citenamefont {Damm}, \citenamefont {Foote}, \citenamefont {Goodman},
  \citenamefont {Hill}, \citenamefont {Hooper}, \citenamefont {Hornady},
  \citenamefont {Hunt}, \citenamefont {Kerr}, \citenamefont {Leppelmeier},
  \citenamefont {Marilleau}, \citenamefont {Moller}, \citenamefont {Molvik},
  \citenamefont {Nexsen}, \citenamefont {Pickles}, \citenamefont {Porter},
  \citenamefont {Poulsen}, \citenamefont {Silver}, \citenamefont {Simonen},
  \citenamefont {Stallard}, \citenamefont {Turner}, \citenamefont {Hsu},
  \citenamefont {Yu}, \citenamefont {Barter}, \citenamefont {Christensen},
  \citenamefont {Dimonte}, \citenamefont {Romesser}, \citenamefont {Ellis},
  \citenamefont {James}, \citenamefont {Lasnier}, \citenamefont {Berzins},
  \citenamefont {Carter}, \citenamefont {Clower}, \citenamefont {Failor},
  \citenamefont {Falabella}, \citenamefont {Flammer},\ and\ \citenamefont
  {Nash}}]{grubb1984thermal}%
  \BibitemOpen
  \bibfield  {author} {\bibinfo {author} {\bibfnamefont {D.~P.}\ \bibnamefont
  {Grubb}}, \bibinfo {author} {\bibfnamefont {S.~L.}\ \bibnamefont {Allen}},
  \bibinfo {author} {\bibfnamefont {T.~A.}\ \bibnamefont {Casper}}, \bibinfo
  {author} {\bibfnamefont {J.~F.}\ \bibnamefont {Clauser}}, \bibinfo {author}
  {\bibfnamefont {F.~H.}\ \bibnamefont {Coensgen}}, \bibinfo {author}
  {\bibfnamefont {D.~L.}\ \bibnamefont {Correll}}, \bibinfo {author}
  {\bibfnamefont {W.~F.}\ \bibnamefont {Cummins}}, \bibinfo {author}
  {\bibfnamefont {C.~C.}\ \bibnamefont {Damm}}, \bibinfo {author}
  {\bibfnamefont {J.~H.}\ \bibnamefont {Foote}}, \bibinfo {author}
  {\bibfnamefont {R.~K.}\ \bibnamefont {Goodman}}, \bibinfo {author}
  {\bibfnamefont {D.~N.}\ \bibnamefont {Hill}}, \bibinfo {author}
  {\bibfnamefont {E.~B.}\ \bibnamefont {Hooper}}, \bibinfo {author}
  {\bibfnamefont {R.~S.}\ \bibnamefont {Hornady}}, \bibinfo {author}
  {\bibfnamefont {A.~L.}\ \bibnamefont {Hunt}}, \bibinfo {author}
  {\bibfnamefont {R.~G.}\ \bibnamefont {Kerr}}, \bibinfo {author}
  {\bibfnamefont {G.~W.}\ \bibnamefont {Leppelmeier}}, \bibinfo {author}
  {\bibfnamefont {J.}~\bibnamefont {Marilleau}}, \bibinfo {author}
  {\bibfnamefont {J.~M.}\ \bibnamefont {Moller}}, \bibinfo {author}
  {\bibfnamefont {A.~W.}\ \bibnamefont {Molvik}}, \bibinfo {author}
  {\bibfnamefont {W.~E.}\ \bibnamefont {Nexsen}}, \bibinfo {author}
  {\bibfnamefont {W.~L.}\ \bibnamefont {Pickles}}, \bibinfo {author}
  {\bibfnamefont {G.~D.}\ \bibnamefont {Porter}}, \bibinfo {author}
  {\bibfnamefont {P.}~\bibnamefont {Poulsen}}, \bibinfo {author} {\bibfnamefont
  {E.~H.}\ \bibnamefont {Silver}}, \bibinfo {author} {\bibfnamefont {T.~C.}\
  \bibnamefont {Simonen}}, \bibinfo {author} {\bibfnamefont {B.~W.}\
  \bibnamefont {Stallard}}, \bibinfo {author} {\bibfnamefont {W.~C.}\
  \bibnamefont {Turner}}, \bibinfo {author} {\bibfnamefont {W.~L.}\
  \bibnamefont {Hsu}}, \bibinfo {author} {\bibfnamefont {T.~L.}\ \bibnamefont
  {Yu}}, \bibinfo {author} {\bibfnamefont {J.~D.}\ \bibnamefont {Barter}},
  \bibinfo {author} {\bibfnamefont {T.}~\bibnamefont {Christensen}}, \bibinfo
  {author} {\bibfnamefont {G.}~\bibnamefont {Dimonte}}, \bibinfo {author}
  {\bibfnamefont {T.~W.}\ \bibnamefont {Romesser}}, \bibinfo {author}
  {\bibfnamefont {R.~F.}\ \bibnamefont {Ellis}}, \bibinfo {author}
  {\bibfnamefont {R.~A.}\ \bibnamefont {James}}, \bibinfo {author}
  {\bibfnamefont {C.~J.}\ \bibnamefont {Lasnier}}, \bibinfo {author}
  {\bibfnamefont {L.~V.}\ \bibnamefont {Berzins}}, \bibinfo {author}
  {\bibfnamefont {M.~R.}\ \bibnamefont {Carter}}, \bibinfo {author}
  {\bibfnamefont {C.~A.}\ \bibnamefont {Clower}}, \bibinfo {author}
  {\bibfnamefont {B.~H.}\ \bibnamefont {Failor}}, \bibinfo {author}
  {\bibfnamefont {S.}~\bibnamefont {Falabella}}, \bibinfo {author}
  {\bibfnamefont {M.}~\bibnamefont {Flammer}}, \ and\ \bibinfo {author}
  {\bibfnamefont {T.}~\bibnamefont {Nash}},\ }\href {\doibase
  https://doi.org/10.1103/PhysRevLett.53.783} {\bibfield  {journal} {\bibinfo
  {journal} {Phys. Rev. Lett.}\ }\textbf {\bibinfo {volume} {53}},\ \bibinfo
  {pages} {783} (\bibinfo {year} {1984})}\BibitemShut {NoStop}%
\bibitem [{\citenamefont {Katanuma}\ \emph {et~al.}(1986)\citenamefont
  {Katanuma}, \citenamefont {Kiwamoto}, \citenamefont {Ishii},\ and\
  \citenamefont {Miyoshi}}]{katanuma1986thermal}%
  \BibitemOpen
  \bibfield  {author} {\bibinfo {author} {\bibfnamefont {I.}~\bibnamefont
  {Katanuma}}, \bibinfo {author} {\bibfnamefont {Y.}~\bibnamefont {Kiwamoto}},
  \bibinfo {author} {\bibfnamefont {K.}~\bibnamefont {Ishii}}, \ and\ \bibinfo
  {author} {\bibfnamefont {S.}~\bibnamefont {Miyoshi}},\ }\href {\doibase
  https://doi.org/10.1063/1.865758} {\bibfield  {journal} {\bibinfo  {journal}
  {Phys. fluids}\ }\textbf {\bibinfo {volume} {29}},\ \bibinfo {pages} {4138}
  (\bibinfo {year} {1986})}\BibitemShut {NoStop}%
\bibitem [{\citenamefont {Pratt}\ and\ \citenamefont
  {Horton}(2006)}]{pratt2006global}%
  \BibitemOpen
  \bibfield  {author} {\bibinfo {author} {\bibfnamefont {J.}~\bibnamefont
  {Pratt}}\ and\ \bibinfo {author} {\bibfnamefont {W.}~\bibnamefont {Horton}},\
  }\href {\doibase 10.1063/1.2188913} {\bibfield  {journal} {\bibinfo
  {journal} {Phys. plasmas}\ }\textbf {\bibinfo {volume} {13}},\ \bibinfo
  {pages} {042513} (\bibinfo {year} {2006})}\BibitemShut {NoStop}%
\bibitem [{\citenamefont {Tamano}(1995)}]{tamano1995tandem}%
  \BibitemOpen
  \bibfield  {author} {\bibinfo {author} {\bibfnamefont {T.}~\bibnamefont
  {Tamano}},\ }\href {\doibase 10.1063/1.871256} {\bibfield  {journal}
  {\bibinfo  {journal} {Phys. Plasmas}\ }\textbf {\bibinfo {volume} {2}},\
  \bibinfo {pages} {2321} (\bibinfo {year} {1995})}\BibitemShut {NoStop}%
\bibitem [{\citenamefont {Ivanov}\ and\ \citenamefont
  {Prikhodko}(2013)}]{ivanov2013gas}%
  \BibitemOpen
  \bibfield  {author} {\bibinfo {author} {\bibfnamefont {A.~A.}\ \bibnamefont
  {Ivanov}}\ and\ \bibinfo {author} {\bibfnamefont {V.}~\bibnamefont
  {Prikhodko}},\ }\href@noop {} {\bibfield  {journal} {\bibinfo  {journal}
  {Plasma Phys. Controlled Fusion}\ }\textbf {\bibinfo {volume} {55}},\
  \bibinfo {pages} {063001} (\bibinfo {year} {2013})}\BibitemShut {NoStop}%
\bibitem [{\citenamefont {Ivanov}\ and\ \citenamefont
  {Prikhodko}(2017)}]{ivanov2017gas}%
  \BibitemOpen
  \bibfield  {author} {\bibinfo {author} {\bibfnamefont {A.~A.}\ \bibnamefont
  {Ivanov}}\ and\ \bibinfo {author} {\bibfnamefont {V.~V.}\ \bibnamefont
  {Prikhodko}},\ }\href@noop {} {\bibfield  {journal} {\bibinfo  {journal}
  {Physics-Uspekhi}\ }\textbf {\bibinfo {volume} {60}},\ \bibinfo {pages} {509}
  (\bibinfo {year} {2017})}\BibitemShut {NoStop}%
\bibitem [{\citenamefont {Anderson}\ \emph {et~al.}(2020)\citenamefont
  {Anderson}, \citenamefont {Clark}, \citenamefont {Forest}, \citenamefont
  {Geiger}, \citenamefont {Mirnov}, \citenamefont {Oliva}, \citenamefont
  {Pizzo}, \citenamefont {Schmitz}, \citenamefont {Wallace}, \citenamefont
  {Kristofek}, \citenamefont {Mumgaaed}, \citenamefont {Peterson},
  \citenamefont {Ram}, \citenamefont {Whyte}, \citenamefont {Wright},
  \citenamefont {Wukitch}, \citenamefont {Green}, \citenamefont {Harvey},
  \citenamefont {Petrov}, \citenamefont {Srinivisan},\ and\ \citenamefont
  {Hakim}}]{anderson2020introducing}%
  \BibitemOpen
  \bibfield  {author} {\bibinfo {author} {\bibfnamefont {J.}~\bibnamefont
  {Anderson}}, \bibinfo {author} {\bibfnamefont {M.}~\bibnamefont {Clark}},
  \bibinfo {author} {\bibfnamefont {C.}~\bibnamefont {Forest}}, \bibinfo
  {author} {\bibfnamefont {B.}~\bibnamefont {Geiger}}, \bibinfo {author}
  {\bibfnamefont {V.}~\bibnamefont {Mirnov}}, \bibinfo {author} {\bibfnamefont
  {S.}~\bibnamefont {Oliva}}, \bibinfo {author} {\bibfnamefont
  {J.}~\bibnamefont {Pizzo}}, \bibinfo {author} {\bibfnamefont
  {O.}~\bibnamefont {Schmitz}}, \bibinfo {author} {\bibfnamefont
  {J.}~\bibnamefont {Wallace}}, \bibinfo {author} {\bibfnamefont
  {G.}~\bibnamefont {Kristofek}}, \bibinfo {author} {\bibfnamefont
  {R.}~\bibnamefont {Mumgaaed}}, \bibinfo {author} {\bibfnamefont
  {E.}~\bibnamefont {Peterson}}, \bibinfo {author} {\bibfnamefont
  {A.}~\bibnamefont {Ram}}, \bibinfo {author} {\bibfnamefont {D.}~\bibnamefont
  {Whyte}}, \bibinfo {author} {\bibfnamefont {J.}~\bibnamefont {Wright}},
  \bibinfo {author} {\bibfnamefont {S.}~\bibnamefont {Wukitch}}, \bibinfo
  {author} {\bibfnamefont {D.}~\bibnamefont {Green}}, \bibinfo {author}
  {\bibfnamefont {R.}~\bibnamefont {Harvey}}, \bibinfo {author} {\bibfnamefont
  {Y.}~\bibnamefont {Petrov}}, \bibinfo {author} {\bibfnamefont
  {B.}~\bibnamefont {Srinivisan}}, \ and\ \bibinfo {author} {\bibfnamefont
  {A.}~\bibnamefont {Hakim}},\ }in\ \href@noop {} {\emph {\bibinfo {booktitle}
  {APS Division of Plasma Physics Meeting Abstracts}}}\ (\bibinfo {year}
  {2020})\BibitemShut {NoStop}%
\bibitem [{\citenamefont {Forest}(2022)}]{forest2022physics}%
  \BibitemOpen
  \bibfield  {author} {\bibinfo {author} {\bibfnamefont {C.}~\bibnamefont
  {Forest}},\ }\href@noop {} {\bibfield  {journal} {\bibinfo  {journal}
  {Bulletin of the American Physical Society}\ } (\bibinfo {year}
  {2022})}\BibitemShut {NoStop}%
\bibitem [{\citenamefont {Egedal}\ \emph {et~al.}(2022)\citenamefont {Egedal},
  \citenamefont {Endrizzi}, \citenamefont {Forest},\ and\ \citenamefont
  {Fowler}}]{egedal2022fusion}%
  \BibitemOpen
  \bibfield  {author} {\bibinfo {author} {\bibfnamefont {J.}~\bibnamefont
  {Egedal}}, \bibinfo {author} {\bibfnamefont {D.}~\bibnamefont {Endrizzi}},
  \bibinfo {author} {\bibfnamefont {C.}~\bibnamefont {Forest}}, \ and\ \bibinfo
  {author} {\bibfnamefont {T.}~\bibnamefont {Fowler}},\ }\href@noop {}
  {\bibfield  {journal} {\bibinfo  {journal} {Nucl. Fusion}\ }\textbf {\bibinfo
  {volume} {62}},\ \bibinfo {pages} {126053} (\bibinfo {year}
  {2022})}\BibitemShut {NoStop}%
\bibitem [{\citenamefont {Beklemishev}(2016)}]{beklemishev2016diamagnetic}%
  \BibitemOpen
  \bibfield  {author} {\bibinfo {author} {\bibfnamefont {A.~D.}\ \bibnamefont
  {Beklemishev}},\ }\href {\doibase 10.1063/1.4960129} {\bibfield  {journal}
  {\bibinfo  {journal} {Phys. Plasmas}\ }\textbf {\bibinfo {volume} {23}},\
  \bibinfo {pages} {082506} (\bibinfo {year} {2016})}\BibitemShut {NoStop}%
\bibitem [{\citenamefont {Kotelnikov}(2020)}]{kotelnikov2020structure}%
  \BibitemOpen
  \bibfield  {author} {\bibinfo {author} {\bibfnamefont {I.}~\bibnamefont
  {Kotelnikov}},\ }\href {\doibase 10.1088/1361-6587/ab8a63} {\bibfield
  {journal} {\bibinfo  {journal} {Plasma Phys. Controlled Fusion}\ }\textbf
  {\bibinfo {volume} {62}},\ \bibinfo {pages} {075002} (\bibinfo {year}
  {2020})}\BibitemShut {NoStop}%
\bibitem [{\citenamefont {Post}(1967)}]{post1967confinement}%
  \BibitemOpen
  \bibfield  {author} {\bibinfo {author} {\bibfnamefont {R.~F.}\ \bibnamefont
  {Post}},\ }\href {\doibase https://doi.org/10.1103/PhysRevLett.18.232}
  {\bibfield  {journal} {\bibinfo  {journal} {Phys. Rev. Lett.}\ }\textbf
  {\bibinfo {volume} {18}},\ \bibinfo {pages} {232} (\bibinfo {year}
  {1967})}\BibitemShut {NoStop}%
\bibitem [{\citenamefont {Logan}\ \emph
  {et~al.}(1972{\natexlab{a}})\citenamefont {Logan}, \citenamefont
  {Lichtenberg}, \citenamefont {Lieberman},\ and\ \citenamefont
  {Makhijani}}]{logan1972multiple}%
  \BibitemOpen
  \bibfield  {author} {\bibinfo {author} {\bibfnamefont {B.~G.}\ \bibnamefont
  {Logan}}, \bibinfo {author} {\bibfnamefont {A.~J.}\ \bibnamefont
  {Lichtenberg}}, \bibinfo {author} {\bibfnamefont {M.~A.}\ \bibnamefont
  {Lieberman}}, \ and\ \bibinfo {author} {\bibfnamefont {A.}~\bibnamefont
  {Makhijani}},\ }\href {\doibase https://doi.org/10.1103/PhysRevLett.28.144}
  {\bibfield  {journal} {\bibinfo  {journal} {Phys. Rev. Lett.}\ }\textbf
  {\bibinfo {volume} {28}},\ \bibinfo {pages} {144} (\bibinfo {year}
  {1972}{\natexlab{a}})}\BibitemShut {NoStop}%
\bibitem [{\citenamefont {Logan}\ \emph
  {et~al.}(1972{\natexlab{b}})\citenamefont {Logan}, \citenamefont {Brown},
  \citenamefont {Lieberman},\ and\ \citenamefont
  {Lichtenberg}}]{logan1972experimental}%
  \BibitemOpen
  \bibfield  {author} {\bibinfo {author} {\bibfnamefont {B.~G.}\ \bibnamefont
  {Logan}}, \bibinfo {author} {\bibfnamefont {I.~G.}\ \bibnamefont {Brown}},
  \bibinfo {author} {\bibfnamefont {M.~A.}\ \bibnamefont {Lieberman}}, \ and\
  \bibinfo {author} {\bibfnamefont {A.~J.}\ \bibnamefont {Lichtenberg}},\
  }\href {\doibase https://doi.org/10.1103/PhysRevLett.29.1435} {\bibfield
  {journal} {\bibinfo  {journal} {Phys. Rev. Lett.}\ }\textbf {\bibinfo
  {volume} {29}},\ \bibinfo {pages} {1435} (\bibinfo {year}
  {1972}{\natexlab{b}})}\BibitemShut {NoStop}%
\bibitem [{\citenamefont {Mirnov}\ and\ \citenamefont
  {Ryutov}(1972)}]{mirnov1972gas}%
  \BibitemOpen
  \bibfield  {author} {\bibinfo {author} {\bibfnamefont {V.~V.}\ \bibnamefont
  {Mirnov}}\ and\ \bibinfo {author} {\bibfnamefont {D.~D.}\ \bibnamefont
  {Ryutov}},\ }\href {\doibase 10.1088/0029-5515/12/6/001} {\bibfield
  {journal} {\bibinfo  {journal} {Nucl. Fusion}\ }\textbf {\bibinfo {volume}
  {12}},\ \bibinfo {pages} {627} (\bibinfo {year} {1972})}\BibitemShut
  {NoStop}%
\bibitem [{\citenamefont {Makhijani}\ \emph {et~al.}(1974)\citenamefont
  {Makhijani}, \citenamefont {Lichtenberg}, \citenamefont {Lieberman},\ and\
  \citenamefont {Logan}}]{makhijani1974plasma}%
  \BibitemOpen
  \bibfield  {author} {\bibinfo {author} {\bibfnamefont {A.}~\bibnamefont
  {Makhijani}}, \bibinfo {author} {\bibfnamefont {A.~J.}\ \bibnamefont
  {Lichtenberg}}, \bibinfo {author} {\bibfnamefont {M.~A.}\ \bibnamefont
  {Lieberman}}, \ and\ \bibinfo {author} {\bibfnamefont {B.~G.}\ \bibnamefont
  {Logan}},\ }\href {\doibase 10.1063/1.1694881} {\bibfield  {journal}
  {\bibinfo  {journal} {Phys. Fluids}\ }\textbf {\bibinfo {volume} {17}},\
  \bibinfo {pages} {1291} (\bibinfo {year} {1974})}\BibitemShut {NoStop}%
\bibitem [{\citenamefont {Tuszewski}, \citenamefont {Lichtenberg},\ and\
  \citenamefont {Eylon}(1977)}]{tuszewski1977transient}%
  \BibitemOpen
  \bibfield  {author} {\bibinfo {author} {\bibfnamefont {M.}~\bibnamefont
  {Tuszewski}}, \bibinfo {author} {\bibfnamefont {A.~J.}\ \bibnamefont
  {Lichtenberg}}, \ and\ \bibinfo {author} {\bibfnamefont {S.}~\bibnamefont
  {Eylon}},\ }\href {\doibase 10.1088/0029-5515/17/5/002} {\bibfield  {journal}
  {\bibinfo  {journal} {Nucl. Fusion}\ }\textbf {\bibinfo {volume} {17}},\
  \bibinfo {pages} {893} (\bibinfo {year} {1977})}\BibitemShut {NoStop}%
\bibitem [{\citenamefont {Burdakov}\ and\ \citenamefont
  {Postupaev}(2016)}]{burdakov2016multiple}%
  \BibitemOpen
  \bibfield  {author} {\bibinfo {author} {\bibfnamefont {A.~V.}\ \bibnamefont
  {Burdakov}}\ and\ \bibinfo {author} {\bibfnamefont {V.~V.}\ \bibnamefont
  {Postupaev}},\ }in\ \href {\doibase 10.1063/1.4964241} {\emph {\bibinfo
  {booktitle} {AIP Conference Proceedings}}},\ Vol.\ \bibinfo {volume} {1771}\
  (\bibinfo {organization} {AIP Publishing LLC},\ \bibinfo {year} {2016})\ p.\
  \bibinfo {pages} {080002}\BibitemShut {NoStop}%
\bibitem [{\citenamefont {Budker}, \citenamefont {Mirnov},\ and\ \citenamefont
  {Ryutov}(1971)}]{budker1971influence}%
  \BibitemOpen
  \bibfield  {author} {\bibinfo {author} {\bibfnamefont {G.~I.}\ \bibnamefont
  {Budker}}, \bibinfo {author} {\bibfnamefont {V.~V.}\ \bibnamefont {Mirnov}},
  \ and\ \bibinfo {author} {\bibfnamefont {D.~D.}\ \bibnamefont {Ryutov}},\
  }\href@noop {} {\bibfield  {journal} {\bibinfo  {journal} {JETP lett}\
  }\textbf {\bibinfo {volume} {14}},\ \bibinfo {pages} {212} (\bibinfo {year}
  {1971})}\BibitemShut {NoStop}%
\bibitem [{\citenamefont {Mirnov}\ and\ \citenamefont
  {Lichtenberg}(1996)}]{mirnov1996multiple}%
  \BibitemOpen
  \bibfield  {author} {\bibinfo {author} {\bibfnamefont {V.~V.}\ \bibnamefont
  {Mirnov}}\ and\ \bibinfo {author} {\bibfnamefont {A.~J.}\ \bibnamefont
  {Lichtenberg}},\ }\href@noop {} {\bibfield  {journal} {\bibinfo  {journal}
  {Rev. of Plasma Phys.}\ }\textbf {\bibinfo {volume} {19}},\ \bibinfo {pages}
  {53} (\bibinfo {year} {1996})}\BibitemShut {NoStop}%
\bibitem [{\citenamefont {Kotelnikov}(2007)}]{kotelnikov2007new}%
  \BibitemOpen
  \bibfield  {author} {\bibinfo {author} {\bibfnamefont {I.~A.}\ \bibnamefont
  {Kotelnikov}},\ }\href {\doibase https://doi.org/10.13182/FST07-A1346}
  {\bibfield  {journal} {\bibinfo  {journal} {Fusion Sci. Technol.}\ }\textbf
  {\bibinfo {volume} {51}},\ \bibinfo {pages} {186} (\bibinfo {year}
  {2007})}\BibitemShut {NoStop}%
\bibitem [{\citenamefont {Tuck}(1968)}]{tuck1968reduction}%
  \BibitemOpen
  \bibfield  {author} {\bibinfo {author} {\bibfnamefont {J.~L.}\ \bibnamefont
  {Tuck}},\ }\href {\doibase 10.1103/PhysRevLett.20.715} {\bibfield  {journal}
  {\bibinfo  {journal} {Phys. Rev. Lett.}\ }\textbf {\bibinfo {volume} {20}},\
  \bibinfo {pages} {715} (\bibinfo {year} {1968})}\BibitemShut {NoStop}%
\bibitem [{\citenamefont {Budker}, \citenamefont {Mironov},\ and\ \citenamefont
  {Ryutov}(1982)}]{budker1982gas}%
  \BibitemOpen
  \bibfield  {author} {\bibinfo {author} {\bibfnamefont {G.~I.}\ \bibnamefont
  {Budker}}, \bibinfo {author} {\bibfnamefont {V.~V.}\ \bibnamefont {Mironov}},
  \ and\ \bibinfo {author} {\bibfnamefont {D.~D.}\ \bibnamefont {Ryutov}},\
  }in\ \href@noop {} {\emph {\bibinfo {booktitle} {Collection of papers}}}\
  (\bibinfo {year} {1982})\BibitemShut {NoStop}%
\bibitem [{\citenamefont {Beklemishev}(2013)}]{beklemishev2013helicoidal}%
  \BibitemOpen
  \bibfield  {author} {\bibinfo {author} {\bibfnamefont {A.~D.}\ \bibnamefont
  {Beklemishev}},\ }\href {\doibase https://doi.org/10.13182/FST13-A16953}
  {\bibfield  {journal} {\bibinfo  {journal} {Fusion Sci. Technol.}\ }\textbf
  {\bibinfo {volume} {63}},\ \bibinfo {pages} {355} (\bibinfo {year}
  {2013})}\BibitemShut {NoStop}%
\bibitem [{\citenamefont {Postupaev}\ \emph {et~al.}(2016)\citenamefont
  {Postupaev}, \citenamefont {Sudnikov}, \citenamefont {Beklemishev},\ and\
  \citenamefont {Ivanov}}]{postupaev2016helical}%
  \BibitemOpen
  \bibfield  {author} {\bibinfo {author} {\bibfnamefont {V.~V.}\ \bibnamefont
  {Postupaev}}, \bibinfo {author} {\bibfnamefont {A.~V.}\ \bibnamefont
  {Sudnikov}}, \bibinfo {author} {\bibfnamefont {A.~D.}\ \bibnamefont
  {Beklemishev}}, \ and\ \bibinfo {author} {\bibfnamefont {I.~A.}\ \bibnamefont
  {Ivanov}},\ }\href {\doibase https://doi.org/10.1016/j.fusengdes.2016.03.029}
  {\bibfield  {journal} {\bibinfo  {journal} {Fusion Engineering and Design}\
  }\textbf {\bibinfo {volume} {106}},\ \bibinfo {pages} {29} (\bibinfo {year}
  {2016})}\BibitemShut {NoStop}%
\bibitem [{\citenamefont {Sudnikov}\ \emph {et~al.}(2019)\citenamefont
  {Sudnikov}, \citenamefont {Beklemishev}, \citenamefont {Postupaev},
  \citenamefont {Ivanov}, \citenamefont {Inzhevatkina}, \citenamefont
  {Sklyarov}, \citenamefont {Burdakov}, \citenamefont {Kuklin}, \citenamefont
  {Rovenskikh},\ and\ \citenamefont {Melnikov}}]{sudnikov2019first}%
  \BibitemOpen
  \bibfield  {author} {\bibinfo {author} {\bibfnamefont {A.~V.}\ \bibnamefont
  {Sudnikov}}, \bibinfo {author} {\bibfnamefont {A.~D.}\ \bibnamefont
  {Beklemishev}}, \bibinfo {author} {\bibfnamefont {V.~V.}\ \bibnamefont
  {Postupaev}}, \bibinfo {author} {\bibfnamefont {I.~A.}\ \bibnamefont
  {Ivanov}}, \bibinfo {author} {\bibfnamefont {A.~A.}\ \bibnamefont
  {Inzhevatkina}}, \bibinfo {author} {\bibfnamefont {V.~F.}\ \bibnamefont
  {Sklyarov}}, \bibinfo {author} {\bibfnamefont {A.~V.}\ \bibnamefont
  {Burdakov}}, \bibinfo {author} {\bibfnamefont {K.~N.}\ \bibnamefont
  {Kuklin}}, \bibinfo {author} {\bibfnamefont {A.~F.}\ \bibnamefont
  {Rovenskikh}}, \ and\ \bibinfo {author} {\bibfnamefont {N.~A.}\ \bibnamefont
  {Melnikov}},\ }\href {\doibase 10.1585/pfr.14.2402023} {\bibfield  {journal}
  {\bibinfo  {journal} {Plasma and Fusion Research}\ }\textbf {\bibinfo
  {volume} {14}},\ \bibinfo {pages} {2402023} (\bibinfo {year}
  {2019})}\BibitemShut {NoStop}%
\bibitem [{\citenamefont {Ivanov}\ \emph {et~al.}(2021)\citenamefont {Ivanov},
  \citenamefont {Ustyuzhanin}, \citenamefont {Sudnikov},\ and\ \citenamefont
  {Inzhevatkina}}]{ivanov2021long}%
  \BibitemOpen
  \bibfield  {author} {\bibinfo {author} {\bibfnamefont {I.~A.}\ \bibnamefont
  {Ivanov}}, \bibinfo {author} {\bibfnamefont {V.}~\bibnamefont {Ustyuzhanin}},
  \bibinfo {author} {\bibfnamefont {A.}~\bibnamefont {Sudnikov}}, \ and\
  \bibinfo {author} {\bibfnamefont {A.}~\bibnamefont {Inzhevatkina}},\
  }\href@noop {} {\bibfield  {journal} {\bibinfo  {journal} {J. of Plasma
  Phys.}\ }\textbf {\bibinfo {volume} {87}} (\bibinfo {year}
  {2021})}\BibitemShut {NoStop}%
\bibitem [{\citenamefont {Sudnikov}\ \emph {et~al.}(2022)\citenamefont
  {Sudnikov}, \citenamefont {Ivanov}, \citenamefont {Inzhevatkina},
  \citenamefont {Larichkin}, \citenamefont {Lomov}, \citenamefont {Postupaev},
  \citenamefont {Tolkachev},\ and\ \citenamefont
  {Ustyuzhanin}}]{sudnikov2022plasma}%
  \BibitemOpen
  \bibfield  {author} {\bibinfo {author} {\bibfnamefont {A.~V.}\ \bibnamefont
  {Sudnikov}}, \bibinfo {author} {\bibfnamefont {I.~A.}\ \bibnamefont
  {Ivanov}}, \bibinfo {author} {\bibfnamefont {A.~A.}\ \bibnamefont
  {Inzhevatkina}}, \bibinfo {author} {\bibfnamefont {M.~V.}\ \bibnamefont
  {Larichkin}}, \bibinfo {author} {\bibfnamefont {K.~A.}\ \bibnamefont
  {Lomov}}, \bibinfo {author} {\bibfnamefont {V.~V.}\ \bibnamefont
  {Postupaev}}, \bibinfo {author} {\bibfnamefont {M.~S.}\ \bibnamefont
  {Tolkachev}}, \ and\ \bibinfo {author} {\bibfnamefont {V.~O.}\ \bibnamefont
  {Ustyuzhanin}},\ }\href@noop {} {\bibfield  {journal} {\bibinfo  {journal}
  {J. of Plasma Phys.}\ }\textbf {\bibinfo {volume} {88}} (\bibinfo {year}
  {2022})}\BibitemShut {NoStop}%
\bibitem [{\citenamefont {Motz}\ and\ \citenamefont
  {Watson}(1967)}]{motz1967radio}%
  \BibitemOpen
  \bibfield  {author} {\bibinfo {author} {\bibfnamefont {H.}~\bibnamefont
  {Motz}}\ and\ \bibinfo {author} {\bibfnamefont {C.}~\bibnamefont {Watson}},\
  }\href@noop {} {\bibfield  {journal} {\bibinfo  {journal} {Advances in
  Electronics and Electron Physics}\ }\textbf {\bibinfo {volume} {23}},\
  \bibinfo {pages} {153} (\bibinfo {year} {1967})}\BibitemShut {NoStop}%
\bibitem [{\citenamefont {Watari}\ \emph {et~al.}(1974)\citenamefont {Watari},
  \citenamefont {Hiroe}, \citenamefont {Sato},\ and\ \citenamefont
  {Ichimaru}}]{watari1974theory}%
  \BibitemOpen
  \bibfield  {author} {\bibinfo {author} {\bibfnamefont {T.}~\bibnamefont
  {Watari}}, \bibinfo {author} {\bibfnamefont {S.}~\bibnamefont {Hiroe}},
  \bibinfo {author} {\bibfnamefont {T.}~\bibnamefont {Sato}}, \ and\ \bibinfo
  {author} {\bibfnamefont {S.}~\bibnamefont {Ichimaru}},\ }\href {\doibase
  https://doi.org/10.1063/1.1694669} {\bibfield  {journal} {\bibinfo  {journal}
  {Phys. Fluids}\ }\textbf {\bibinfo {volume} {17}},\ \bibinfo {pages} {2107}
  (\bibinfo {year} {1974})}\BibitemShut {NoStop}%
\bibitem [{\citenamefont {Hatori}\ and\ \citenamefont
  {Watanabe}(1975)}]{hatori1975critical}%
  \BibitemOpen
  \bibfield  {author} {\bibinfo {author} {\bibfnamefont {T.}~\bibnamefont
  {Hatori}}\ and\ \bibinfo {author} {\bibfnamefont {T.}~\bibnamefont
  {Watanabe}},\ }\href {\doibase https://doi.org/10.1088/0029-5515/15/1/015}
  {\bibfield  {journal} {\bibinfo  {journal} {Nucl. Fusion}\ }\textbf {\bibinfo
  {volume} {15}},\ \bibinfo {pages} {143} (\bibinfo {year} {1975})}\BibitemShut
  {NoStop}%
\bibitem [{\citenamefont {Watari}\ \emph {et~al.}(1978)\citenamefont {Watari},
  \citenamefont {Hatori}, \citenamefont {Kumazawa}, \citenamefont {Hidekuma},
  \citenamefont {Aoki}, \citenamefont {Kawamoto}, \citenamefont {Inutake},
  \citenamefont {Hiroe}, \citenamefont {Nishizawa}, \citenamefont {Adati},
  \citenamefont {Sato}, \citenamefont {Watanabe}, \citenamefont {Obayashi},\
  and\ \citenamefont {Takayama}}]{watari1978radio}%
  \BibitemOpen
  \bibfield  {author} {\bibinfo {author} {\bibfnamefont {T.}~\bibnamefont
  {Watari}}, \bibinfo {author} {\bibfnamefont {T.}~\bibnamefont {Hatori}},
  \bibinfo {author} {\bibfnamefont {R.}~\bibnamefont {Kumazawa}}, \bibinfo
  {author} {\bibfnamefont {S.}~\bibnamefont {Hidekuma}}, \bibinfo {author}
  {\bibfnamefont {T.}~\bibnamefont {Aoki}}, \bibinfo {author} {\bibfnamefont
  {T.}~\bibnamefont {Kawamoto}}, \bibinfo {author} {\bibfnamefont
  {M.}~\bibnamefont {Inutake}}, \bibinfo {author} {\bibfnamefont
  {S.}~\bibnamefont {Hiroe}}, \bibinfo {author} {\bibfnamefont
  {A.}~\bibnamefont {Nishizawa}}, \bibinfo {author} {\bibfnamefont
  {K.}~\bibnamefont {Adati}}, \bibinfo {author} {\bibfnamefont
  {T.}~\bibnamefont {Sato}}, \bibinfo {author} {\bibfnamefont {T.}~\bibnamefont
  {Watanabe}}, \bibinfo {author} {\bibfnamefont {H.}~\bibnamefont {Obayashi}},
  \ and\ \bibinfo {author} {\bibfnamefont {K.}~\bibnamefont {Takayama}},\
  }\href {\doibase https://doi.org/10.1063/1.862153} {\bibfield  {journal}
  {\bibinfo  {journal} {Phys. Fluids}\ }\textbf {\bibinfo {volume} {21}},\
  \bibinfo {pages} {2076} (\bibinfo {year} {1978})}\BibitemShut {NoStop}%
\bibitem [{\citenamefont {Uehara}\ and\ \citenamefont
  {Hagiwara}(1978)}]{uehara1978radio}%
  \BibitemOpen
  \bibfield  {author} {\bibinfo {author} {\bibfnamefont {K.}~\bibnamefont
  {Uehara}}\ and\ \bibinfo {author} {\bibfnamefont {S.}~\bibnamefont
  {Hagiwara}},\ }\href {\doibase https://doi.org/10.1063/1.862182} {\bibfield
  {journal} {\bibinfo  {journal} {Phys. Fluids}\ }\textbf {\bibinfo {volume}
  {21}},\ \bibinfo {pages} {2316} (\bibinfo {year} {1978})}\BibitemShut
  {NoStop}%
\bibitem [{\citenamefont {Hiroe}\ \emph {et~al.}(1978)\citenamefont {Hiroe},
  \citenamefont {Sato}, \citenamefont {Watari}, \citenamefont {Hidekuma},
  \citenamefont {Kumazawa}, \citenamefont {Adati}, \citenamefont {Shoji},
  \citenamefont {Kawasaki}, \citenamefont {Miyahara}, \citenamefont {Akaishi},
  \citenamefont {Kubota}, \citenamefont {Watanabe},\ and\ \citenamefont
  {Miyake}}]{hiroe1978experiment}%
  \BibitemOpen
  \bibfield  {author} {\bibinfo {author} {\bibfnamefont {S.}~\bibnamefont
  {Hiroe}}, \bibinfo {author} {\bibfnamefont {T.}~\bibnamefont {Sato}},
  \bibinfo {author} {\bibfnamefont {T.}~\bibnamefont {Watari}}, \bibinfo
  {author} {\bibfnamefont {S.}~\bibnamefont {Hidekuma}}, \bibinfo {author}
  {\bibfnamefont {R.}~\bibnamefont {Kumazawa}}, \bibinfo {author}
  {\bibfnamefont {K.}~\bibnamefont {Adati}}, \bibinfo {author} {\bibfnamefont
  {T.}~\bibnamefont {Shoji}}, \bibinfo {author} {\bibfnamefont
  {S.}~\bibnamefont {Kawasaki}}, \bibinfo {author} {\bibfnamefont
  {A.}~\bibnamefont {Miyahara}}, \bibinfo {author} {\bibfnamefont
  {K.}~\bibnamefont {Akaishi}}, \bibinfo {author} {\bibfnamefont
  {Y.}~\bibnamefont {Kubota}}, \bibinfo {author} {\bibfnamefont
  {N.}~\bibnamefont {Watanabe}}, \ and\ \bibinfo {author} {\bibfnamefont
  {S.}~\bibnamefont {Miyake}},\ }\href {\doibase
  https://doi.org/10.1063/1.862276} {\bibfield  {journal} {\bibinfo  {journal}
  {Phys. Fluids}\ }\textbf {\bibinfo {volume} {21}},\ \bibinfo {pages} {676}
  (\bibinfo {year} {1978})}\BibitemShut {NoStop}%
\bibitem [{\citenamefont {Fader}\ \emph {et~al.}(1981)\citenamefont {Fader},
  \citenamefont {Jong}, \citenamefont {Stufflebeam},\ and\ \citenamefont
  {Sziklas}}]{fader1981rf}%
  \BibitemOpen
  \bibfield  {author} {\bibinfo {author} {\bibfnamefont {W.}~\bibnamefont
  {Fader}}, \bibinfo {author} {\bibfnamefont {R.}~\bibnamefont {Jong}},
  \bibinfo {author} {\bibfnamefont {J.}~\bibnamefont {Stufflebeam}}, \ and\
  \bibinfo {author} {\bibfnamefont {E.}~\bibnamefont {Sziklas}},\ }\href
  {\doibase https://doi.org/10.1103/PhysRevLett.46.999} {\bibfield  {journal}
  {\bibinfo  {journal} {Phys. Rev. Lett.}\ }\textbf {\bibinfo {volume} {46}},\
  \bibinfo {pages} {999} (\bibinfo {year} {1981})}\BibitemShut {NoStop}%
\bibitem [{\citenamefont {Fisch}, \citenamefont {Rax},\ and\ \citenamefont
  {Dodin}(2003)}]{fisch2003current}%
  \BibitemOpen
  \bibfield  {author} {\bibinfo {author} {\bibfnamefont {N.}~\bibnamefont
  {Fisch}}, \bibinfo {author} {\bibfnamefont {J.}~\bibnamefont {Rax}}, \ and\
  \bibinfo {author} {\bibfnamefont {I.}~\bibnamefont {Dodin}},\ }\href@noop {}
  {\bibfield  {journal} {\bibinfo  {journal} {Phys. rev. lett.}\ }\textbf
  {\bibinfo {volume} {91}},\ \bibinfo {pages} {205004} (\bibinfo {year}
  {2003})}\BibitemShut {NoStop}%
\bibitem [{\citenamefont {Dodin}, \citenamefont {Fisch},\ and\ \citenamefont
  {Rax}(2004)}]{dodin2004ponderomotive}%
  \BibitemOpen
  \bibfield  {author} {\bibinfo {author} {\bibfnamefont {I.~Y.}\ \bibnamefont
  {Dodin}}, \bibinfo {author} {\bibfnamefont {N.}~\bibnamefont {Fisch}}, \ and\
  \bibinfo {author} {\bibfnamefont {J.}~\bibnamefont {Rax}},\ }\href@noop {}
  {\bibfield  {journal} {\bibinfo  {journal} {Phys. plasmas}\ }\textbf
  {\bibinfo {volume} {11}},\ \bibinfo {pages} {5046} (\bibinfo {year}
  {2004})}\BibitemShut {NoStop}%
\bibitem [{\citenamefont {Dodin}\ and\ \citenamefont
  {Fisch}(2005)}]{dodin2005ponderomotive}%
  \BibitemOpen
  \bibfield  {author} {\bibinfo {author} {\bibfnamefont {I.}~\bibnamefont
  {Dodin}}\ and\ \bibinfo {author} {\bibfnamefont {N.}~\bibnamefont {Fisch}},\
  }\href@noop {} {\bibfield  {journal} {\bibinfo  {journal} {Phys. Rev. E}\
  }\textbf {\bibinfo {volume} {72}},\ \bibinfo {pages} {046602} (\bibinfo
  {year} {2005})}\BibitemShut {NoStop}%
\bibitem [{\citenamefont {Shi}, \citenamefont {Ren},\ and\ \citenamefont
  {Sun}(2019)}]{shi2019magnetic}%
  \BibitemOpen
  \bibfield  {author} {\bibinfo {author} {\bibfnamefont {P.}~\bibnamefont
  {Shi}}, \bibinfo {author} {\bibfnamefont {B.}~\bibnamefont {Ren}}, \ and\
  \bibinfo {author} {\bibfnamefont {X.}~\bibnamefont {Sun}},\ }\href@noop {}
  {\bibfield  {journal} {\bibinfo  {journal} {Phys. Plasmas}\ }\textbf
  {\bibinfo {volume} {26}},\ \bibinfo {pages} {102501} (\bibinfo {year}
  {2019})}\BibitemShut {NoStop}%
\bibitem [{\citenamefont {{Miller}}, \citenamefont {{Be'ery}},\ and\
  \citenamefont {{Barth}}(2021)}]{miller2021rate}%
  \BibitemOpen
  \bibfield  {author} {\bibinfo {author} {\bibfnamefont {T.}~\bibnamefont
  {{Miller}}}, \bibinfo {author} {\bibfnamefont {I.}~\bibnamefont {{Be'ery}}},
  \ and\ \bibinfo {author} {\bibfnamefont {I.}~\bibnamefont {{Barth}}},\ }\href
  {\doibase 10.1063/5.0056665} {\bibfield  {journal} {\bibinfo  {journal}
  {Phys. Plasmas}\ }\textbf {\bibinfo {volume} {28}},\ \bibinfo {eid} {112506}
  (\bibinfo {year} {2021})}\BibitemShut {NoStop}%
\bibitem [{\citenamefont {Lawson}(1957)}]{lawson1957some}%
  \BibitemOpen
  \bibfield  {author} {\bibinfo {author} {\bibfnamefont {J.~D.}\ \bibnamefont
  {Lawson}},\ }\href {\doibase 10.1088/0370-1301/70/1/303} {\bibfield
  {journal} {\bibinfo  {journal} {Proceedings of the physical society. Section
  B}\ }\textbf {\bibinfo {volume} {70}},\ \bibinfo {pages} {6} (\bibinfo {year}
  {1957})}\BibitemShut {NoStop}%
\bibitem [{\citenamefont {Wurzel}\ and\ \citenamefont
  {Hsu}(2022)}]{wurzel2022progress}%
  \BibitemOpen
  \bibfield  {author} {\bibinfo {author} {\bibfnamefont {S.~E.}\ \bibnamefont
  {Wurzel}}\ and\ \bibinfo {author} {\bibfnamefont {S.~C.}\ \bibnamefont
  {Hsu}},\ }\href@noop {} {\bibfield  {journal} {\bibinfo  {journal} {Phys.
  Plasmas}\ }\textbf {\bibinfo {volume} {29}},\ \bibinfo {pages} {062103}
  (\bibinfo {year} {2022})}\BibitemShut {NoStop}%
\bibitem [{\citenamefont {Guo}\ \emph {et~al.}(2005)\citenamefont {Guo},
  \citenamefont {Hoffman}, \citenamefont {Milroy}, \citenamefont {Miller},\
  and\ \citenamefont {Votroubek}}]{guo2005stabilization}%
  \BibitemOpen
  \bibfield  {author} {\bibinfo {author} {\bibfnamefont {H.}~\bibnamefont
  {Guo}}, \bibinfo {author} {\bibfnamefont {A.}~\bibnamefont {Hoffman}},
  \bibinfo {author} {\bibfnamefont {R.}~\bibnamefont {Milroy}}, \bibinfo
  {author} {\bibfnamefont {K.}~\bibnamefont {Miller}}, \ and\ \bibinfo {author}
  {\bibfnamefont {G.}~\bibnamefont {Votroubek}},\ }\href@noop {} {\bibfield
  {journal} {\bibinfo  {journal} {Phys. Rev. Lett.}\ }\textbf {\bibinfo
  {volume} {94}},\ \bibinfo {pages} {185001} (\bibinfo {year}
  {2005})}\BibitemShut {NoStop}%
\bibitem [{\citenamefont {Jhang}\ \emph {et~al.}(2005)\citenamefont {Jhang},
  \citenamefont {Kim}, \citenamefont {Lee}, \citenamefont {Bak},\ and\
  \citenamefont {Park}}]{jhang2005influence}%
  \BibitemOpen
  \bibfield  {author} {\bibinfo {author} {\bibfnamefont {H.}~\bibnamefont
  {Jhang}}, \bibinfo {author} {\bibfnamefont {S.}~\bibnamefont {Kim}}, \bibinfo
  {author} {\bibfnamefont {S.}~\bibnamefont {Lee}}, \bibinfo {author}
  {\bibfnamefont {J.}~\bibnamefont {Bak}}, \ and\ \bibinfo {author}
  {\bibfnamefont {B.}~\bibnamefont {Park}},\ }\href {\doibase
  10.1088/0029-5515/45/9/011} {\bibfield  {journal} {\bibinfo  {journal} {Nucl.
  Fusion}\ }\textbf {\bibinfo {volume} {45}},\ \bibinfo {pages} {1109}
  (\bibinfo {year} {2005})}\BibitemShut {NoStop}%
\bibitem [{\citenamefont {Kwon}\ \emph {et~al.}(2005)\citenamefont {Kwon},
  \citenamefont {Bak}, \citenamefont {Choh}, \citenamefont {Choi},
  \citenamefont {Choi}, \citenamefont {England}, \citenamefont {Hong},
  \citenamefont {Jhang}, \citenamefont {Kim}, \citenamefont {Kim},
  \citenamefont {Ko}, \citenamefont {Lee}, \citenamefont {Lee}, \citenamefont
  {Lho}, \citenamefont {Na}, \citenamefont {Park}, \citenamefont {Park},
  \citenamefont {Seo}, \citenamefont {Seo}, \citenamefont {Yeom}, \citenamefont
  {Yoo}, \citenamefont {You}, \citenamefont {Yoon},\ and\ \citenamefont
  {Yoon}}]{kwon2005progress}%
  \BibitemOpen
  \bibfield  {author} {\bibinfo {author} {\bibfnamefont {M.}~\bibnamefont
  {Kwon}}, \bibinfo {author} {\bibfnamefont {J.}~\bibnamefont {Bak}}, \bibinfo
  {author} {\bibfnamefont {K.}~\bibnamefont {Choh}}, \bibinfo {author}
  {\bibfnamefont {J.}~\bibnamefont {Choi}}, \bibinfo {author} {\bibfnamefont
  {J.}~\bibnamefont {Choi}}, \bibinfo {author} {\bibfnamefont {A.}~\bibnamefont
  {England}}, \bibinfo {author} {\bibfnamefont {J.}~\bibnamefont {Hong}},
  \bibinfo {author} {\bibfnamefont {H.}~\bibnamefont {Jhang}}, \bibinfo
  {author} {\bibfnamefont {J.}~\bibnamefont {Kim}}, \bibinfo {author}
  {\bibfnamefont {S.}~\bibnamefont {Kim}}, \bibinfo {author} {\bibfnamefont
  {D.}~\bibnamefont {Ko}, \bibfnamefont {WH~Lee}}, \bibinfo {author}
  {\bibfnamefont {S.}~\bibnamefont {Lee}, \bibfnamefont {JH~Lee}}, \bibinfo
  {author} {\bibfnamefont {H.}~\bibnamefont {Lee}}, \bibinfo {author}
  {\bibfnamefont {T.}~\bibnamefont {Lho}}, \bibinfo {author} {\bibfnamefont
  {H.}~\bibnamefont {Na}}, \bibinfo {author} {\bibfnamefont {B.}~\bibnamefont
  {Park}}, \bibinfo {author} {\bibfnamefont {M.}~\bibnamefont {Park}}, \bibinfo
  {author} {\bibfnamefont {D.}~\bibnamefont {Seo}}, \bibinfo {author}
  {\bibfnamefont {S.}~\bibnamefont {Seo}}, \bibinfo {author} {\bibfnamefont
  {J.}~\bibnamefont {Yeom}}, \bibinfo {author} {\bibfnamefont {S.}~\bibnamefont
  {Yoo}}, \bibinfo {author} {\bibfnamefont {K.}~\bibnamefont {You}}, \bibinfo
  {author} {\bibfnamefont {J.}~\bibnamefont {Yoon}}, \ and\ \bibinfo {author}
  {\bibfnamefont {S.}~\bibnamefont {Yoon}},\ }\href@noop {} {\bibfield
  {journal} {\bibinfo  {journal} {Fusion sci. Technol.}\ }\textbf {\bibinfo
  {volume} {47}},\ \bibinfo {pages} {17} (\bibinfo {year} {2005})}\BibitemShut
  {NoStop}%
\bibitem [{\citenamefont {Zhu}\ \emph {et~al.}(2019)\citenamefont {Zhu},
  \citenamefont {Shi}, \citenamefont {Yang}, \citenamefont {Zheng},
  \citenamefont {Luo}, \citenamefont {Ying},\ and\ \citenamefont
  {Sun}}]{zhu2019new}%
  \BibitemOpen
  \bibfield  {author} {\bibinfo {author} {\bibfnamefont {G.}~\bibnamefont
  {Zhu}}, \bibinfo {author} {\bibfnamefont {P.}~\bibnamefont {Shi}}, \bibinfo
  {author} {\bibfnamefont {Z.}~\bibnamefont {Yang}}, \bibinfo {author}
  {\bibfnamefont {J.}~\bibnamefont {Zheng}}, \bibinfo {author} {\bibfnamefont
  {M.}~\bibnamefont {Luo}}, \bibinfo {author} {\bibfnamefont {J.}~\bibnamefont
  {Ying}}, \ and\ \bibinfo {author} {\bibfnamefont {X.}~\bibnamefont {Sun}},\
  }\href@noop {} {\bibfield  {journal} {\bibinfo  {journal} {Phys. Plasmas}\
  }\textbf {\bibinfo {volume} {26}},\ \bibinfo {pages} {042107} (\bibinfo
  {year} {2019})}\BibitemShut {NoStop}%
\bibitem [{\citenamefont {Akimune}\ \emph {et~al.}(1981)\citenamefont
  {Akimune}, \citenamefont {Ikeda}, \citenamefont {Hirata},\ and\ \citenamefont
  {Okamoto}}]{akimune1981dynamic}%
  \BibitemOpen
  \bibfield  {author} {\bibinfo {author} {\bibfnamefont {H.}~\bibnamefont
  {Akimune}}, \bibinfo {author} {\bibfnamefont {I.}~\bibnamefont {Ikeda}},
  \bibinfo {author} {\bibfnamefont {T.}~\bibnamefont {Hirata}}, \ and\ \bibinfo
  {author} {\bibfnamefont {F.}~\bibnamefont {Okamoto}},\ }\href@noop {}
  {\bibfield  {journal} {\bibinfo  {journal} {J. of the Physical Society of
  Japan}\ }\textbf {\bibinfo {volume} {50}},\ \bibinfo {pages} {2729} (\bibinfo
  {year} {1981})}\BibitemShut {NoStop}%
\bibitem [{\citenamefont {Majeski}\ \emph {et~al.}(1987)\citenamefont
  {Majeski}, \citenamefont {Browning}, \citenamefont {Meassick}, \citenamefont
  {Hershkowitz}, \citenamefont {Intrator},\ and\ \citenamefont
  {Ferron}}]{majeski1987effect}%
  \BibitemOpen
  \bibfield  {author} {\bibinfo {author} {\bibfnamefont {R.}~\bibnamefont
  {Majeski}}, \bibinfo {author} {\bibfnamefont {J.}~\bibnamefont {Browning}},
  \bibinfo {author} {\bibfnamefont {S.}~\bibnamefont {Meassick}}, \bibinfo
  {author} {\bibfnamefont {N.}~\bibnamefont {Hershkowitz}}, \bibinfo {author}
  {\bibfnamefont {T.}~\bibnamefont {Intrator}}, \ and\ \bibinfo {author}
  {\bibfnamefont {J.}~\bibnamefont {Ferron}},\ }\href@noop {} {\bibfield
  {journal} {\bibinfo  {journal} {Phys. Rev. Lett.}\ }\textbf {\bibinfo
  {volume} {59}},\ \bibinfo {pages} {206} (\bibinfo {year} {1987})}\BibitemShut
  {NoStop}%
\bibitem [{\citenamefont {Yasaka}\ and\ \citenamefont
  {Itatani}(1984)}]{yasaka1984rf}%
  \BibitemOpen
  \bibfield  {author} {\bibinfo {author} {\bibfnamefont {Y.}~\bibnamefont
  {Yasaka}}\ and\ \bibinfo {author} {\bibfnamefont {R.}~\bibnamefont
  {Itatani}},\ }\href@noop {} {\bibfield  {journal} {\bibinfo  {journal} {Nucl.
  Fusion}\ }\textbf {\bibinfo {volume} {24}},\ \bibinfo {pages} {445} (\bibinfo
  {year} {1984})}\BibitemShut {NoStop}%
\bibitem [{\citenamefont {Yasaka}\ and\ \citenamefont
  {Itatani}(1985)}]{yasaka1985rf}%
  \BibitemOpen
  \bibfield  {author} {\bibinfo {author} {\bibfnamefont {Y.}~\bibnamefont
  {Yasaka}}\ and\ \bibinfo {author} {\bibfnamefont {R.}~\bibnamefont
  {Itatani}},\ }\href@noop {} {\bibfield  {journal} {\bibinfo  {journal} {Nucl.
  Fusion}\ }\textbf {\bibinfo {volume} {25}},\ \bibinfo {pages} {29} (\bibinfo
  {year} {1985})}\BibitemShut {NoStop}%
\bibitem [{\citenamefont {D'ippolito}, \citenamefont {Myra},\ and\
  \citenamefont {Francis}(1987)}]{dippolito1987rf}%
  \BibitemOpen
  \bibfield  {author} {\bibinfo {author} {\bibfnamefont {D.}~\bibnamefont
  {D'ippolito}}, \bibinfo {author} {\bibfnamefont {J.}~\bibnamefont {Myra}}, \
  and\ \bibinfo {author} {\bibfnamefont {G.}~\bibnamefont {Francis}},\
  }\href@noop {} {\bibfield  {journal} {\bibinfo  {journal} {Phys. Rev. Lett.}\
  }\textbf {\bibinfo {volume} {58}},\ \bibinfo {pages} {2216} (\bibinfo {year}
  {1987})}\BibitemShut {NoStop}%
\bibitem [{\citenamefont {Knowlton}, \citenamefont {Porkolab},\ and\
  \citenamefont {Takase}(1988)}]{knowlton1988stabilization}%
  \BibitemOpen
  \bibfield  {author} {\bibinfo {author} {\bibfnamefont {S.}~\bibnamefont
  {Knowlton}}, \bibinfo {author} {\bibfnamefont {M.}~\bibnamefont {Porkolab}},
  \ and\ \bibinfo {author} {\bibfnamefont {Y.}~\bibnamefont {Takase}},\
  }\href@noop {} {\bibfield  {journal} {\bibinfo  {journal} {Nucl. Fusion}\
  }\textbf {\bibinfo {volume} {28}},\ \bibinfo {pages} {99} (\bibinfo {year}
  {1988})}\BibitemShut {NoStop}%
\bibitem [{\citenamefont {Jin}, \citenamefont {Fisch},\ and\ \citenamefont
  {Reiman}(2020)}]{jin2020pulsed}%
  \BibitemOpen
  \bibfield  {author} {\bibinfo {author} {\bibfnamefont {S.}~\bibnamefont
  {Jin}}, \bibinfo {author} {\bibfnamefont {N.~J.}\ \bibnamefont {Fisch}}, \
  and\ \bibinfo {author} {\bibfnamefont {A.~H.}\ \bibnamefont {Reiman}},\
  }\href@noop {} {\bibfield  {journal} {\bibinfo  {journal} {Phys. Plasmas}\
  }\textbf {\bibinfo {volume} {27}},\ \bibinfo {pages} {062508} (\bibinfo
  {year} {2020})}\BibitemShut {NoStop}%
\bibitem [{\citenamefont {England}\ \emph {et~al.}(1989)\citenamefont
  {England}, \citenamefont {Eldridge}, \citenamefont {Knowlton}, \citenamefont
  {Porkolab},\ and\ \citenamefont {Wilson}}]{england1989power}%
  \BibitemOpen
  \bibfield  {author} {\bibinfo {author} {\bibfnamefont {A.}~\bibnamefont
  {England}}, \bibinfo {author} {\bibfnamefont {O.}~\bibnamefont {Eldridge}},
  \bibinfo {author} {\bibfnamefont {S.}~\bibnamefont {Knowlton}}, \bibinfo
  {author} {\bibfnamefont {M.}~\bibnamefont {Porkolab}}, \ and\ \bibinfo
  {author} {\bibfnamefont {J.}~\bibnamefont {Wilson}},\ }\href@noop {}
  {\bibfield  {journal} {\bibinfo  {journal} {Nucl. Fusion}\ }\textbf {\bibinfo
  {volume} {29}},\ \bibinfo {pages} {1527} (\bibinfo {year}
  {1989})}\BibitemShut {NoStop}%
\bibitem [{\citenamefont {D{\i}az}(2001)}]{diaz2001overview}%
  \BibitemOpen
  \bibfield  {author} {\bibinfo {author} {\bibfnamefont {F.~C.}\ \bibnamefont
  {D{\i}az}},\ }in\ \href@noop {} {\emph {\bibinfo {booktitle} {AIP Conference
  Proceedings}}},\ Vol.\ \bibinfo {volume} {595}\ (\bibinfo {organization}
  {American Institute of Physics},\ \bibinfo {year} {2001})\ pp.\ \bibinfo
  {pages} {3--15}\BibitemShut {NoStop}%
\bibitem [{\citenamefont {Li}\ and\ \citenamefont {Wan}(2011)}]{li2011recent}%
  \BibitemOpen
  \bibfield  {author} {\bibinfo {author} {\bibfnamefont {J.}~\bibnamefont
  {Li}}\ and\ \bibinfo {author} {\bibfnamefont {B.}~\bibnamefont {Wan}},\
  }\href@noop {} {\bibfield  {journal} {\bibinfo  {journal} {Nucl. Fusion}\
  }\textbf {\bibinfo {volume} {51}},\ \bibinfo {pages} {094007} (\bibinfo
  {year} {2011})}\BibitemShut {NoStop}%
\bibitem [{\citenamefont {Hooke}(1984)}]{hooke1984review}%
  \BibitemOpen
  \bibfield  {author} {\bibinfo {author} {\bibfnamefont {W.}~\bibnamefont
  {Hooke}},\ }\href@noop {} {\bibfield  {journal} {\bibinfo  {journal} {Plasma
  Phys. Controlled Fusion}\ }\textbf {\bibinfo {volume} {26}},\ \bibinfo
  {pages} {133} (\bibinfo {year} {1984})}\BibitemShut {NoStop}%
\bibitem [{\citenamefont {Fisch}(1987)}]{fisch1987theory}%
  \BibitemOpen
  \bibfield  {author} {\bibinfo {author} {\bibfnamefont {N.~J.}\ \bibnamefont
  {Fisch}},\ }\href@noop {} {\bibfield  {journal} {\bibinfo  {journal} {Rev.
  Mod. Phys.}\ }\textbf {\bibinfo {volume} {59}},\ \bibinfo {pages} {175}
  (\bibinfo {year} {1987})}\BibitemShut {NoStop}%
\bibitem [{\citenamefont {Malkin}, \citenamefont {Shvets},\ and\ \citenamefont
  {Fisch}(2000)}]{malkin2000detuned}%
  \BibitemOpen
  \bibfield  {author} {\bibinfo {author} {\bibfnamefont {V.}~\bibnamefont
  {Malkin}}, \bibinfo {author} {\bibfnamefont {G.}~\bibnamefont {Shvets}}, \
  and\ \bibinfo {author} {\bibfnamefont {N.}~\bibnamefont {Fisch}},\
  }\href@noop {} {\bibfield  {journal} {\bibinfo  {journal} {Phys. Rev. Lett.}\
  }\textbf {\bibinfo {volume} {84}},\ \bibinfo {pages} {1208} (\bibinfo {year}
  {2000})}\BibitemShut {NoStop}%
\bibitem [{\citenamefont {Light}\ and\ \citenamefont
  {Chen}(1995)}]{light1995helicon}%
  \BibitemOpen
  \bibfield  {author} {\bibinfo {author} {\bibfnamefont {M.}~\bibnamefont
  {Light}}\ and\ \bibinfo {author} {\bibfnamefont {F.~F.}\ \bibnamefont
  {Chen}},\ }\href@noop {} {\bibfield  {journal} {\bibinfo  {journal} {Phys.
  Plasmas}\ }\textbf {\bibinfo {volume} {2}},\ \bibinfo {pages} {1084}
  (\bibinfo {year} {1995})}\BibitemShut {NoStop}%
\bibitem [{\citenamefont {Miljak}\ and\ \citenamefont
  {Chen}(1998)}]{miljak1998helicon}%
  \BibitemOpen
  \bibfield  {author} {\bibinfo {author} {\bibfnamefont {D.~G.}\ \bibnamefont
  {Miljak}}\ and\ \bibinfo {author} {\bibfnamefont {F.~F.}\ \bibnamefont
  {Chen}},\ }\href@noop {} {\bibfield  {journal} {\bibinfo  {journal} {Plasma
  Sources Science and Technology}\ }\textbf {\bibinfo {volume} {7}},\ \bibinfo
  {pages} {61} (\bibinfo {year} {1998})}\BibitemShut {NoStop}%
\bibitem [{\citenamefont {He}\ \emph {et~al.}(2015)\citenamefont {He},
  \citenamefont {Sun}, \citenamefont {Liu},\ and\ \citenamefont
  {Qin}}]{he2015volume}%
  \BibitemOpen
  \bibfield  {author} {\bibinfo {author} {\bibfnamefont {Y.}~\bibnamefont
  {He}}, \bibinfo {author} {\bibfnamefont {Y.}~\bibnamefont {Sun}}, \bibinfo
  {author} {\bibfnamefont {J.}~\bibnamefont {Liu}}, \ and\ \bibinfo {author}
  {\bibfnamefont {H.}~\bibnamefont {Qin}},\ }\href {\doibase
  https://doi.org/10.1016/j.jcp.2014.10.032} {\bibfield  {journal} {\bibinfo
  {journal} {J. Comp. Phys.}\ }\textbf {\bibinfo {volume} {281}},\ \bibinfo
  {pages} {135} (\bibinfo {year} {2015})}\BibitemShut {NoStop}%
\bibitem [{\citenamefont {Fowler}, \citenamefont {Moir},\ and\ \citenamefont
  {Simonen}(2017)}]{fowler2017new}%
  \BibitemOpen
  \bibfield  {author} {\bibinfo {author} {\bibfnamefont {T.}~\bibnamefont
  {Fowler}}, \bibinfo {author} {\bibfnamefont {R.}~\bibnamefont {Moir}}, \ and\
  \bibinfo {author} {\bibfnamefont {T.}~\bibnamefont {Simonen}},\ }\href@noop
  {} {\bibfield  {journal} {\bibinfo  {journal} {Nucl. Fusion}\ }\textbf
  {\bibinfo {volume} {57}},\ \bibinfo {pages} {056014} (\bibinfo {year}
  {2017})}\BibitemShut {NoStop}%
\end{thebibliography}%
